\documentclass[11pt,a4paper]{article}
\usepackage{jheppub}
\usepackage{amsmath,amstext,amssymb}
\usepackage{graphicx}
\usepackage{feynmp}
\usepackage{tikz}
\usepackage{epstopdf}
\usetikzlibrary{matrix,shapes,arrows,calc}
\usepackage{caption}
\usepackage{subcaption}
\usepackage{float}
\usepackage{comment}
\usepackage{rotating}
\usepackage{verbatim}
\def\be{\begin{equation}}
\def\ee{\end{equation}}
\def\ba{\begin{eqnarray}}
\def\ea{\end{eqnarray}}

\usepackage{graphics}
\usepackage{graphicx}% Include figure files
\usepackage{dcolumn}% Align table columns on decimal point
\usepackage{bm}% bold math
\usepackage{epsfig}
\usepackage{graphicx}
\usepackage{multirow}
\usepackage{dcolumn}% Align table columns on decimal point
\usepackage{graphicx,epsfig}%

\begin{document}

\preprint{DESY 14-171}

\title{Analytic structure of the $n = 7$ scattering amplitude in $\mathcal{N}=4$ SYM theory in multi-Regge kinematics: Conformal Regge cut contribution}

\author[a]{Jochen Bartels,}%,\note{Corresponding author.}}
\author[a,b]{Andrey Kormilitzin,}
\author[a,c,d]{Lev N.Lipatov\footnote{This work has been supported by the grant RFBR 13-02-01246a}}%\note{Also at Some University.}}

\affiliation[a]{II. Institut f\"{u}r Theoretische Physik, Universit\"{a}t Hamburg, Luruper Chaussee 149,\\
D-22761 Hamburg, Germany}
%\affiliation[b]{Mathematical Institute, Andrew Wiles Building, Woodstock Road, Oxford, OX2 6CG, United Kingdom}
\affiliation[b]{Oxford-Man Institute, University of Oxford, Eagle House, Walton Well Road, Oxford, OX2 6ED, United Kingdom}
\affiliation[c]{Physics Department, St.Petersburg State University, Ulyanovskaya 3, St.Petersburg 198504,Russia}
\affiliation[c]{Theoretical Physics Department, Petersburg Nuclear Physics Institute, Orlova Roscha, Gatchina,
188300, St. Petersburg, Russia} 

% e-mail addresses: one for each author, in the same order as the authors
\emailAdd{jochen.bartels@desy.de}
\emailAdd{andrey.kormilitzin@desy.de}
\emailAdd{lipatov@mail.desy.de}

\date{\today}

\abstract
{In this second part  of our investigation \cite{Bartels:2013jna} of the analytic structure of the $2\to5$ scattering amplitude in the planar limit of $\mathcal{N}=4$ SYM in multi-Regge kinematics  we compute, in all kinematic regions, the Regge cut contributions in leading order. The results are infrared finite and conformally invariant.}

%\pacs{12.38.Cy, 12.38.Aw,12.40.Vv}

\maketitle

\section{Introduction}

It is now well established that the Bern-Dixon-Smirnow (BDS) conjecture \cite{Bern:2005iz} for the MHV n-point scattering amplitude in the planar limit of the $\mathcal{N}=4$ SYM theory is incomplete for $n \ge 6$. 
One of the first indications for this was found  in \cite{Bartels:2008ce,Bartels:2008sc} and in \cite{Alday:2007he}.  
Corrections to the BDS-formula have been named "remainder functions" $R_n$, and in recent years major efforts have been made for determining these remainder functions, in particular the remainder function $R_6$ for the case $n=6$. The function $R_6$ has been calculated for two, three loops \cite{Goncharov:2010jf,DelDuca:2009au,DelDuca:2010zg, Dixon:2011pw, Dixon:2011nj, Dixon:2012yy,Pennington:2012zj,Dixon:2013eka,Lipstein:2013xra,Golden:2013xva,DelDuca:2013lma,Dixon:2014xca,Dixon:2014iba}, four loops \cite{Dixon:2014voa} and even several attempts have been made for $n=7$ case up to two loops \cite{Golden:2014xqf,Golden:2014xqa,Basso:2014koa,Basso:2014nra}.

%Recently the group of Spradlin and Volovich have introduced a new method which allows to compute the $n=7$ remainder function at two-loop accuracy $R^{(2)}_7$ from cluster algebra approach \cite{Golden:2014xqf,Golden:2014xqa}.

When trying to go beyond this loop expansion, it has turned out to be useful to consider a special kinematic limit, in particular the multi-Regge limit. For the $n=6$ point amplitude the comparison of the BDS conjecture with the leading logarithmic approximation which extends over all orders of the coupling constant,  has shown that BDS formula fails in two major aspects:\\
1) the Regge pole contributions do not have the correct phase structure in all kinematic regions,\\
2) it does not contain the Regge cut contributions which are predicted by leading-log calculations.
Therefore, it is the remainder function which contains Regge cut contributions.  

A careful analysis has shown that this cut contribution vanishes both in the Euclidean region and in the physical region where all energies are positive. It is nonzero only in special kinematic regions, named "Mandelstam regions": these are physical regions where some of the energy variables are positive, others negative ("mixed regions": the precise definition will be given later on). These results have been generalized also beyond the leading logarithmic approximation, and there is no doubt that the multi-Regge limit plays a key role for the determination of the remainder functions.

To construct the remainder function in the multi-Regge limit it is therefore necessary to consider all 
possible kinematic regions and to find the correct structure of Regge pole and Regge cut contributions. The first step 
is the analysis of the Regge pole contributions. It is well known that in non-abelian gauge theories the gauge bosons reggeize, and in the leading approximation the $2 \to n+1$ production amplitudes can be written in a simple factorizing form with exchange of reggeized gluons in all $t$-channels. Beyond the leading approximation this factorizing form of the Regge pole contribution remains valid in the region of all energies being positive, but the production vertices become complex-valued functions.
%in agreement with the results of Regge theory derived from dual models  \cite{Weis:1972ir,Weis:1972tn,Brower:1974yv} or scalar theories \cite{Drummond:1969ft}. 
This factorizing representation is equivalent to another representation, in which the scattering amplitude is written as a sum of $k_n$ different terms\footnote{The numbers $k_n$ coincide with the Catalan numbers
$C_n$ with $C_n=1,1,2,5,14,42,...$ for $n=0,1,2,3,4,5,...$. They satisfy the recurrence relation $C_{n+1} = \sum_{i=0}^{i=n}C_iC_{n-i}$.}, where each of them has a distinct set of non-vanishing simultaneous energy discontinuities: in this representation the agreement with the Steinmann relations is explicit. 

When applying these results to the planar amplitudes of $\mathcal{N}=4$ SYM theory, an important difference between 
planar and fully signatured amplitudes was discovered \cite{Bartels:2008ce,Bartels:2008sc,Lipatov:2010qf}. Namely,  the simple factorized form of the Regge pole contributions is valid in the physical region with all energy variables being positive (and also in the Euclidean region), but it takes a different form in other regions, in particular in the Mandelstam regions mentioned before. In the latter region the Regge pole contribution has a term which contains an unphysical singularity and requires the existence of Regge cut contributions with the same phase structure. In the sum, the singular terms contained in the Regge poles and in the Regge cut contributions cancel, leading to a sum of IR finite and conformal invariant pole and cut contributions.   
    
In a recent paper \cite{Bartels:2013jna}  we have started a systematic study of these Regge pole and Regge cut contributions. We found it instructive to first return to the 6-point case, and then developed tools which allow us to extend to higher order scattering amplitudes, in particular to the 7-point amplitude.  As the first step we have analyzed the Regge pole contribution. Particular attention has been given to the appearance of unphysical pole singularities, and we have outlined, for the $2\to 5$ scattering amplitudes, that these pole singularities have to be canceled by Regge cut contributions. As a result, we have found that, in all kinematic regions, the scattering amplitude  can be written as a sum of conformal invariant Regge pole contributions and Regge cut amplitudes (a brief summary is presented in Appendix B). Whereas our construction was designed to find explicit conformal expressions for the Regge pole contributions (valid to all orders in the coupling constant), we did not determine the explicit expressions of the Regge cut contribution. It is the purpose of the present paper, to complete our program by computing the Regge cut contribution. To this end we have to develop a slightly different strategy which allows to 
compute, from energy discontinuities, Regge cut contributions. At present we will restrict ourselves to the weak coupling limit, but a NLO calculation is within reach. Again, our main focus is in the 7-point amplitude. The extension to the 8-point case is under the way.        

It may be useful to make a few preparatory remarks on our tools. Our calculations will make use of the analytic structure of scattering amplitudes in multi-Regge kinematics, and we will compute, via unitarity integrals, energy discontinuities. To be a bit more specific, we first write the scattering amplitude as a sum of several terms: for the six point case we have five terms, for the 7-point amplitude 14 terms, for the 8-point cases 42 terms and so on. Each term is written as a multiple Sommerfeld-Watson integral, where the integrand consists of a product of complex energy factors and a real-valued coefficient function, the partial wave, which depends upon the angular momentum variables, the squared momentum transfers and the Toller angles. The phase structure is contained in the energy factors only. The partial waves are written as sums of the Regge contributions, Regge poles and Regge cuts. Whereas the pole contributions have been analyzed in our previous paper, the focus of this paper will be on the Regge cut singularities: we will compute them from energy discontinuities, i.e. our calculations will boil down to unitarity integrals. This is the point where, at present,  we restrict ourselves to the weak coupling approximation, since 
inside the unitarity integrals we will insert the leading-log expressions of the scattering amplitudes. 

An important ingredient into this construction is the observation that Regge pole and Regge cut contributions come with products of trigonometric factors which have to be determined before the energy discontinuities can be addressed.
The origin of these trigonometric factors is the factorization of the Regge pole contribution which, in the case of 
planar scattering amplitudes, leads to the appearance of unphysical pole singularities. As we had discussed already in our previous paper, these singularities must cancel in the scattering amplitude, i.e. in  the sum of the partial waves contributions. This is the place where the existence of Regge cuts becomes mandatory. As an important part of our calculations we will find a systematic way of computing these trigonometric factors.            

There exists an extensive literature on Regge theory, mainly on Regge poles \cite{Brower:1974yv,White:1976qm}. One of the key concepts is the introduction of signature: in order to define proper analytic continuation in angular momentum plane, one has to define combinations of amplitudes which are even or odd under crossing. Many general results in Regge theory (e.g. signature conservation rules) cannot be considered without signature. In the context of AdS/CFT duality we consider the limit of large $N_c$ and are thus led to planar amplitudes to which signature does not apply. {\it A priori}, therefore, it is not clear to what extent results from the literature can be used\footnote{We thank A.White for a helpful discussion on this point}. Nevertheless, in our calculations we will adopt results of Regge theory, and we have to view them as assumptions: their validity has to be justified by the results. The key features which we consider as "proof of consistency" are:\\ 
1) agreement with perturbation theory, wherever results on multiparticle scattering amplitudes are available.\\   
2) after removing those IR-singular pieces which are part of the BDS formula, the remainder function has to be IR finite and\\ 
3) conformal invariant. 
For the $2 \to 4$ and $2 \to 5$ scattering amplitudes our construction has been completed and satisfies the constraints, and for the $2 \to 6$ scattering amplitude results will be published soon.

Our paper will be organized as follows. We begin (Sec. II) with  the $2 \to 5$ scattering amplitude, define our ansatz (the sum of 14 terms), and we list the trigonometric factors for the Regge pole contributions. We then discuss these factors for the Regge cut contributions and formulate rules which can be used also for higher point amplitudes.
In order to illustrate our strategy of using energy discontinuities we make a digression (Sec. IV) and complete the construction of Regge cuts in the $2\to 4$ case. In Sec. V we return to the $2 \to 5$ case and calculate, via energy discontinuities, the Regge cut contribution. Finally, in Sec. VI we list our predictions for the scattering amplitude in different kinematic regions. A few details of our calculations of the $2\to 5$ scattering amplitude are presented in Appendix A and a table, and a brief summary of the results of our previous paper is given in Appendix B.

\section{Analytic structure and trigonometric coefficients}
We begin with the analytic structure of the 7-point amplitude. In multi-Regge kinematics the scattering amplitude can be written as a sum of 14 terms which we will name "analytic decomposition":
\be
T= \sum T_{ijk},
\label{decomp}
\ee
where each subscript "i,j,k" is related to a production vertex and takes the values $L$ (left) or $R$ (right). In the planar approximation for the $2\to5$ amplitude each term belongs to a maximal set of nonoverlapping energy discontinuities \footnote{for signatured amplitudes there exist additional nonplanar contributions
\cite{White:1990ch}. Some of them can be derived from configurations which are planar in a crossed channel.}: 
\begin{figure}[H]
\centering
\includegraphics[scale=0.8]{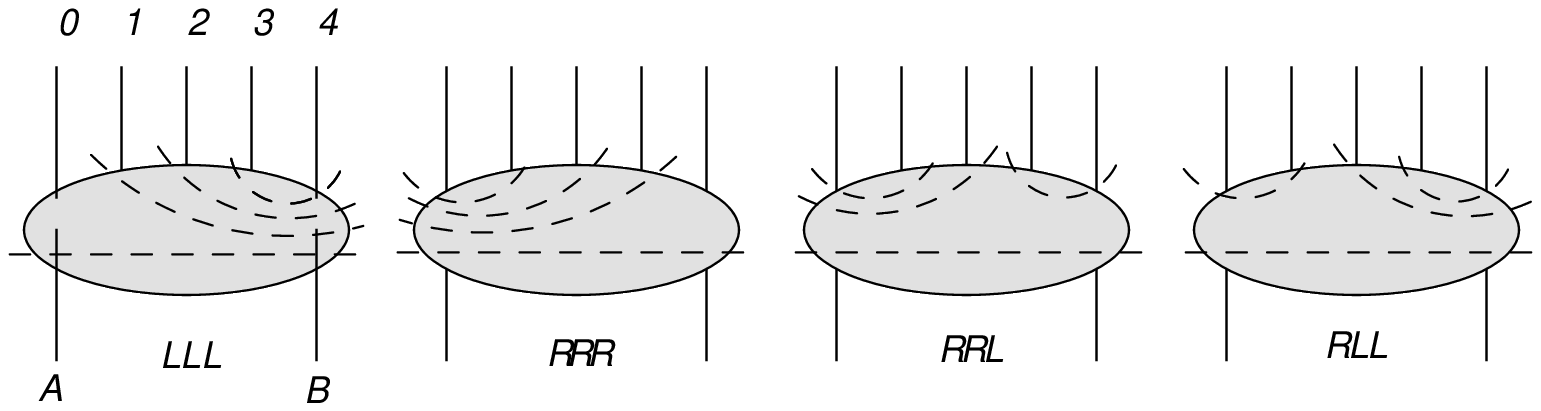}
\caption{Terms without Regge cuts. For the produced particles we also use the labels $a,b,c$.}
\label{f3_to_w3}
\end{figure}
\begin{figure}[H]
\centering
\includegraphics[scale=0.8]{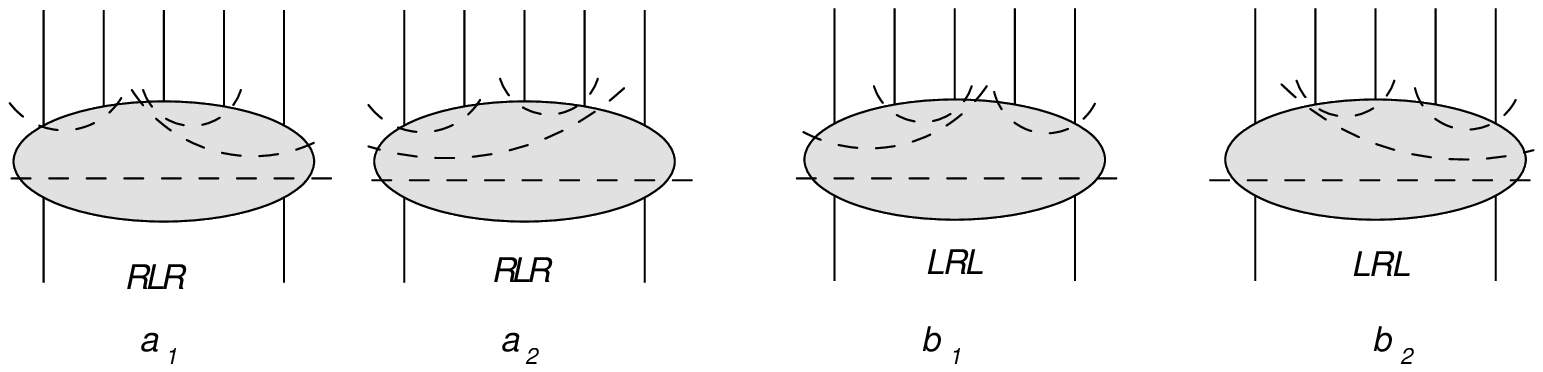}\\
\caption{Terms which contain Regge cut contributions: two doublets (a) and (b)}
\label{f3_to_w3doublets}
\end{figure}
\begin{figure}[H]
\centering
\includegraphics[scale=0.8]{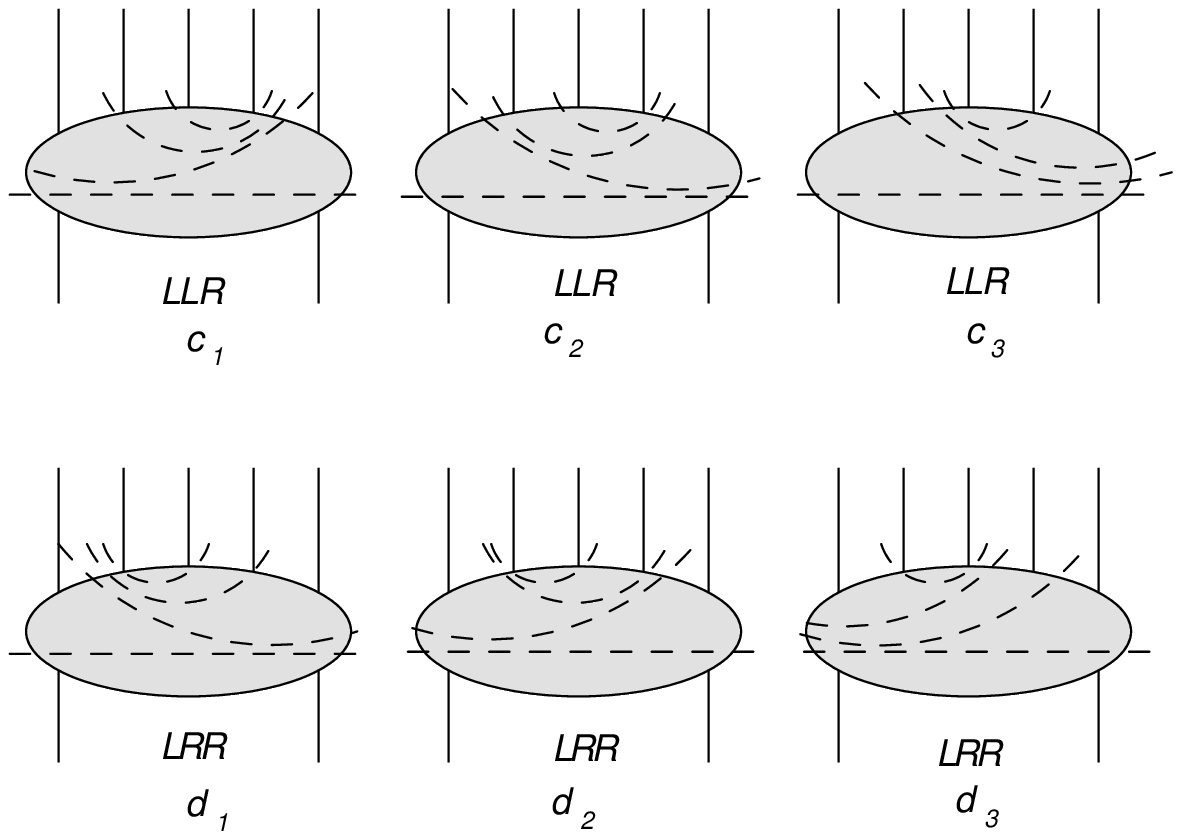}
\caption{Terms which contain Regge cut contributions: two triplets (c) and (d)}
\label{f3_to_w3triplets}
\end{figure}
\noindent
We write each term as a multiple Sommerfeld-Watson integral, where the integrand consists of a product of energy factors and of a real-valued partial wave which depends upon momentum transfers $t_i=-\vec{q}_i^2$ and angular momenta $\omega_i=j_i-1$ and contains the singularities in the angular momentum plane. As an example, the first term reads as follows:
\ba
T_{LLL} = s \int \int \int \int \frac{d \omega'_1 d \omega'_2 d \omega_3'd \omega'_4} {(2\pi i)^4} (-s_{34})^{\omega'_{43}} 
(-s_{234})^{\omega'_{32}}(-s_{1234})^{\omega'_{21}}(-s)^{\omega'_{1}} \times
\nonumber\\\times F_{LLL}(t_1,t_2,t_3,t_4;\omega'_1,\omega'_2,\omega'_3,\omega'_4) .
\label{struct-LLL}
\ea
We denote these partial waves by $F_{ijk}$. As we have said already, the subscripts take the values $R$ or $L$, and their origin is discussed in Appendix A. Each partial wave may consist of several contributions which contain Regge pole or Regge cut singularities:
\be
F_{ijk} = F_{ijk}^{pole} + F_{ijk}^{\text{Regge cut 1}}+ F_{ijk}^{\text{Regge  cut 2}} + ... \,.
\ee 
In particular, all 14 terms contain a Regge pole piece. A Regge cut in the $t_3$ channel is contained in all those terms which contain the discontinuity in $s_3$ ($RLR$ and $LLR$), and a Regge cut in  the $t_2$-channel in the terms with a nonvanishing discontinuity in $s_2$ ($LRL$ and  $LRR$). Finally, the long Regge cut extending over the $t_2$ and $t_3$ channels is contained in the first two terms of the triplets  $LLR$ and $LRR$: they all have 
the discontinuity in $s_{123}$.   

From the decomposition in (\ref{decomp}) we derive the scattering amplitudes in different kinematic regions. Following the notations introduced in \cite{Bartels:2013jna}, we will label the different kinematic regions by products $\tau_i\tau_j...$. Each factor $\tau_i$ stands for a "twist" of the corresponding $t_i$-channel state and takes us into a "crossed" channel.  For example, the configuration $\tau_1\tau_4$ has twists in the $t_1$ and $t_4$-channels and denotes the kinematic region where the three produced particles have become "incoming particles". Further examples can be found in \cite{Bartels:2013jna}. The choice of the kinematic region determines the phases of the energy factors after their analytic continuation. Each of the 14 terms, therefore, comes with a certain phase, and in their sum cancellations may occur. Prominent examples are the region where all squared energies are positive (each positive energy $s_i$ comes with a phase  $e^{-i\pi}$) and the Euclidean region where all energies are negative (each negative energy has a factor $1$). In both regions, all terms containing Regge cuts sum up to zero, and only Regge pole contributions remain.   

To understand the existence of Regge cuts it is necessary to say a few words about the connection between 
the decomposition (\ref{decomp}) and Feynman diagrams. In multi-Regge kinematics, the sum of relevant Feynman amplitudes for a $2 \to n+1$ multiparticle production process can be decomposed according to the analytic structure, and it can  be written as a sum of multiple dispersion integrals in the energy variables; in Regge theory these dispersion integrals can be used to define Froissart-Gribov partial wave projections which contain the Regge singularities. This leads to the decomposition (\ref{decomp}). However, the existence of Regge cuts in the scattering amplitude can most easily be understood if we go back to the Feynman amplitudes (i.e. prior to the decomposition (\ref{decomp})). As an example, let us return to the Regge cut in the planar $2 \to 4$ amplitude (Fig.\ref{Mandelstam}):
\begin{figure}[H]
\centering
\includegraphics[scale=0.8]{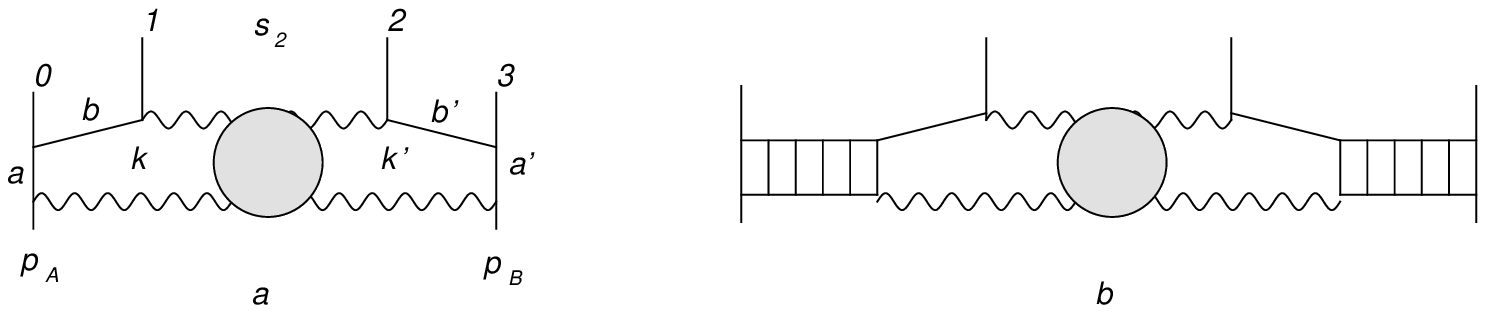}
\caption{Mandelstam criterion for the Regge cut in the $2\to4$ scattering amplitude (wavy lines denote reggeons, straight lines scalar particles). \\
(a) the simplest diagram illustrating the Mandelstam criterion\\ (b) a generalization (enhanced diagram) in which the propagators $a$ and $a'$ are replaced by sets of ladder diagrams (reggeons).}
\label{Mandelstam}
\end{figure}
\noindent   
Let us consider the kinematic region where all energies are positive. It is well-known \cite{Mandelstam:1963cw} that,
as a rule,  Regge cuts cancel in planar diagrams. In Fig.\ref{Mandelstam}a this cancellation is easily seen \cite{Lipatov:2010qf}: introducing Sudakov variables $k=\beta p_A + \alpha p_B + k_{\perp}$ and concentrating on the $\alpha$-integral of the $k$ loop momentum in the left hand part of the diagram, the two singularities coming from the poles of line "a" and "b" lie on the same side of the integration contour and thus lead to a vanishing integral. However, when analytically continuing into the kinematic region $s,s_2>0$, $s_1,s_{012},s_3,s_{123}<0$ the poles of the lines b and b' move and on the other sides of the integration contours and the Regge cut remains. In fact, the particles "1" and "2" are in the initial state, and the diagram becomes "physically nonplanar". This leads to the following Mandelstam condition:  in order to have a nonvanishing Regge cut contribution, one needs, at both ends of the two reggeon cut, nonplanar $\alpha$ ($\beta$) integrals. For the $2\to4$ production amplitude this is achieved by analytically continuing in $s_1$ and $s_3$, i.e. by twisting the $t_1$ and $t_3$ channels. This Mandelstam condition can easily be applied to more general $2 \to n$ scattering processes with $n>4$. 

Returning to the decomposition (\ref{decomp}), we have already stated that Feynman diagrams in the multi-Regge limit may contribute to several terms in this decomposition. In each term, the content of Regge singularity of the partial waves $F_{ijk}$ is independent of the kinematic region.  The vanishing of a Regge cut contribution in {\it Feynman diagrams}  (so-called Amati-Fubini-Stanghellini (AFS) cancellation), in the decomposition (\ref{decomp}) therefore translates into a cancellation between different terms. Applying the Mandelstam criterion to the kinematic region of all energies being positive or negative, we immediately see that Regge cuts must cancel for all $2\to n$ processes.  In contrast, there exist Mandelstam regions ("mixed" regions) where some energies are positive, others negative. As an example, for $2\to5$, we have the Mandelstam region  $s_1=s_{01}, s_{012}, s_{0123}, s_4=s_{45}, s_{345}, s_{2345} <0; s_{2},s_{3},s_{234},s>0$ (in our notation: $\tau_1\tau_4$). In this region, a Regge cut contribution extending over the $t_2$ and $t_3$ channels exists. 

Further details on the decomposition are presented in Appendix A. Here we list the energy factors which determine the phases of the scattering amplitudes:  
\begin{eqnarray}
LLL: \hspace{0.5cm}  &\left(-s_{4}\right)^{\omega_{43}} \left(-s_{345}\right)^{\omega_{32}}\left(-s_{2345}\right)^{\omega_{21}}\left(-s\right)^{\omega_1}
\label{phases-LLL}\\ 
RRR: \hspace{0.5cm}&\left(-s_{1}\right)^{\omega_{12}} \left(-s_{012}\right)^{\omega_{23}}\left(-s_{0123}\right)^{\omega_{34}}\left(-s\right)^{\omega_4} \label{phases-RRR}
\\
RRL: \hspace{0.5cm}& \left(-s_{4}\right)^{\omega_{43}} \left(-s_{1}\right)^{\omega_{12}}\left(-s_{012}\right)^{\omega_{23}}\left(-s\right)^{\omega_3}\label{phases-RRL}\\
RLL: \hspace{0.5cm}&\left(-s_{1}\right)^{\omega_{12}} \left(-s_{4}\right)^{\omega_{43}}\left(-s_{345}\right)^{\omega_{32}}\left(-s\right)^{\omega_2}\label{phases-RLL}.
\end{eqnarray}
Next the doublets:
\begin{eqnarray}
\label{phases-RLR}
RLR: \hspace{0.5cm}  a_1\;&=&\;\left(-s_{1}\right)^{\omega_{12}} \left(-s_{3}\right)^{\omega_{34}}\left(-s_{234}\right)^{\omega_{42}}\left(-s\right)^{\omega_2}\\ 
a_2\;&=&\;\left(-s_{1}\right)^{\omega_{12}} \left(-s_{3}\right)^{\omega_{32}}\left(-s_{0123}\right)^{\omega_{24}}\left(-s\right)^{\omega_4} \nonumber
\end{eqnarray}
and
\begin{eqnarray}
\label{phases-LRL}
LRL: \hspace{0.5cm}b_1\;&=&\;\left(-s_{2}\right)^{\omega_{21}} \left(-s_{012}\right)^{\omega_{13}}\left(-s_{4}\right)^{\omega_{43}}\left(-s\right)^{\omega_3}\\ 
b_2\;&=&\;\left(-s_{2}\right)^{\omega_{23}} \left(-s_{4}\right)^{\omega_{43}}\left(-s_{1234}\right)^{\omega_{31}}\left(-s\right)^{\omega_1}.\nonumber
\end{eqnarray}
Similarly for the two triplets:
\begin{eqnarray}
\label{phases-LLR}
LLR: \hspace{0.5cm}c_1\;&=&\;\left(-s_{3}\right)^{\omega_{32}} \left(-s_{123}\right)^{\omega_{21}}\left(-s_{0123}\right)^{\omega_{14}}\left(-s\right)^{\omega_4}\\ \nonumber
c_2\;&=&\;\left(-s_{3}\right)^{\omega_{32}} \left(-s_{123}\right)^{\omega_{24}}\left(-s_{1234}\right)^{\omega_{41}}\left(-s\right)^{\omega_1}\\ \nonumber
c_3\;&=&\;\left(-s_{3}\right)^{\omega_{34}} \left(-s_{234}\right)^{\omega_{42}}\left(-s_{1234}\right)^{\omega_{21}}\left(-s\right)^{\omega_1} \nonumber
\end{eqnarray}
and 
\begin{eqnarray}\label{phases-LRR}
LRR: \hspace{0.5cm}d_1\;&=&\;\left(-s_{2}\right)^{\omega_{23}} \left(-s_{123}\right)^{\omega_{34}}\left(-s_{1234}\right)^{\omega_{41}}\left(-s\right)^{\omega_1}\\ \nonumber
d_2\;&=&\;\left(-s_{2}\right)^{\omega_{23}} \left(-s_{123}\right)^{\omega_{31}}\left(-s_{0123}\right)^{\omega_{14}}\left(-s\right)^{\omega_4}\\ \nonumber
d_3\;&=&\;\left(-s_{2}\right)^{\omega_{21}} \left(-s_{012}\right)^{\omega_{13}}\left(-s_{0123}\right)^{\omega_{34}}\left(-s\right)^{\omega_4}. \nonumber
\end{eqnarray}
It should be noted that in these expressions, for simplicity, we have disregarded $\kappa$ factors as well as energy scales. Details are described in \cite{Bartels:2013jna}. Depending on  the kinematic regions, these energy factors lead to different phases. A complete list of phases  in the different kinematic regions is presented in the Appendix B. 

Let us now discuss the form of the partial waves. Regge pole contributions are contained in all partial waves, whereas Regge cuts can be contained only in those partial waves which have nonvanishing energy discontinuities along the Regge cut. In detail, partial waves which have a cut in the energy $s_2$ ($s_3$) are expected to have a short Regge cut in the $t_2$-($t_3$) channel: $LRL$ and $LRR$  ($RLR$ and $LLR$). The long Regge cut in the $\omega_2$ and $\omega_3$ channel can contribute only to the partial waves with a nonvanishing energy discontinuity in the $s_{123}$: $LLR$ and $LRR$.
    
Let us go through the partial waves. Simplest are the Regge pole contributions. In most of the partial waves, these Regge pole contributions contain trigonometric factors which are closely related to Regge factorization. For the case of signatured $2\to4$ scattering amplitudes, it has been shown in  \cite{Brower:1974yv} that the property of Regge factorization and the analytic decomposition (\ref{decomp}) are compatible only if the Regge poles contain special combinations of trigonometric factors. One of our tasks is to generalize this and to find the corresponding factors for the case $2 \to 5$ 
(and for higher $n>5$). This will be done in  Appendix A where we formulate general rules for computing these factors.  

One of the peculiar features of these factors is that, in certain kinematic regions, they contain unphysical singularities of the type $\sim 1/\sin \pi \omega_2$ which should not be present in scattering amplitudes (and certainly do not appear in perturbation theory). In our previous paper  \cite{Bartels:2013jna} we have discussed these singular terms in detail: starting from the Regge-factorized form of the $2 \to n+1$ scattering amplitude we have calculated the Regge pole contributions to scattering amplitude in all different kinematic regions. In particular,  for the $2 \to 4$, and for the $2 \to 5$ cases we presented a full list of these singular terms and of the kinematic regions where they appear.  Here it is important to note the difference between signatured and planar amplitudes. In planar amplitudes, these singular terms coming from the Regge poles have to cancel against Regge cut contributions: in fact, in  \cite{Bartels:2013jna} we have already derived the phase structure of the Regge cut which allows to absorb and cancel these singular terms. In the present  paper we will complete this discussion by computing the full Regge cut amplitudes. This leads to the conclusion that, for the planar amplitudes, the existence of Regge cut contributions is necessary for obtaining scattering amplitudes which are free from  unphysical singularities.   

For signatured amplitudes the situation is slightly different. In order to obtain signatured amplitudes we form even or odd combinations of different kinematic regions. As an example, we return to the simplest case of the $2 \to 4$ amplitude, and list the two regions with singular terms: 
\ba
\tau_1\tau_3:&  &   e^{-i\pi\omega_2}\left[e^{i\pi\left(\omega_a+\omega_b\right)} - 2ie^{i\pi\omega_2}\frac{\Omega_a \Omega_b}{\Omega_2} \right]\nonumber\\
&=&e^{-i\pi\omega_2}\left[ \cos \pi \omega_{ab} + i \sin \pi( \omega_a+\omega_b) - 2i \frac{ \cos\pi \omega_{2} 
\Omega_a \Omega_b}{\Omega_2} \right] \\
\tau_1\tau_2\tau_3:& & -\left[e^{-i\pi\left(\omega_a+\omega_b\right)} + 2ie^{-i\pi\omega_2}\frac{\Omega_a \Omega_b}{\Omega_2} \right]\nonumber\\
&=& - \left[ \cos \pi \omega_{ab} - i \sin \pi( \omega_a+\omega_b) + 2i \frac{ \cos\pi \omega_{2} \Omega_a \Omega_b}{\Omega_2} \right].
\ea  
In the signatured amplitude the singular terms appear in the following combination:
\be
- 2i \tau_1\tau_3 \left( e^{-i\pi\omega_2} + \tau_2\right) \frac{ \cos\pi \omega_{2}  \Omega_a \Omega_b}{\Omega_2} ,
\ee
i.e. the singularities cancel for odd signature $\tau_2=-1$, and there is no need for a Regge cut in the $t_2$ channel.  At the same time, because of signature conservation, the signatured amplitude cannot contain in the $t_2$ channel a Regge cut composed of two (odd signatured) gluons. The situation will be different for even signature $\tau_2=+1$: the singularity in the Regge pole is present, and signature conservation admits the two reggeon cut.      

Let us now address the trigonometric factors for the $2\to 5$ scattering amplitude. With the notations 
\be
\Omega_i = \sin \pi \omega_i,\;\;
\Omega_{ij} = \sin \pi (\omega_i -\omega_j),\;\; \omega_{ij}=\omega_i-\omega_j
\ee
and 
\be
\omega_i= -\frac{\gamma_K}{4} \ln \frac{|q_i|^2}{\lambda^2}, \hspace{0.5cm}
\omega_a= - \frac{\gamma_K}{8} \ln \frac{|q_1|^2 |q_2|^2}{|q_1-q_2|^2 \lambda^2}, \hspace{0.5cm} 
\omega_b= - \frac{\gamma_K}{8} \ln \frac{|q_2|^2 |q_3|^2}{|q_2-q_3|^2 \lambda^2}
\ee
the results for the first four partial waves are\footnote{In the following it will be 
understood that our expressions for the partial waves have to be multiplied with the Born
amplitudes and with the Regge pole propagators, e.g.  $1/  (\omega'_1 -\omega_1)$ etc.
The Born amplitude carries a factor $s$: its sign will be included when we 
presents results for Regge pole and Regge cut contributions for the different kinamatic regions.} :
\be
F_{LLL}^{pole}=\frac{V_L(a)}{\Omega_{21}}\frac{V_L(b)}{\Omega_{32}}
\frac{V_L(c)}{\Omega_{43}}
\label{eq:poleLLL}
\ee
\be
F_{RRR}^{pole}=\frac{V_R(a)}{\Omega_{12}}\frac{V_R(b)}{\Omega_{23}}
\frac{V_R(c)}{\Omega_{34}}
\label{eq:poleRRR}
\ee
\be
F_{RRL}^{pole}=\frac{V_R(a)}{\Omega_{12}}\frac{V_R(b)}{\Omega_{23}}
\frac{V_L(c)}{\Omega_{43}}
\label{eq:poleRRL}
\ee
\be
F_{RLL}^{pole}=\frac{V_R(a)}{\Omega_{12}}\frac{V_L(b)}{\Omega_{32}}
\frac{V_L(c)}{\Omega_{43}}.
\label{eq:poleRLL}
\ee  
Here the vertex functions are given by:
\ba 
V_R(a) &=& \sin \pi(\omega_1 - \omega_a) = \Omega_{1a}\nonumber\\
V_L(a) &=& \sin \pi(\omega_2 - \omega_a) = \Omega_{2a}.
\ea
Next we consider the two doublets which contain  Regge poles and cuts. We write:
\be 
F_{RLR(1)}=F_{RLR(1)}^{pole} + F_{RLR(1)}^{\omega_3-cut}. 
\ee
\be 
F_{RLR(2)}=F_{RLR(2)}^{pole} + F_{RLR(2)}^{\omega_3-cut}. 
\ee
and
\be 
F_{LRL(1)}=F_{LRL(1)}^{pole} + F_{LRL(1)}^{\omega_2-cut}. 
\ee
\be 
F_{LRL(2)}=F_{LRL(2)}^{pole} + F_{LRL(2)}^{\omega_2-cut}. 
\ee
The pole and cut terms differ by their singularities in the angular momentum planes $\omega_1$,  $\omega_2$,
and $\omega_3$. However, later on we will see that the cut pieces will contain subtractions related to the 
Regge pole terms. For the pole terms we have the trigonometric factors (see Appendix A):
\be
\label{RLR(1)-pole}
F_{RLR(1)}^{pole}=
\frac{\Omega_{4}}{\Omega_3}\frac{\Omega_{32}}{\Omega_{42}}\;
\frac{V_R(a)}{\Omega_{12}}\frac{V_L(b)}{\Omega_{32}}
\frac{V_R(c)}{\Omega_{34}}
\ee
\be
F_{RLR(2)}^{pole}=
\frac{\Omega_{2}}{\Omega_3}\frac{\Omega_{34}}{\Omega_{24}}\; 
\frac{V_R(a)}{\Omega_{12}}
\frac{V_L(b)}{\Omega_{32}}
\frac{V_R(c)}{\Omega_{34}}
\ee
and
\be
F_{LRL(1)}^{pole}=
\frac{\Omega_{1}}{\Omega_2}\frac{\Omega_{23}}{\Omega_{13}}\;
\frac{V_L(a)}{\Omega_{21}}\frac{V_R(b)}{\Omega_{23}}
\frac{V_L(c)}{\Omega_{43}}.
\ee
\be         
F_{LRL(2)}^{pole}= \frac{\Omega_{3}}{\Omega_2}\frac{\Omega_{21}}{\Omega_{31}}\; 
\frac{V_L(a)}{\Omega_{21}}\frac{V_R(b)}{\Omega_{23}}
\frac{V_L(c)}{\Omega_{43}}.
\ee
Next we have to find the trigonometric factors of the $\omega_3$-cut contribution. We first observe that, in the region where all energies are positive, the two partial waves $F_{RLR(1)}$ and $F_{RLR(2)}$ come with the same phase (Appendix A). The absence of the Regge cuts implies that they must be opposite equal. We make the ansatz:
\ba
F_{RLR(1)}^{ \omega_3-cut}= \frac{V_R(a)}{\Omega_{12}} \frac{W_{\omega_3;RLR}}{\Omega_{42}} \nonumber\\
F_{RLR(2)}^{ \omega_3-cut}=\frac{V_R(a)}{\Omega_{12}} \frac{W_{\omega_3;RLR}}{\Omega_{24}}. 
\label{RLR-cut}
\ea
The form of the first factor, $\frac{V_R(a)}{\Omega_{21}}$ follows from the requirement of Regge factorization, whereas 
the existence of the denominator $1/\Omega_{24}$ can be deduced from a study of the kinematic region $\tau_2\tau_4$.
Namely, in this region the Regge cut is expected to be present,  and the amplitude has to be free from unphysical singularities.
From the energy factors of $F_{RLR(1)}$ and  $F_{RLR(2)}$ we have the phases  
$e^{-i\pi(\omega_3+\omega_1-\omega_2)} e^{-i\pi(\omega_2-\omega_4)} $ and $e^{-i\pi(\omega_3+\omega_1-\omega_2)} e^{-i\pi(\omega_4-\omega_2)}$, i.e. in the difference we find a factor $2i \sin \pi(\omega_2-\omega_4)$
which just cancels this denominator.  
With a similar argument for the partial waves $F_{LRL(1)}$ and $F_{LRL(2)}$ we put:
\ba
F_{LRL(1)}^{ \omega_2-cut}= \frac{W_{\omega_2;LRL}}{\Omega_{13}} \frac{V_L(c)}{\Omega_{43}} \nonumber\\
F_{LRL(2)}^{ \omega_2-cut}= \frac{W_{\omega_2;LRL}}{\Omega_{31}} \frac{V_L(c)}{\Omega_{43}}.
\ea
Finally we turn to the triplets which contain Regge poles and two types of cuts, a "short" one and a "long" one. In detail:
\ba
F_{LLR(1)} &=&  F_{LLR(1)}^{pole} + F_{LLR(1)}^{\omega_3-cut} + F_{LLR(1)}^{\omega_2- \omega_3-cut}
\nonumber\\
F_{LLR(2)} &=&  F_{LLR(2)}^{pole} + F_{LLR(2)}^{\omega_3-cut} + F_{LLR(2)}^{\omega_2- \omega_3-cut}
\nonumber\\
F_{LLR(3)} &=&  F_{LLR(3)}^{pole} + F_{LLR(3)}^{\omega_3-cut}.
\ea
Similarly
\ba
F_{LRR(1)} &=&  F_{LRR(1)}^{pole} + F_{LRR(1)}^{\omega_2-cut} + F_{LRR(1)}^{\omega_2- \omega_3-cut}
\nonumber\\
F_{LRR(2)} &=&  F_{LRR(2)}^{pole} + F_{LRR(2)}^{\omega_2-cut} + F_{LRR(2)}^{\omega_2- \omega_3-cut}
\nonumber\\
F_{LRR(3)} &=&  F_{LRR(3)}^{pole} + F_{LRR(3)}^{\omega_2-cut}.
\ea
Again, the pole and cut terms differ by their singularities in the complex angular momentum planes.  The Regge pole terms are (Appendix A):

\ba
F_{LLR(1)}^{pole}&=&\frac{\Omega_{1}}{\Omega_{3}}\frac{\Omega_{34}}{\Omega_{14}} \qquad \frac{V_L(a)}{\Omega_{21}}\frac{V_L(b)}{\Omega_{32}}\frac{V_R(c)}{\Omega_{34}}\\
F_{LLR(2)}^{pole}&=&\frac{\Omega_{4}}{\Omega_{3}}\frac{\Omega_{34} \Omega_{21}}{\Omega_{24}\Omega_{41}} \;\; \frac{V_L(a)}{\Omega_{21}} \frac{V_L(b)}{\Omega_{32}}\frac{V_R(c)}{\Omega_{34}},\\
F_{LLR(3)}^{pole}&=&\frac{\Omega_{4}}{\Omega_{3}}\frac{\Omega_{32}}{\Omega_{42}} \qquad \frac{V_L(a)}{\Omega_{21}}\frac{V_L(b)}{\Omega_{32}}\frac{V_R(c)}{\Omega_{34}}.
\ea
For the second triplet:
\ba
F_{LRR(1)}^{pole}&=&\frac{\Omega_{4}}{\Omega_{2}}\frac{\Omega_{21}}{\Omega_{41}} \qquad \frac{V_L(a)}{\Omega_{21}}\frac{V_R(b)}{\Omega_{23}}\frac{V_R(c)}{\Omega_{34}},\\
F_{LRR(2)}^{pole}&=&\frac{\Omega_1}{\Omega_{2}}\frac{\Omega_{34}\Omega_{21}} {\Omega_{31}\Omega_{14}}\;\; \frac{V_L(a)}{\Omega_{21}} \frac{V_R(b)}{\Omega_{23}} \frac{V_R(c)}{\Omega_{34}},\\
F_{LRR(3)}^{pole}&=&\frac{\Omega_{1}}{\Omega_{2}} \frac{\Omega_{23}}{\Omega_{13}}\qquad \frac{V_L(2)}{\Omega_{21}}\; \frac{V_R(3)} {\Omega_{23}}\frac{V_R(4)}{\Omega_{34}}.
\ea
%
\begin{comment}
\be
F_{LLR(1)}^{pole}=\frac{\Omega_{1}}{\Omega_{3}}\frac{\Omega_{34}}{\Omega_{14}}\;
\frac{V_L(a)}{\Omega_{21}}\frac{V_L(b)}{\Omega_{32}}\frac{V_R(c)}{\Omega_{34}}
\ee
\be
F_{LLR(2)}^{pole}=\frac{\Omega_{4}}{\Omega_{3}}\frac{\Omega_{34} \Omega_{21}}{\Omega_{24}\Omega_{41}}\;
\frac{V_L(a)}{\Omega_{21}} \frac{V_L(b)}{\Omega_{32}}\frac{V_R(c)}{\Omega_{34}},
\ee
\be
F_{LLR(3)}^{pole}=\frac{\Omega_{4}}{\Omega_{3}}\frac{\Omega_{32}}{\Omega_{42}}\;
\frac{V_L(a)}{\Omega_{21}} \frac{V_L(b)}{\Omega_{32}}\frac{V_R(c)}{\Omega_{34}}.
\ee
For the second triplet:
\be
F_{LRR(1)}^{pole}=\frac{\Omega_{4}}{\Omega_{2}}\frac{\Omega_{21}}{\Omega_{41}}\;
\frac{V_L(a)}{\Omega_{21}}\frac{V_R(b)}{\Omega_{23}}\frac{V_R(c)}{\Omega_{34}}
\ee
\be
F_{LRR(2)}^{pole}=\frac{\Omega_1}{\Omega_{2}}\frac{\Omega_{34}\Omega_{21}} {\Omega_{31}\Omega_{14}}\;
\frac{V_L(a)}{\Omega_{21}} \frac{V_R(b)}{\Omega_{23}} \frac{V_R(c)}{\Omega_{34}}
\ee
\be
F_{LRR(3)}^{pole}=\frac{\Omega_{1}}{\Omega_{2}} \frac{\Omega_{23}}{\Omega_{13}} \frac{V_L(2)}{\Omega_{21}}\; \frac{V_R(3)} {\Omega_{23}}\frac{V_R(4)}{\Omega_{34}}.
\ee
\end{comment}
For the short cut in the $\omega_3$-channel  we observe that, for positive energies, the absence of the 
Regge cut requires the cancellation of the three partial waves $F_{LLR(1)}$, $F_{LLR(2)}$, and $F_{LLR(3)}$. Therefore, in the ansatz:
\ba
F_{LLR(1)}^{\omega_3-cut}&&= x_1
\frac{V_L(a)}{\Omega_{21}} \frac{W_{\omega_3;LLR}}{\Omega_{24}}\nonumber\\
F_{LLR(2)}^{\omega_3-cut}&&= x_2
 \frac{V_L(a)}{\Omega_{21}} \frac{W_{\omega_3;LLR}}{\Omega_{24}}\nonumber\\
F_{LLR(3)}^{\omega_3-cut}&&=  x_3\frac{V_L(a)}{\Omega_{21}} \frac{W_{\omega_3;LLR}}{\Omega_{24}}
%\label{LLR-cut}
\ea
the sum of the coefficents $x_i$ must be zero:
\be
x_1+x_2+x_3=0.
\ee
In order to obtain more information for the $x_i$, we compute the contribution of the $\omega_3$ -cut to the 
scattering amplitude. Namely, in the region $\tau_2\tau_4$ where this Regge cut is expected to be present,
we have:
\ba
F_{RLR(1)}^{\omega_3-cut} + F_{RLR(2)}^{\omega_3-cut}   &\rightarrow& 2i e^{-i\pi(\omega_1+\omega_3)} e^{i\pi \omega_2} \frac{V_R(a)}{\Omega_{12}} W_{\omega_3;RLR}
\ea   
and 
\ba
F_{LLR(1)}^{\omega_3-cut} + F_{LLR(2)}^{\omega_3-cut}+ F_{LLR(3)}^{\omega_3-cut}  &\rightarrow &  
-2i x_3 \frac{V_L(a)}{\Omega_{21}} e^{-i\pi \omega_3} W_{\omega_3;LLR},
\ea
where we have used $x_1+x_2=-x_3$. Taking the sum of the last two equations and observing the Regge factorization formula
\be
\frac{V_R(a)}{\Omega_{12}} e^{i\pi \omega_2}+ \frac{V_L(a)}{\Omega_{21}} e^{i\pi \omega_1} = e^{i\pi \omega_a},
\label{id-1}
\ee
we are led to the identifications
\be
x_3=-1
\ee
and 
\be
W_{\omega_3;RLR} = W_{\omega_3;LRL} = W_{\omega_3}.
\ee
The result for the sum of all five terms  then becomes:   
\be
RLR(1) + RLR(2) + LLR(1) + LLR(2)+ LLR(3) = 2i e^{-i\pi(\omega_1+\omega_3)} e^{i\pi \omega_a} W_{\omega_3}.
%\label{cut3tau24}
\ee  
Next we consider the region $\tau_1\tau_2\tau_4$.  We find:
\ba
\tau_1 \tau_2 \tau_4: & - e^{-i\pi \omega_3}  \frac{V_L(a)}{\Omega_{21}\Omega_{24}} 
\left( x_1 e^{-i\pi (\omega_1-\omega_4-\omega_2)} +x_2  e^{-i\pi (\omega_4- \omega_1-\omega_2 )} +x_3  e^{-i\pi (\omega_2- \omega_1-\omega_4)} \right)  W_{\omega_3}\nonumber\\
&=-2i e^{-i\pi \omega_3} V_L(a) \left( x_1 \frac{e^{i\pi \omega_4}}{\Omega_{24}} +x_2 
\frac{e^{i\pi \omega_1}}{\Omega_{21}}\right)W_{\omega_3}.
\ea
The coefficients $x_1$ and $x_2$ must be chosen to cancel the unphysical singularities $\sim 1/\Omega_{21}$, $\sim 1/\Omega_{24}$; furthermore, they must  satisfy $x_1+x_2=-x_3=1$. The solution of these conditions is 
\ba
 x_1&=&\frac{\Omega_1}{\Omega_2} \frac{\Omega_{24}}{\Omega_{14}}\\
 x_2&=&\frac{\Omega_4}{\Omega_2} \frac{\Omega_{21}}{\Omega_{41}}.
 \ea
With these findings the trigonometric factors of the $\omega_3$-cut become:   
\ba
F_{LLR(1)}^{\omega_3-cut}&&= \frac{\Omega_1}{\Omega_2} \frac{\Omega_{24}}{\Omega_{14}}
\frac{V_L(a)}{\Omega_{21}} \frac{W_{\omega_3}}{\Omega_{24}}\nonumber\\
F_{LLR(2)}^{\omega_3-cut}&&= \frac{\Omega_4}{\Omega_2} \frac{\Omega_{21}}{\Omega_{41}}
 \frac{V_L(a)}{\Omega_{21}} \frac{W_{\omega_3}}{\Omega_{24}}\nonumber\\
F_{LLR(3)}^{\omega_3-cut}&&= \hspace{1.2cm} \frac{V_L(a)}{\Omega_{21}} \frac{W_{\omega_3}}{\Omega_{42}}.
\label{LLR-cut}
\ea
An analogous argument applies to the short cut in the $\omega_2$-channel, and the trigonometric factors become:
\ba
F_{LRR(1)}^{\omega_2-cut}&&= \frac{\Omega_4}{\Omega_3} \frac{\Omega_{31}}{\Omega_{41}}
 \frac{W_{\omega_2}}{\Omega_{31}}\frac{V_R(c)}{\Omega_{34}}\nonumber\\
F_{LRR(2)}^{\omega_2-cut}&&= \frac{\Omega_1}{\Omega_3} \frac{\Omega_{34}}{\Omega_{14}}
  \frac{W_{\omega_2}}{\Omega_{31}}\frac{V_R(c)}{\Omega_{34}}\nonumber\\
F_{LRR(3)}^{\omega_2-cut}&&= \hspace{1.2cm}  \frac{W_{\omega_2}}{\Omega_{13}}\frac{V_R(c)}{\Omega_{34}}.
\label{LRR-cut}
\ea
Finally, the long cut term has the form: 
\ba
F_{LLR(1)}^{\omega_2- \omega_3-cut} &=& \frac{W_{\omega_2\omega_3;L}}{\Omega_{32} \Omega_{14}}\nonumber\\
F_{LLR(2)}^{\omega_2- \omega_3-cut} &=& \frac{W_{\omega_2\omega_3;L}}{\Omega_{32} \Omega_{41}}
\ea
and
\ba
F_{LRR(1)}^{\omega_2- \omega_3-cut} &=& \frac{W_{\omega_2\omega_3;R}}{\Omega_{23} \Omega_{41}}\nonumber\\
F_{LRR(2)}^{\omega_2- \omega_3-cut} &=& \frac{W_{\omega_2\omega_3;R}}{\Omega_{23} \Omega_{14}}.
\ea
As discussed before, the absence of this Regge cut in the kinematic region where all energies are positive requires 
the cancellation of  the two partial waves $F_{LLR(1)}^{\omega_2- \omega_3-cut}$, $F_{LLR(2)}^{\omega_2- \omega_3-cut}$; the same argument applies to  $F_{LRR(1)}^{\omega_2- \omega_3-cut}$ and  $F_{LRR(2)}^{\omega_2- \omega_3-cut}$.

This completes our derivation of the trigonometric factors for Regge poles and Regge cuts of all 14 terms. Our construction of the trigonometric factors for the Regge cuts has followed the line of arguments given  in our previous paper\cite{Bartels:2013jna}: we required that in all kinematic regions the scattering amplitudes satisfy Regge factorization and are free from unphysical singularities. In the Appendix A we make use of these results and formulate rules for the Regge cuts which generalize those of the Regge poles. These rules can also be used for the $2\to 6$ amplitude. At present we do not know how to "derive" these rules; as we have said before, the justification will come from the IR finiteness and conformal invariance of our final results.  

We conclude this section with a few comments on the Regge pole contributions. First, in our previous paper \cite{Bartels:2013jna} our discussion of Regge pole contributions has started from the factorizing expression. This representation is equivalent to the decomposition (\ref{decomp}). To illustrate this,  we go into the region of positive energies. With the identity
\be
\frac{\Omega_{4}}{\Omega_{3}} \frac{\Omega_{32}}{\Omega_{42}} +
\frac{\Omega_{2}}{\Omega_{3}} \frac{\Omega_{34}}{\Omega_{24}} =1
\ee
it is easy to see that the sum of two pole terms,  $F_{RLR(1)}^{pole}$ and $F_{RLR(2)}^{pole}$, can be written as: 
\be
F_{RLR(1)}^{pole} + F_{RLR(2)}^{pole} \rightarrow
\frac{V_R(a)}{\Omega_{12}}
\frac{V_L(b)}{\Omega_{32}} \frac{V_R(c )}{\Omega_{34}},
\label{eq:poleRLR}
\ee
where the arrow indicates that we have multiplied the partial waves with their phases. Similarly,
\be
F_{LRL(1)}^{pole} + F_{LRL(2)}^{pole}\rightarrow
\frac{V_L(a)}{\Omega_{21}}
\frac{V_R(b)}{\Omega_{23}} \frac{V_L(c)}{\Omega_{43}}.
\label{eq:poleLRL}
\ee
For the triplets we need the identities:
\be
\frac{\Omega_{4}}{\Omega_{3}} \frac{\Omega_{21}\Omega_{34}}
{\Omega_{24}\Omega_{41}} 
+ \frac{\Omega_{1}}{\Omega_{3}} \frac{\Omega_{34}}{\Omega_{14}} 
+ \frac{\Omega_{4}}{\Omega_{3}} \frac{\Omega_{32}}{\Omega_{42}}
=1
\ee
and
\be
\frac{\Omega_{1}}{\Omega_{2}} \frac{\Omega_{21}\Omega_{34}}
{\Omega_{31}\Omega_{14}} 
+ \frac{\Omega_{1}}{\Omega_{2}} \frac{\Omega_{23}}{\Omega_{13}} 
+ \frac{\Omega_{4}}{\Omega_{2}} \frac{\Omega_{21}}{\Omega_{41}}
=1
\ee
and  obtain
\be
F_{LLR(1)}^{pole}+F_{LLR(2)}^{pole}+F_{LLR(3)}^{pole} \rightarrow  \frac{V_L(a)}{\Omega_{21}}\frac{V_L(b)}{\Omega_{32}}
\frac{V_R(c)}{\Omega_{34}}.
\label{eq:poleLLR}
\ee 
and
\be
F_{LRR(1)}^{pole} + F_{LRR(2)}^{pole}+ F_{LRR(3)}^{pole} \rightarrow
\frac{V_L(a)}{\Omega_{21}}\frac{V_R(b)}{\Omega_{23}} \frac{V_R(c)}{\Omega_{34}}.
\label{eq:poleLRR}
\ee
When combining the results of all 14 partial waves, it is convenient to use identities such as (\ref{id-1}). 
In this way one obtains, for the sum of all 14 terms, the factorizing expression: 
\be
\frac{T^{pole}}{ \Gamma(t_1) |s_1|^{\omega_1}  |s_2|^{\omega_2}|s_3|^{\omega_3}|s_4|^{\omega_4} \Gamma(t_4)}
= e^{i\pi (\omega _a + \omega_b+\omega_c)} e^{-i\pi(\omega_1+\omega_2+\omega_3+\omega_4)}.
\ee
As far as the other kinematic regions are concerned, it is possible - but much more tedious - to perform similar calculations for the other kinematic regions and to arrive at the same results as those listed in \cite{Bartels:2013jna}. 

\section{A digression: the $2\to4$ scattering amplitude}
To illustrate our future strategy we briefly return to the well-studied case of the $2 \to 4$ scattering \cite{Bartels:2013jna,Lipatov:2010qf}.
We begin with the ansatz consisting of five terms. We write:
\be
T_{2 \to 4} = T_{LL}+T_{RR}+T_{RL}+T_{LR(1)} +T_{LR(2)}.
\label{five-terms}
\ee
Each term has an energy factor which - depending on the kinematic region - determines the phase: 
\ba
LL: \hspace{0.5cm} & (-s_3)^{\omega_{32}}(-s_{123})^{\omega_{21}}(-s)^{\omega_{1}} \\
RR: \hspace{0.5cm} &(-s_1)^{\omega_{12}}(-s_{012})^{\omega_{23}}(-s)^{\omega_{3}}\\
RL: \hspace{0.5cm} &(-s_1)^{\omega_{12}}(-s_{3})^{\omega_{32}}(-s)^{\omega_{2}}\\
LR(1): \hspace{0.5cm} &(-s_2)^{\omega_{21}}(-s_{012})^{\omega_{13}}(-s)^{\omega_{3}}\nonumber\\
LR(2): \hspace{0.5cm} &(-s_2)^{\omega_{23}}(-s_{123})^{\omega_{31}}(-s)^{\omega_{1}}.
\label{2to4phases}
\ea
The first three terms have Regge poles only. For the last two partial waves we write a sum of 
Regge pole and Regge cut contributions. We have from \cite{Bartels:2008ce}:
\ba
F_{LL}^{pole} &=& \frac{V_L(a)}{\Omega_{21}} \frac{V_L(b)}{\Omega_{32}}\\
F_{RR}^{pole} &=& \frac{V_R(a)}{\Omega_{12}} \frac{V_R(b)}{\Omega_{23}}\\
F_{RL}^{pole} &=& \frac{V_R(a)}{\Omega_{12}} \frac{V_L(b)}{\Omega_{32}}\\
F_{LR(1)} &=& F_{LR(1)}^{pole} +F_{LR(1)}^{cut} 
\nonumber\\
&=& \frac{V_L(a)}{\Omega_{21}} \frac{V_R(b)}{\Omega_{23}}\;\frac{\Omega_1}{\Omega_2} \frac{\Omega_{23}}{\Omega_{13}} + \frac{W_{\omega_2}}{\Omega_{13}}
\label{2to4-partialwavesLR1}
\\
F_{LR(2)} &=& F_{LR(2)}^{pole} +F_{LR(2)}^{cut}\nonumber\\
&=&\frac{V_L(a)}{\Omega_{21}} \frac{V_R(b)}{\Omega_{23}}\; \frac{\Omega_3}{\Omega_2} \frac{\Omega_{21}}{\Omega_{31}}+ \frac{W_{\omega_2}}{\Omega_{31}}.
\label{2to4-partialwavesLR2}
\ea    
In the next step we describe the derivation of the function $W_{\omega_2}$. To this end, we consider single-energy discontinuities\footnote{We define $disc_x f(x) = \frac{1}{2i} \left( f(x+i \epsilon) - f(x-i\epsilon)\right)$; $disc_s (-s)^{\omega} = - |s|^{\omega} \sin \pi \omega = - |s|^{\omega}\Omega$.}  of the full scattering amplitude. It is important to observe that, when calculating discontinuities,  we have to take into account all five terms in (\ref{five-terms}). Furthermore,  in each term there may be different Regge contributions,  Regge poles and Regge cuts. The former ones are known, and they contain singular terms  $\sim 1/\Omega_2$. We will find that these singular pieces in the Regge poles  will  be "inherited" also by  the Regge cut contributions. Only at the end, when the full scattering amplitude is computed, we will show that these singularities completely cancel in all kinematic regions.
  
First we consider, in the kinematic region of positive energies, the discontinuity in $s_2$ which is contained only in the two partial waves $F_{LR}$. From the Regge pole we obtain:
\be
 - e^{-i(\omega_1 + \omega_3)} \frac{V_L(a) V_R(b)}{\Omega_2},
\ee 
whereas the contribution of the Regge cut reads:
\be
  e^{-i(\omega_1 + \omega_3)} e^{i\pi \omega_2} W_{\omega_2}.
\ee     
The discontinuity of the full scattering amplitude  therefore becomes:
\be
 disc_{12}\; T_{2 \to 4} = ... \Delta_{12},\hspace{1cm}  \Delta_{12} =   e^{-i(\omega_1 + \omega_3)} \left(-  \frac{V_L(a) V_R(b)}{\Omega_2} + e^{i\pi \omega_2} W_{\omega_2} \right),
%\label{6point-s12-pos-energies}
\ee  
where the {\bf dots} indicate that we have left out the integration symbols and the energy factors. Here the important result is the singular term $\sim 1/\Omega_2$: since the lhs is computed from the unitarity integral and has no singularity, the Regge cut amplitude $W_{\omega_2}$ on the lhs must contain a singular term with cancels the singularity from the Regge pole.   

So far our results are valid to all orders in the coupling constant. The energy discontinuities have to be calculated from unitarity integrals, and at this stage the restriction in accuracy enters. In \cite{Bartels:2008sc} the discontinuities in $s_{12}$ has  been calculated in the leading approximation. Restricting ourselves to this approximation we can neglect the phases and obtain:
\ba
W_{\omega_2}  &&=   \Delta_{12}   + \pi  \frac{\omega_{2a} \omega_{2b}}{\omega_2} \nonumber\\
&&=  \Delta_{12}   + \pi \left( \omega_2 -\omega_a - \omega_b + \frac{\omega_a \omega_b}{\omega_2}
\right).
 \label{6point-s12-pos-energies}
\ea  
This is not yet our final result. In \cite{Bartels:2008sc} we have discussed  that the RPRR vertex consists
of a "local" and a "nonlocal" piece (Fig.\ref{RPRR-vertex}a). The former piece does not couple to a Regge cut. It satisfies the bootstrap condition in the $t_2$-channel: in leading order this condition implies that the vertex does not depend upon the momenta $k$ and $q_2-k$ separately, but only upon the sum $q_2$. As a result, when multiplying the production with the BFKL  color octet Green's function, the local term reggeizes, whereas the second one leads to the reggeon cut (Fig.\ref{RPRR-vertex}b):
\begin{figure}[H]
\centering 
\epsfig{file=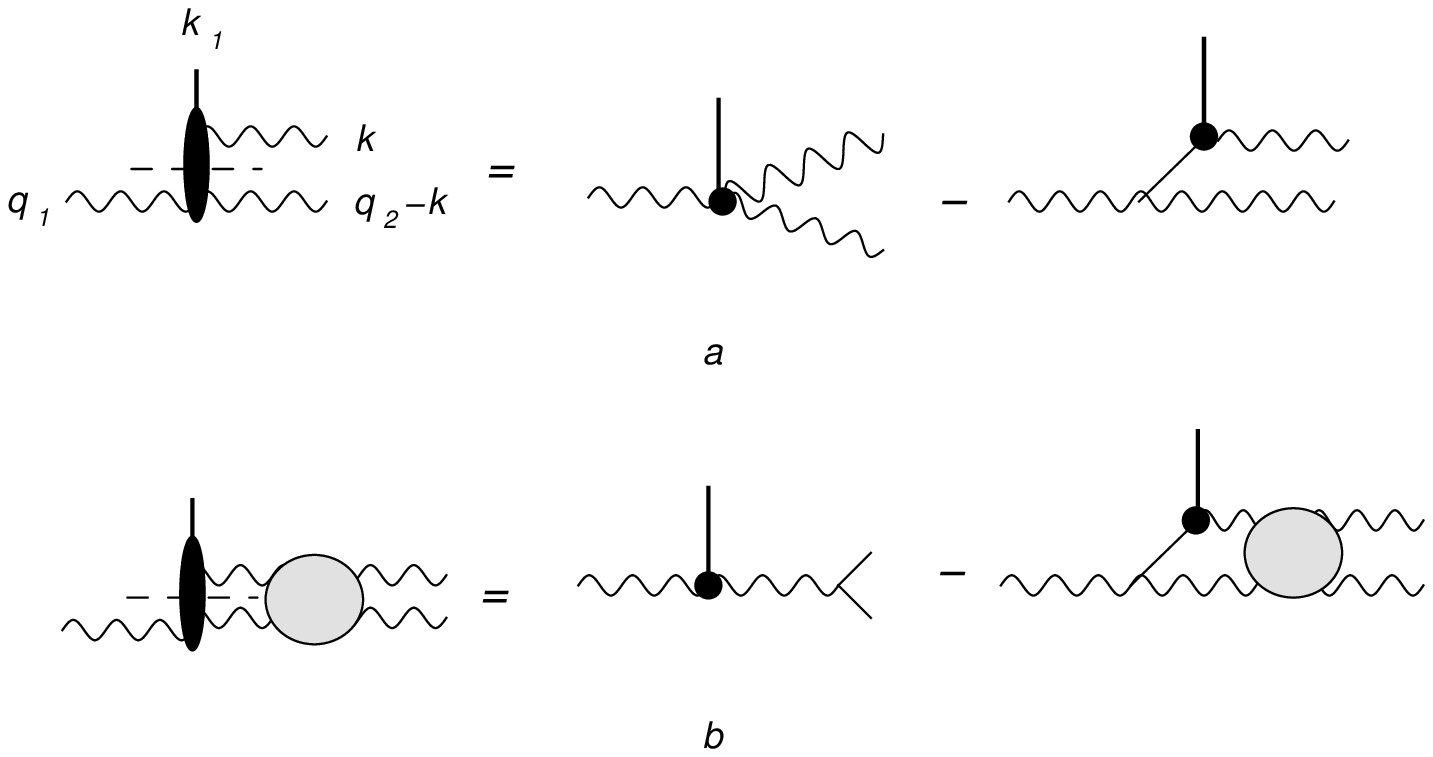,scale=0.8}
\caption{The RPRR production vertex}
\label{RPRR-vertex} 
\end{figure}   
\noindent        
Finally, when the two production vertices are combined, we arrive at the terms illustrated in Fig.\ref{s12-discontinuity}: 
\begin{figure}[H]
\centering
\epsfig{file=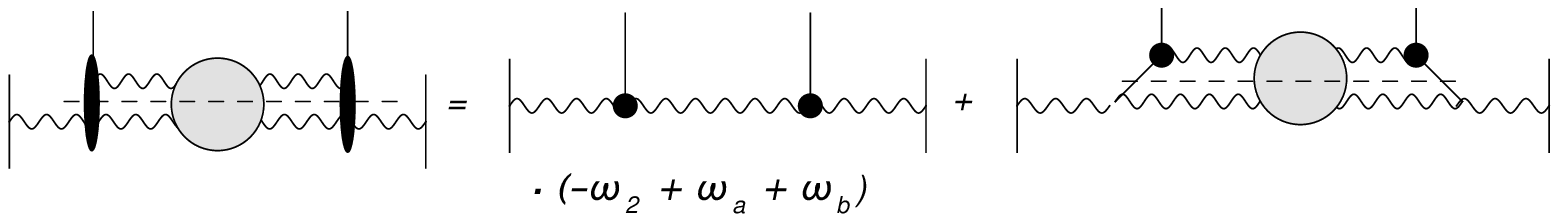,scale=0.8}\\ 
\caption{the $s_{12}$-discontinuity $\Delta_{12}$ of the $2 \to 4$ scattering amplitude} 
\label{s12-discontinuity}
\end{figure} 
\noindent 
Inserting this into (\ref{6point-s12-pos-energies}), the terms $\sim   \omega_2 -\omega_a - \omega_b$ cancel. What is left is the Regge cut piece (second diagram in Fig.\ref{s12-discontinuity}):   we separate the infrared divergent lowest order (one loop) term\footnote{We follow the notation of \cite{Bartels:2013jna}} from the infrared finite cut amplitude which we denote by $f_{\omega_2}$ and obtain.
\be
\Delta_{12} = - \pi(\omega_2 -\omega_a-\omega_b) + \frac{\pi}{2} V_{13} +f_{\omega_2},
\ee
where
\be 
V_{ik}=   \frac{\gamma_K}{4} \ln \frac{|q_i|^2 |q_k|^2}{|q_i-q_k|^2 \lambda^2}.
\ee
Since the term $V_{13}$ is neither infrared finite nor conformal invariant, we introduce 
the phase  
\be
\delta_{13}=\pi \left(  V_{13} +  \omega_a + \omega_b\right)  = \pi \frac{\gamma_K}{4}
\ln \frac {|q_1| |q_3| |k_a| |k_b|}{|k_a+k_b|^2 |q_2|^2}
\ee
 and write 
\be
\Delta_{12} = - \pi(\omega_2 -\omega_a-\omega_b) - \frac{\pi(\omega + \omega_b)}{2 }+ \frac{\delta_{13} }{2}  +f_{\omega_2},
\ee
The resulting Regge cut amplitude,  $f_{\omega_2}$, is defined to begin with at least one iteration of the color octet BFKL kernel and is given by:
\ba
 f_{\omega_2}=
\frac{g^2 N_c}{16\pi^2} \sum_n (-1)^n \int \frac{d\nu}{\nu^2 +\frac{n^2}{4}} \left( \left(-\frac{s_2}{s_0}\right)^{\omega(\nu,n)} - 1\right)
\left(\frac{q_3^*k_a^*}{k_b^* q_1^*}\right)^{i\nu-\frac{n}{2}}  \left(\frac{q_3k_a}{k_b q_1}\right)^{i\nu+\frac{n}{2}}
\label{W-omega2-cut}
\ea
(here we have included one of the $\omega$-integrals from the Sommerfeld-Watson integral representation).
The calculation of the unitarity integral and impact factors which leads to this expression has been described in \cite{Bartels:2008sc} and will not 
be repeated here. We still write the expression for $f_{\omega_2}$ in a slightly more general form \cite{Bartels:2010ej,Fadin:2011we} which also specifies the energy scale $s_0$. 
We introduce the anharmonic ratios:
\be
u_1=\frac{(-s)(-s_2)}{(-s_{012})(-s_{123})},\;\;\;\; u_2=\frac{(-s_3)(-t_1)}{(-s_{123})(-t_2)},\;\;\;\;u_3=\frac{(-s_1)(-t_3)}{(-s_{012})(-t_2)},
\ee 
and the complex-valued variable $w$:
\be
w=\frac{q_3 k_a}{q_1 k_b}.
\ee
We write:
\be
f_{\omega_2}=
\frac{g^2 N_c}{16\pi^2} \sum_n (-1)^n  \left( \frac{w}{w*}\right)^{\frac{n}{2}} \int \frac{d\nu}{2\pi i} \Phi_{\nu,n}^*  \left[ \left( -\sqrt{u_2u_3}\right)^ {-\omega(\nu,n)} - 1 \right] \Phi_{\nu,n}
|w|^{2i\nu}  .
\ee
It is important to note that this expression is conformal invariant. To summarize the construction of $f_{\omega_2}$, the impact factor  $\Phi_{\nu,n}$ has its origin in the "nonlocal" piece of the $RPRR$ production vertex only (i.e. from the full $RPRR$ production vertex we first have to remove the "local" term),  and from the BFKL Green's function we remove the one-loop contribution. Our final result for $W_{\omega_2}$ thus becomes: 
\be
W_{\omega_2} =   \pi \frac{\omega_a\omega_b}{\omega_2} 
- \frac{\pi(\omega_a + \omega_b)}{2 }+ \frac{\delta_{13} }{2}  +f_{\omega_2}.
\label{W-omega2}
\ee
At the end of this section we will show that the first two terms, which coincide with leading approximation of the "subtraction" defined in \cite{Bartels:2013jna},  will cancel parts of the Regge pole, leaving what we call "conformal Regge pole". The last two terms are conformally invariant, and $f_{\omega_2}$ defines the "conformally invariant Regge cut" amplitude 

Before we conclude this digression on the $2 \to 4$ amplitude, we want to make several comments.
First, our choice of computing the discontinuity in the kinematic region where all energies are positive was
not unique. Alternatively, we could compute the discontinuity in $s_2$ also in another kinematic region, e.g. in the region $\tau_1 \tau_3$:
\be
\tau_1\tau_3: \hspace{1cm} disc_{12}\; T_{2 \to 4}^{\tau_1\tau_3} =... \Delta_{12}^{\tau_1\tau_3}, \hspace{0.5cm} \Delta_{12}^{\tau_1\tau_3}= 
   - \frac{V_L(a) V_R(b)}{\Omega_2} + e^{-i\pi \omega_2} W_{\omega_2} .
\label{6point-s12-tau1-3}
\ee  
The phases are different from those of the positive energy result in (\ref{6point-s12-pos-energies}). They reflect the fact that, in the unitarity integral, the amplitudes on both sides of the unitarity integral, have their phases, and they clearly depend upon the kinematic region where the unitarity integral is computed. In the weak coupling limit, these phases can be neglected and  we obtain the same result for  $W_{\omega_2}$.

Next, it is instructive to consider also other discontinuities, e.g. in the total energy $s$. In the region of positive energies 
we obtain:
\ba
disc_{s}\; T_{2 \to 4} &=&... \Delta_{s} ,\nonumber\\ \Delta_s& =& - \frac{\Omega_2 V_R(a) e^{-i\pi \omega_{12}} -  \Omega_1 V_L(a) e^{-i\pi \omega_{21}} }{\Omega_{12}} 
\frac{\Omega_3 V_R(b) e^{-i\pi \omega_{23}} -  \Omega_2 V_L(b) e^{-i\pi \omega_{32}} }{\Omega_{23}} \nonumber\\
&&+ e^{i(\omega_1 -  \omega_3)} e^{i\pi \omega_2} W_{\omega_2},
\label{6point-s-pos-energies}
\ea  
whereas in the region $\tau_1 \tau_3$ the result is much simpler:
\be
\tau_1\tau_3: \hspace{1cm} \Delta_{s}^{\tau_1\tau_3} = - \frac{\Omega_a\Omega_b}{\Omega_2} + e^{-i\pi \omega_2} W_{\omega_2}.
\label{6point-s-tau1-3}
\ee  
Let us verify that these different expressions for $W_{\omega_2}$ all coincide in the weak coupling limit.  
To see this in detail, we first note that, after neglecting the phases, $\Delta_{12}$ and $\Delta_{12}^{\tau_1\tau_3}$
coincide:
\be
\Delta_{12} = - \pi \frac{\omega_{2a} \omega_{2b}}{\omega_2} + W_{\omega_2}.
\ee
We are thus led to compare the two equations:
\ba
W_{\omega_2} &&=   \Delta_{12} + \pi  \frac{\omega_{2a} \omega_{2b}}{\omega_2}\nonumber\\
&&=  \Delta_{12}  +  \pi \left( \omega_2 - \omega_a - \omega_b + \frac{\omega_a \omega_b}{\omega_2}\right)
\label{W_omega_2_Delta_12}
\ea
and 
\ba
W_{\omega_2} 
=    \Delta_{s} +  \pi  \frac{\omega_a \omega_b}{\omega_2}.
\label{W_omega_2_Delta_s}
\ea 
The reason why these seemingly different expression for $W_{\omega_2}$ coincide, lies in application of the bootstrap relations. As we have explained above, for the discontinuity in $s_2$, $\Delta_{12}$, we have applied the bootstrap 
condition in the $t_2$ channel which, in (\ref{6point-s12-pos-energies}), leads to the cancellation of the terms $\sim   \omega_2 -\omega_a - \omega_b$. In contrast, for the discontinuity in $s$, namely $\Delta_s$, we apply the bootstrap condition to the $t_1$ and to the $t_3$ channel: we illustrate the result in Fig.\ref{s-discontinuity}: 
\begin{figure}[H]
\centering
\epsfig{file=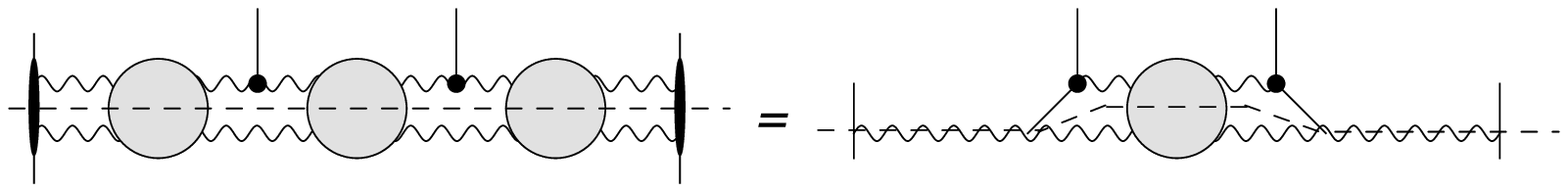,scale=0.8}
\caption{the $s$-discontinuity $\Delta_s$ of the $2 \to 4$ scattering amplitude}
\label{s-discontinuity}   
\end{figure} 
\noindent 
This leads directly to the second piece on the rhs of Fig.\ref{RPRR-vertex}, which results from the nonlocal piece of the $RPRR$ production vertex. With this observation, $W_{\omega_2}$ in (\ref{W_omega_2_Delta_s}), agrees with (\ref{W_omega_2_Delta_12}), i.e. both energy discontinuities lead to the same answer. This equality can be seen also directly by comparing the second term on the rhs of Fig.\ref{s-discontinuity} and the rhs of Fig.\ref{s12-discontinuity}: for the $\alpha$ integral on the lhs (and for the $\beta$ integral on the rhs) of the Green's function there are two ways of closing the integration contour which give the same answer.    

As we have explained above, in this last part of our discussion, we had to restrict ourselves to the leading logarithmic approximation. This was because, in evaluating the energy discontinuities via unitarity integrals, so far we have used only the leading approximation for the scattering amplitudes. Fortunately, all building blocks for a NLO calculation are known:  the $RPR$ production vertex, the $RPRR$ vertex, the gluon trajectory function, and the bootstrap condition. So it is possible to verify that our construction of the Regge cut piece can be done also 
in NLO.   

In the final step we put the pieces together and compute, for different kinematic regions, the full scattering amplitude. In \cite{Bartels:2013jna} we have presented a full list of the Regge pole contributions. They can be derived from the phases in (\ref{2to4phases}) and the pole pieces in (\ref{2to4-partialwavesLR2}); in \cite{Bartels:2013jna} we found a slightly simpler way of calculation. For the two most interesting regions $\tau_1\tau_3$ and $\tau_1\tau_2\tau_3$, we found the following results:
\ba
\tau_1\tau_3:&  &   e^{-i\pi\omega_2}\left[e^{i\pi\left(\omega_a+\omega_b\right)} - 2ie^{i\pi\omega_2}\frac{\Omega_a \Omega_b}{\Omega_2} \right]\nonumber\\
&=&e^{-i\pi\omega_2}\left[ \cos \pi \omega_{ab} + i \sin \pi( \omega_a+\omega_b) - 2i \frac{ \cos\pi \omega_{2} 
\Omega_a \Omega_b}{\Omega_2} \right] \\
\tau_1\tau_2\tau_3:& & -\left[e^{-i\pi\left(\omega_a+\omega_b\right)} + 2ie^{-i\pi\omega_2}\frac{\omega_a \Omega_b}{\Omega_2} \right]\nonumber\\
&=& - \left[ \cos \pi \omega_{ab} - i \sin \pi( \omega_a+\omega_b) + 2i \frac{ \cos\pi \omega_{2} \Omega_a \Omega_b}{\Omega_2} \right].
\ea  
These regions contain the Regge cut contributions. From (\ref{2to4phases}), (\ref{2to4-partialwavesLR1}), and (\ref{2to4-partialwavesLR2}) we derive the phase structure of the cut contributions:
\ba
\tau_1\tau_3:  &    2i e^{-i\pi \omega_2} W_{\omega_2} \\
\tau_1\tau_2\tau_3: &   2i W_{\omega_2}.
\ea     
Combining poles and cuts we arrive at:
\ba
\tau_1\tau_3:
&e^{-i\pi\omega_2}\left[ \cos \pi \omega_{ab} + i \sin \pi( \omega_a+\omega_b) - 2i \frac{ \cos\pi \omega_{2} 
\Omega_a \Omega_b}{\Omega_2}  +2i W_{\omega_2}\right] 
\label{amp-tau1-tau3}\\
\tau_1\tau_2\tau_3: & - \left[ \cos \pi \omega_{ab} - i \sin \pi( \omega_a+\omega_b) + 2i \frac{ \cos\pi \omega_{2} \Omega_a \Omega_b}{\Omega_2} - 2i W_{\omega_2}\right].
\label{amp-tau1-tau2-tau3}
\ea  
These results are valid for all orders. We remind that  $W_{\omega_2}$ is a real-valued function and contains no further phases. 

Since, for the Regge cut contribution $W_{\omega_2}$, we have only the leading logarithmic result we can approximate 
(\ref{amp-tau1-tau3}) and (\ref{amp-tau1-tau2-tau3}): 
\ba
\tau_1\tau_3:
&e^{-i\pi\omega_2}\left[\cos \pi \omega_{ab}+i \pi (\omega_a+\omega_b)  - 2i \pi\frac{ 
\omega_a \omega_b}{\omega_2}  +2i W_{\omega_2}\right] 
\label{amp-tau1-tau3LL}\\
\tau_1\tau_2\tau_3: & - \left[ \cos \pi \omega_{ab} - i \pi( \omega_a+\omega_b) + 2i\pi \frac{  \omega_a \omega_b}{\omega_2} - 2i W_{\omega_2}\right].
\label{amp-tau1-tau2-tau3LL}
\ea
When inserting the result (\ref{W-omega2}) for the Regge-cut amplitude  
into (\ref{amp-tau1-tau3LL}) we immediately notice the cancellation of the singular terms,
$\frac{\omega_a\omega_b}{\omega_2}$ and of the terms $\sim \omega_a+\omega_b$. 
We obtain:
\ba
\tau_1\tau_3:
&e^{-i\pi\omega_2}\left[ \cos \pi \omega_{ab} + i \delta_{13}  +  2 i f_{\omega_2} \right]  
\label{amp-tau1-tau3-final}\\
\tau_1\tau_2\tau_3: & - \left[ \cos \pi \omega_{ab} - i \delta_{13}  -2 i  f_{\omega_2}\right] 
\label{amp-tau1-tau2-tau3-final}
\ea
Our phase $\delta_{13}$ coincides with the phase contained in th BDS amplitude \cite{Bartels:2008ce}.

We thus have found that the scattering amplitude can be written  as a sum of the conformal invariant Regge pole term, $\cos \pi \omega_{ab}$, and a conformal invariant and infrared finite Regge cut term \cite{Lipatov:2010qf}. Whereas the pole term is given by an all-order expression, the derivation of the cut term  
has been presented here only in the leading approximation:

\section{$2\to5$: computing Regge cut contributions from energy discontinuities}

Returning now to the $2 \to 5$ scattering amplitude, we proceed in the same way as in the $2 \to 4$ case.
Our ansatz, a sum of 14 terms, has already been described in Sec. II, and we have listed the trigonometric factors. 
In this section we calculate the Regge cut contributions via energy discontinuities.

\subsection{Short Regge cuts: discontinuity in  $s_3$}  

We begin with the discontinuity  in $s_3=s_{34}$; it receives contributions from the doublet $RLR$ and the triplet $LLR$.  These are the five partial waves which contain the short cut in $\omega_3$.  For simplicity we chose the kinematic region where all energies are positive. Together with the phases listed in Appendix B, the Regge pole terms of the two partial waves $RLR(1)$ and $RLR(2)$ are found to lead to 
\ba
  disc_{3}\; (T_{RLR(1)}^{pole}  +T_{RLR(2)}^{pole}) =... -\frac{e^{-i\pi (\omega_1+\omega_4)}}{\Omega_3} \frac{V_R(a)}{\Omega_{12}} V_L(b) V_R(c),
\ea
where, as before, the dots indicate that we have left out the integration symbols and the energy factors. From the three partial waves $LLR(1)$,  $LLR(2)$, and  $LLR(3)$ we obtain:
\ba
  disc_{s_3}\; (T_{LLR(1)}^{pole} +T_{LLR(2)}^{pole} +T_{LLR(3)}^{pole}) =... -\frac{e^{-i\pi (\omega_2+\omega_4)}}{\Omega_3} \frac{V_L(a)}{\Omega_{21}} V_L(b) V_R(c).
\ea
Their sum equals:
\ba
disc_{3}\; ( T_{RLR(1)}^{pole} +T_{RLR(2)}^{pole}+ T_{LLR(1)}^{pole} +T_{LLR(2)} ^{pole}+T_{LLR(3)}^{pole}) \nonumber\\ 
=...- e^{-i\pi(\omega_1+\omega_2+\omega_4)} e^{i \pi \omega_a} \frac{V_L(b)  V_R(c)}{\Omega_3}.
\ea 
Next we consider the contributions of the Regge cut term. From the doublet $RLR$ we have
\ba
  disc_{s_3}\; (T_{RLR(1)}^{\omega_3-cut} +T_{RLR(2)}^{\omega_3-cut})  = ...  e^{-i\pi (\omega_1 + \omega_2+\omega_4)}
 e^{i\pi \omega_3} \frac{V_R(a)}{\Omega_{12}}e^{i\pi \omega_2} W_{\omega_3},
\ea 
whereas the triplet $LLR$ yields:
\ba
  disc_{3} \;(T_{LLR(1)}^{\omega_3-cut} +T_{LLR(2)}^{\omega_3-cut} +T_{LLR(3)}^{\omega_3-cut}) =... e^{-i\pi (\omega_1 + \omega_2+\omega_4)}
 e^{i\pi \omega_3} \frac{V_L(a)}{\Omega_{21}}e^{i\pi \omega_1} W_{\omega_3}.
\ea
Their sum equals:
\ba
disc_{3}\; ( T_{RLR(1)}^{\omega_3-cut} +T_{RLR(2)}^{\omega_3-cut}+ T_{LLR(1)}^{\omega_3-cut}+T_{LLR(2)}^{\omega_3-cut} +T_{LLR(3)}^{\omega_3-cut} )  \nonumber\\ =... 
 e^{-i\pi(\omega_1+\omega_2+\omega_4)} e^{i \pi \omega_a} e^{i\pi \omega_3} W_{\omega_3}.
\ea 
We finally note that the long cut pieces in  $LLR(1)$ and  $LLR(2)$ cancel 
against each other and do not contribute to the $s_3$ discontinuity in the positive energy region.

As a result, the discontinuity in $s_3=s_{34}$ of the full scattering amplitude $T_{2\to5}$ in the region of only positive energies equals:
\be
\Delta_{34} = e^{-i\pi (\omega_1 +\omega_2+ \omega_4)} e^{i \pi \omega_a} 
\left( - \frac{V_L(b) V_R(c)} {\Omega_3} + e^{i\pi \omega_3} W_{\omega_3}\right).
\label{2to5disc34}
\ee
 We mention that in other kinematic regions the results are similar, e.g.
\ba
\tau_2\tau_4:\hspace{0.5cm}
\Delta_{34}^{\tau_2\tau_4} = e^{-i\pi \omega_1} e^{i \pi \omega_a} 
\left(- \frac{V_L(b) V_R(c)} {\Omega_3} + e^{-i\pi \omega_3} W_{\omega_3}\right).
\ea
The important feature of these expressions is the singular term from the Regge pole contribution: similar to the $2\to4$ case,
on the lhs the energy discontinuity is computed from unitarity integrals and thus is free from the unphysical 
pole $\sim 1/ \sin \pi \omega_3$. Hence, on the rhs, the function  $W_{\omega_3}$ must compensate the 
singularity.

Let us now evaluate (\ref{2to5disc34}) in the weak coupling approximation. 
We proceed exactly in the same way as we have described for the $2\to4$ case and obtain:
\be
W_{\omega_3}= \Delta_{34} +\pi \left( \omega_3 -\omega_b-\omega_c + \frac{\omega_a \omega_b}{\omega_3}\right).
\label{W-omega3}
\ee
For the computation of the discontinuity on the lhs we, as before,  decompose the production vertices  into "local" and "nonlocal" 
pieces and make use of the bootstrap equation. This removes, on the rhs of (\ref{W-omega3}), the terms 
$\omega_3$, $\omega_b$, and  $\omega_c$, and we arrive at the analogue of (\ref{W-omega2}):
\ba
W_{\omega_3} =\pi \frac{\omega_b \omega_c}{\omega_3}  -\frac{\pi}{2} (\omega_b + \omega_c) +
\frac{\delta_{24}}{2}   +  f_{\omega_3}
\label{W_omega_3},
\ea
where
\be
\delta_{24}= \pi (V_{24} + \omega_b+\omega_c)
\ee
The integral representation for $f_{\omega_3}$ is easily derived from (\ref{W-omega2-cut}). As expected, the partial wave in (\ref{W_omega_3}) consists of a "subtraction" (first two terms) which will be shown to cancel against the unwanted parts of the Regge pole terms. It agrees with the result obtained in our previous paper [c.f. Appendix B].
The piece "$\frac{1}{2} \delta_{24} + f_{\omega_3}$" represents the conformal invariant and infrared finite Regge-cut amplitude.

\subsection{Long cut: discontinuity in $s_{123}$}

Let us now turn to the long Regge cut contribution which is contained in the two triplets: $LLR(1)$,  $LLR(2)$,
$LRR(1)$,  $LRR(2)$. In order to determine $W_{\omega_2\omega_3}$ we consider the discontinuity in $s_{123}$.
For simplicity we again take all energies to be positive. We begin with the Regge pole contribution:
\be
disc_{123} \left( T^{pole}_{LLR(1)} + T^{pole}_{LLR(2)} \right) = ... -e^{-i\pi (\omega_1 + \omega_4)} V_L(a) 
e^{-i\pi \omega_{32}} \frac{V_L(b)}{\Omega_3 \Omega_{32}} V_R(c)
\ee
and
\be
disc_{123} \left( T^{pole}_{LRR(1)} + T^{pole}_{LRR(2)} \right) = ... -e^{-i\pi (\omega_1 + \omega_4)} V_L(a) 
e^{-i\pi \omega_{23}} \frac{V_R(b)}{\Omega_2 \Omega_{23}} V_R(c).
\ee
Next we consider the short cuts in $\omega_3$ and $\omega_2$. We find: 
\be
disc_{123} \left( T^{\omega_3-cut}_{LLR(1)} + T^{\omega_3-cut}_{LLR(2)} \right) = ...  
 -e^{-i\pi (\omega_1 + \omega_4)} e^{-i\pi \omega_{32}} V_L(a) \frac{W_{\omega_3}}{\Omega_2} 
\ee
and
\be
disc_{123} \left( T^{\omega_2-cut}_{LRR(1)} + T^{\omega_2-cut}_{LRR(2)} \right) = ...  
 -e^{-i\pi (\omega_1 + \omega_4)} e^{-i\pi \omega_{23}} \frac{W_{\omega_2}}{\Omega_3} V_R(c). 
\ee
Finally the long cut:
\be
disc_{123} \left( T^{\omega_2 \omega_3-cut}_{LLR(1)} + T^{\omega_2\omega_3-cut}_{LLR(2)} \right) = ...
 e^{-i\pi (\omega_1 + \omega_4)} e^{-i\pi \omega_{32}} e^{i\pi \omega_2} \frac{W_{\omega_2\omega_3;L}}{\Omega_{32}}
 \ee
and 
\be
disc_{123} \left( T^{\omega_2 \omega_3-cut}_{LRR(1)} + T^{\omega_2\omega_3-cut}_{LRR(2)} \right) = ...
 e^{-i\pi (\omega_1 + \omega_4)} e^{-i\pi \omega_{23}} e^{i\pi \omega_3} \frac{W_{\omega_2\omega_3;R}}{\Omega_{23}}.
 \ee
The sum of all terms equals:
\ba
\Delta_{123} = e^{-i\pi(\omega_1+\omega_4)}
\Big[ e^{-i\pi \omega_{32}} \left( e^{i\pi \omega_2} \frac{W_{\omega_2\omega_3;L}}{\Omega_{32}}
-\frac{V_L(a) W_{\omega_3}}{\Omega_2} - \frac{V_L(a) V_L(b)V_R(c)}{\Omega_3 \Omega_{32}}\right)\nonumber\\
+  e^{-i\pi \omega_{23}} \left( e^{i\pi \omega_3} \frac{W_{\omega_2\omega_3;R}}{\Omega_{32}}
-\frac{W_{\omega_2}V_R(c)}{\Omega_3} - \frac{V_L(a) V_R(b)V_R(c)}{\Omega_2 \Omega_{23}}\right)
\Big]
\ea
So far the results for the discontinuity are valid to all orders. In the weak coupling limit we find for the sum of all terms:
\be
\label{disc123LL} 
\Delta_{123}  =   - \pi \frac{\omega_{2a} \omega_b \omega_{3c}}{\omega_2\omega_3}- 
\frac{\omega_{2a}}{\omega_2} W_{\omega_3} - W_{\omega_2} \frac{\omega_{3c}}{\omega_3} +
\frac{W_{\omega_2\omega_3;L} - W_{\omega_2\omega_3;R}}{\pi \omega_{32}}  .
\ee
Let us first evaluate  the discontinuity on the lhs which we illustrate in Fig.\ref{s123-discontinuity-2to5}. Similarly to the $2 \to 4$ case we decompose the structure of the $RPRR$ and the $RRPR$ production vertices:
\begin{figure}[H]
\centering
\epsfig{file=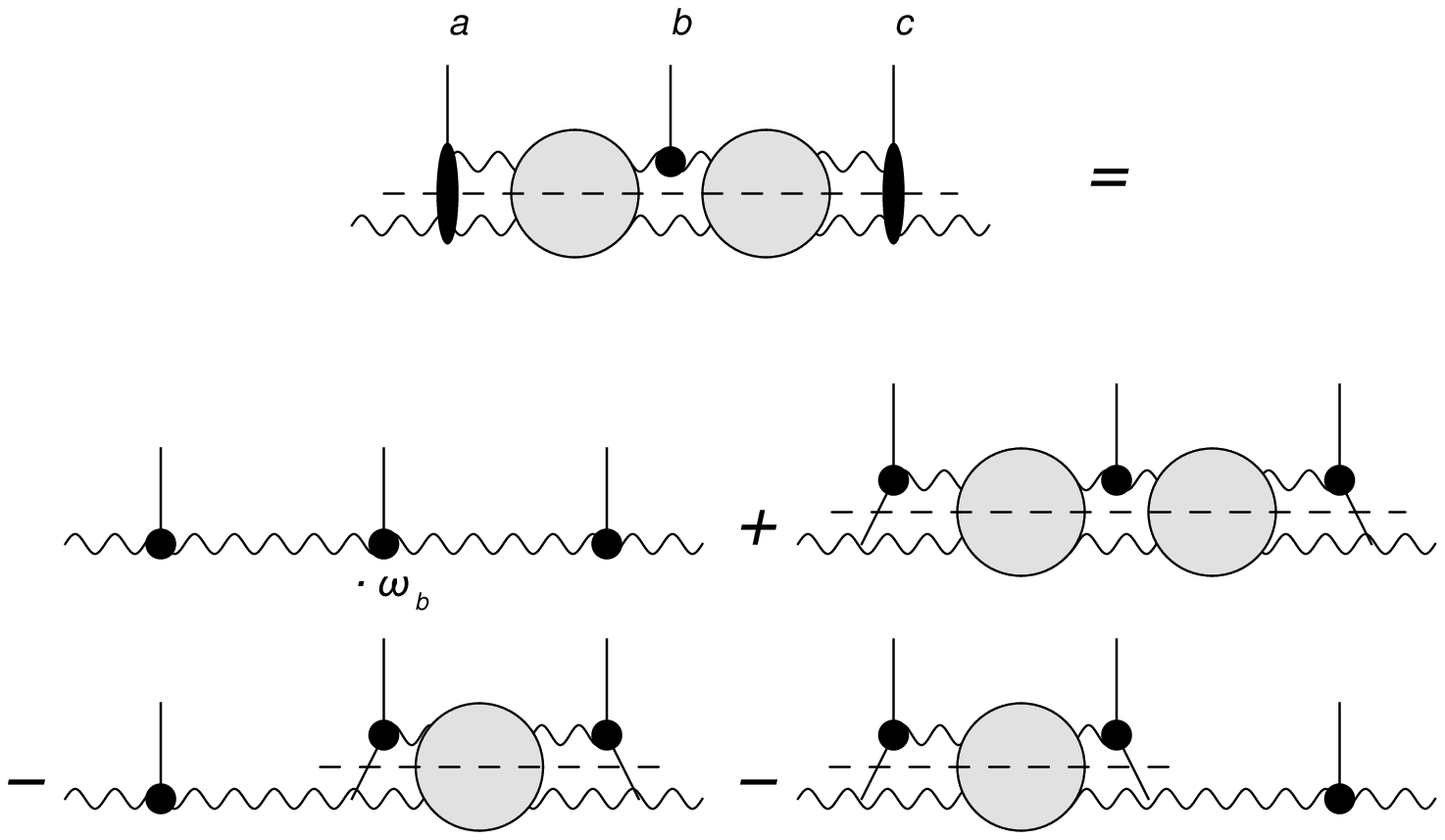, scale=0.8}
\caption {Illustration of the discontinuity $\Delta_{123}$ in the weak coupling limit}
\label{s123-discontinuity-2to5}
\end{figure} 
\noindent
After separating the one loop contributions we obtain:
\be
\label{disc123bootstrapa}
\Delta_{123}= -\pi \omega_b + \frac{\pi}{2} V_{14} + f_{\omega_2\omega_3} - \frac{\pi}{2} V_{13} - \frac{\pi}{2} V_{24} -
f_{\omega_3} - f_{\omega_2} ,
 \ee 
which we can also write in the form
\be
\label{disc123bootstrapb}
\Delta_{123}= \frac{1}{2} (\delta_{14} - \delta_{24} -\delta_{13}) +f_{\omega_2\omega_3}
-f_{\omega_2} - f_{\omega_3}
\ee
with
\be
\delta_{14}=\pi(V_{14} +\omega_a + \omega_c).  
\ee
For $f_{\omega_2\omega_3}$ we have the integral representation 
\cite{Bartels:2011ge}:
\ba
\label{f-long}
f_{\omega_2\omega_3}= &&\frac{a}{2} \sum_{n_1,n_2} (-1)^{n_1+n_2}\int\int \frac{d\nu_1 d\nu_2}{(2\pi)^2}
 \frac{1}{i\nu_1+\frac{n_1}{2}} \left( \frac{k_a^*q_3^*}{q_1^* k_b^*} \right)^{i\nu_1-\frac{n_1}{2}}
\left( \frac{k_a q_3}{q_1 k_b} \right)^{i\nu_1+\frac{n_1}{2}} 
\left( \frac{-s_2}{s_{02}} \right)^{\omega(\nu_1,n_1)}\nonumber\\
&&C(\nu_1,\nu_2,n_1,n_2) \left( \frac{-s_3}{s_{03}} \right)^{\omega(\nu_2,n_2 )} 
\left( \frac{k_b^*q_4^*}{q_2^* k_c^*} \right)^{i\nu_2-\frac{n_2}{2}}
\left( \frac{k_b q_4}{q_2 k_c} \right)^{i\nu_2+\frac{n_2}{2}}  \frac{1}{i\nu_2- \frac{n_2}{2}}|_{\text{sub}},
\ea
%\ba
%\label{f-long}
%f_{\omega_2\omega_3} = &&\frac{a}{2} \sum_{n_1,n_2} (-1)^{n_1+n_2}\int\int \frac{d\nu_1 d\nu_2}{(2\pi)^2}\nonumber \\
%&&\times \Phi(\nu_1,n_1)   C(\nu_1,n_1;\nu_2,n_2) \Phi(\nu_2,n_2)
%\times |w_1|^{i\nu_1} e^{-i n_1\phi_1}
%\left((1-u_{11}) \frac{|w_1|}{|1+w_1|^2}\right)^{-\omega(\nu_1,n_1)} \nonumber\\
%&&\times \left( (1-u_{12}) \frac{|w_2|}{|1+w_2|^2}\right)^{-\omega(\nu_2,n_2)} |w_2|^{i\nu_2} e^{-i n_2\phi_2}.
%\ea
where the subscript "sub" indicates that we have subtracted the one loop contribution, and the function $C(\nu_1,\nu_2,n_1,n_2)$ is the "central emission vertex" as defined in \cite{Bartels:2011ge} [cf. Eq.(19)]. 

It may be useful to write this Regge cut amplitude also in terms of anharmonic ratios. We introduce the six anharmonic ratios:
\ba
u_{11}=\frac{(-s_{0123})(-s_2)}{(-s_{012})(-s_{123})}, \;
 u_{21}=\frac{(-s_{234})(-t_1)}{(-s_{1234})(-t_2)},\;u_{31}=\frac{(-s_1)(-t_3)}{(-s_{012})(-t_2)},\nonumber\\
u_{12}=\frac{(-s_{1234})(-s_3)}{(-s_{123})(-s_{234})}, \; u_{22}=\frac{(-s_4)(-t_2)}{(-s_{234})(-t_3)},\;u_{32}=\frac{(-s_{012})(-t_4)}{(-s_{0123})(-t_3)},
\ea 
and the complex-valued variables $w_\sigma$:
\be
w_1=\frac{q_3 k_a}{q_1 k_b},\; w_2=\frac{q_4 k_b}{q_2 k_c}.
\ee
The integral representation becomes: 
\ba
\label{f-long-conform}
f_{\omega_2\omega_3}= &&\frac{a}{2} \sum_{n_1,n_2} (-1)^{n_1+n_2}
\left(\frac{w_1}{w_1^*}\right)^{n_1} \left(\frac{w_2}{w_2^*}\right)^{n_2} 
\int\int \frac{d\nu_1 d\nu_2}{(2\pi)^2}
 \Phi(\nu_1,n_1)^*  |w_1|^{2i\nu_1}
\left( -\sqrt{u_{21} u_{31}}
 \right)
^{-\omega(\nu_1,n_1)}
\nonumber\\
&&C(\nu_1,\nu_2,n_1,n_2) \left( -\sqrt{u_{22} u_{32}}
 \right)
^{-\omega(\nu_2,n_2)}
|w_2|^{2i\nu_2} \Phi(\nu_2,n_2) |_{\text{sub}}.
\ea
Returning to the energy discontinuity we  insert (\ref{disc123bootstrapb}) into the (\ref{disc123LL}) and obtain:
\be
\frac{W_{\omega_2\omega_3;L} - W_{\omega_2\omega_3;R}}{\pi \omega_{32}}
 +\frac{\omega_a}{\omega_2} W_{\omega_3} + W_{\omega_2} \frac{\omega_c}{\omega_3} =  \pi \frac{\omega_a\omega_b\omega_c}{\omega_2\omega_3} -\frac{\pi}{2}(\omega_a+\omega_c) +\frac{1}{2} \delta_{14} + f_{\omega_2\omega_3}.
\label{W-long-L-R1}
\ee 
We notice that the single discontinuity is not sufficient to determine $W_{\omega_2\omega_3;L}$ and $W_{\omega_2\omega_3;R}$ separately. However, it fixes the combination which appears in the leading approximation of the scattering amplitude. On the rhs of (\ref{W-long-L-R1}) we again find the subtraction terms which will be canceled by the Regge pole  contributions, and the conformal invariant Regge cut contribution. The subtraction term agrees with the result of our previous paper (Appendix B).
  
For comparison, we consider also another discontinuity in the total energy $s$ in the kinematic region $\tau_1\tau_4$. First, one has to write the Regge pole contribution. After some algebra we find for the sum of all partial waves
\be
disc_s \sum T_{ijk}^{pole} = ... - \frac{\Omega_a \Omega_b \Omega_c}{\Omega_2 \Omega_3},
\ee  
where, again, the dots stand for the $\omega$-integrals and energy factors. For later purposes it will be convenient to use the identity 
\be
\frac{\Omega_a \Omega_b \Omega_c}{\Omega_2 \Omega_3}=\Omega_a \left( e^{-i\pi \omega_3} \frac{\Omega_b}{\Omega_3 \Omega_{23}} + e^{-i\pi \omega_2} \frac{\Omega_b}{\Omega_2 \Omega_{32}} \right) \Omega_c.
\ee 
Next we consider the short $\omega_3$-cut, contained in the doublet $RLR$ 
\be
disc_s \left( T_{RLR(1)}^{\omega_3-cut} + T_{RLR(2)}^{\omega_3-cut}\right) =... e^{-i\pi \omega_3} \frac {V_R(a)}{\Omega_{12}} W_{\omega_3} 
\ee
and in the triplet $LLR$:
\ba
disc_s \left( T_{LLR(1)}^{\omega_3-cut}+T_{LLR(2)}^{\omega_3-cut}+T_{LLR(3)}^{\omega_3-cut} \right) = \nonumber\\
 =... e^{-i\pi \omega_3} \frac{\Omega_{1}}{\Omega_{2}} \frac {V_L(a)}{\Omega_{21}} W_{\omega_3}. 
\ea
Their sum  equals:
\ba
disc_s \left( T_{RLR(1)}^{\omega_3-cut} + T_{RLR(2)}^{\omega_3-cut}+ T_{LLR(1)}^{\omega_3-cut}+T_{LLR(2)}^{\omega_3-cut}+T_{LLR(3)}^{\omega_3-cut}\right) \nonumber\\
=... e^{-i\pi \omega_3}  \frac{\Omega_a}{\Omega_2} W_{\omega_3}.
\ea
%It is important to note that the coefficient of  $W_{\omega_3}$ is the same as in (\ref{cut3tau14}).
An analogous result holds for the short cut contained in the doublet $LRL$ and in the triplet $LRR$. Finally, the contributions of the long cut are:
\be
disc_s \left( T_{LLR(1)}^{\omega_2-\omega_3-cut} +T_{LLR(2)}^{\omega_2-\omega_3-cut} \right) = ...   e^{-i\pi \omega_3} \frac{W_{\omega_2\omega_3;L}}
{\Omega_{32}}
\ee  
and 
\be
disc_s \left( T_{LRR(1)}^{\omega_2-\omega_3-cut} +T_{LRR(2)}^{\omega_3-\omega_2-cut} \right) = ...  e^{-i\pi \omega_2} \frac{W_{\omega_2\omega_3;R}}
{\Omega_{23}}.
\ee
The full discontinuity in $s$ becomes:
\ba
\Delta_s^{\tau_1\tau_4} = 
  e^{-i\pi(\omega_2+\omega_3)}\frac{ \Omega_a\Omega_b\Omega_c}{\Omega_{2} \Omega_3} 
  +\frac{e^{i\pi \omega_2}W_{\omega_2\omega_3;L} -e^{i\pi \omega_3} W_{\omega_2\omega_3;R}}{\Omega_{32}}
\nonumber\\ +    e^{+i\pi \omega_2} 
\frac{\Omega_a}{\Omega_2} W_{\omega_3} + e^{i\pi \omega_3}  W_{\omega_2}
\frac{\Omega_c}{\Omega_3}  .
\ea
To proceed further let us restrict ourselves to the leading logarithmic approximation. We obtain:
\be
\Delta_s^{\tau_1\tau_4} =  \frac{
W_{\omega_2\omega_3;L} - W_{\omega_2\omega_3;R}}{\pi \omega_{32}} + \frac{\omega_a}{\omega_2} W_{\omega_3}
+ \frac{\omega_c}{\omega_3} W_{\omega_2} - \pi \frac{\omega_a \omega_b \omega_c}{\omega_2\omega_3}.
\label{Delta_s-long}
\ee
For the lhs we use the bootstrap relations in the $t_1$ and $t_4$ channels and obtain the result illustrated Fig.\ref{s-discontinuity-2to5}:
\begin{figure}[H]
\centering
\epsfig{file=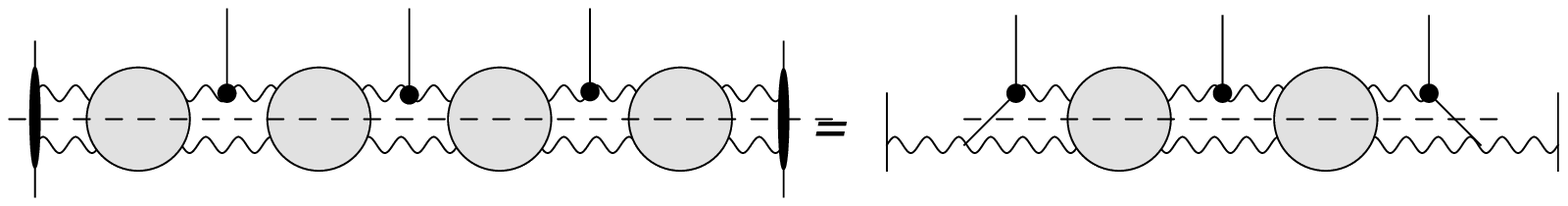,scale=0.8}
\caption{the $s$-discontinuity $\Delta_s$ of the $2 \to 5$ scattering amplitude}  
\label{s-discontinuity-2to5}
\end{figure} 
\noindent 
As we did before, we isolate on the rhs the IR-divergent one loop term:
\be
\label{Delta_s}  
 \Delta_s^{\tau_1\tau_4} = \frac{1}{2} \delta_{14} - \frac{\pi}{2}(\omega_a + \omega_c) + f_{\omega_2\omega_3},
\ee
where $f_{\omega_2\omega_3}$ is given in (\ref{f-long}). We thus find for $W_{\omega_2\omega_3}$ in the leading logarithmic approximation:
\be
\frac{W_{\omega_2\omega_3;L} - W_{\omega_2\omega_3;R}}{\pi \omega_{32}}
+\frac{\omega_a}{\omega_2}W_{\omega_3} +W_{\omega_2} \frac{\omega_c}{\omega_3} 
 = \pi \frac{\omega_a\omega_b\omega_c}{\omega_2\omega_3}- \frac{\pi}{2}(\omega_a+ \omega_c)
+  \frac{1}{2} \delta_{14} + f_{\omega_2\omega_3},
\ee  
which agrees with our previous result (\ref{W-long-L-R1}).

\section{The $2 \to 5$ scattering amplitudes in different kinematic regions}

In this final section we put pieces together and compute the scattering amplitudes.  As we have mentioned before, in the region of all energies being positive all Regge cut contributions cancel, and we are left with the Regge pole terms only. They have been computed in \cite{Bartels:2013jna}. Most important, in some kinematic regions singular terms appear, e.g. $\sim 1/\Omega_2$. We will show that the Regge cut contributions will remove all these unwanted singularities. 

We begin with the kinematic region $\tau_2\tau_4$. In this region, only the short Regge cut in the $\omega_3$ channel is nonzero, whereas both the short cut in the $\omega_2$ channel and the long cut vanish. From eqs.(\ref{phases-LLL}) - (\ref{phases-LRR}) we derive, for the product of the partial waves and their phase factors the following contributions to the scattering amplitude: 
\ba
F_{RLR(1)}^{\omega_3-cut} + F_{RLR(2)}^{\omega_3-cut}   &\rightarrow& 2i e^{-i\pi(\omega_1+\omega_3)} e^{i\pi \omega_2} \frac{V_R(a)}{\Omega_{12}} W_{\omega_3}
\ea   
and 
\ba
F_{LLR(1)}^{\omega_3-cut} + F_{LLR(2)}^{\omega_3-cut}+ F_{LLR(3)}^{\omega_3-cut}  &\rightarrow &  
2i  \frac{V_L(a)}{\Omega_{21}} e^{-i\pi \omega_3} W_{\omega_3}.
\ea
Taking the sum of the last two equations and observing the identity  (\ref{id-1}), 
we obtain the result:
\be
RLR(1) + RLR(2) + LLR(1) + LLR(2)+ LLR(3) = 2i e^{-i\pi(\omega_1+\omega_3)} e^{i\pi \omega_a} W_{\omega_3}.
\label{cut3tau24}
\ee
Similarly for the region $\tau_2\tau_3\tau_4$:
\be
\tau_2\tau_3\tau_4:  2i e^{-i\pi \omega_1} e^{i\pi \omega_a} W_{\omega_3}.
\label{cut3tau234}
\ee
We combine these Regge cut results with the Regge poles which are  taken from \cite{Bartels:2013jna}:  
\ba
\tau_2\tau_4: & e^{-i\pi\left(\omega_1+\omega_3\right)}e^{i\pi\omega_a}\left[e^{i\pi\left(\omega_b+\omega_c\right)} - 2ie^{i\pi\omega_3}\frac{\Omega_b \Omega_c}{\Omega_3} \right] \nonumber\\
&= e^{-i\pi\left(\omega_1+\omega_3\right)}e^{i\pi\omega_a}
\left[\cos \pi(\omega_b-\omega_c) \right.\nonumber\\ &\left. 
+i \sin \pi(\omega_b+\omega_c)- 2i\frac{\cos \pi \omega_3\Omega_b\Omega_c}{\Omega_3} \right] \label{poletau24}\\
\tau_2\tau_3\tau_4: &-e^{-i\pi\omega_1}e^{i\pi\omega_a}\left[e^{-i\pi\left(\omega_b+\omega_c\right)} + 2ie^{-i\pi\omega_3}\frac{\Omega_b \Omega_c}{\Omega_3} \right]\nonumber\\
&= -e^{-i\pi\omega_1}e^{i\pi\omega_a}\left[\cos \pi(\omega_b-\omega_c)\right.\nonumber\\ &\left.
 -i \sin \pi(\omega_b+\omega_c)+ 2i\frac{\cos \pi \omega_3\Omega_b\Omega_c}{\Omega_3} \right].
\label{poletau234}
\ea
When combining these Regge pole expressions with the Regge cuts in (\ref{cut3tau24}) and (\ref{cut3tau234}) (with $W_{\omega_3}$) from (\ref{W_omega_3})) , one easily verifies the cancellation between the subtraction terms  in $W_{\omega_3}$ and parts of the Regge pole contributions. The results for the scattering amplitudes are:
\ba
\label{final-24}
\tau_2\tau_4: & e^{-i\pi\left(\omega_1+\omega_3\right)}e^{i\pi\omega_a}\left[  \cos \pi \omega_{bc} + i \delta_{24}  +2 i f_{\omega_3}\right]\\
\label{final-234}
\tau_2\tau_3\tau_4: &-e^{-i\pi\omega_1}e^{i\pi\omega_a}\left[  \cos \pi \omega_{bc} - i \delta_{24}  - 2i f_{\omega_3}\right].
\ea
The expressions for the two regions $\tau_1\tau_3$ and $\tau_1\tau_2\tau_3$ can easily be obtained by symmetry considerations.   

For the remaining kinematic regions we have to calculate the contributions of the short cuts and of the long cut.
First we complete our calculations of the short cut in $\omega_3$:
\ba 
\tau_1 \tau_4: &2i e^{-i\pi \omega_3} \frac{\Omega_a}{\Omega_2} W_{\omega_3}\label{cut3tau14}\\
\tau_1 \tau_2 \tau_4: &  -2i e^{-i\pi \omega_3}  \frac{V_L(a)}{\Omega_2}W_{\omega_3}\label{cut3tau124} \\
\tau_1 \tau_3 \tau_4: & 2i  \frac{\Omega_a}{\Omega_2} W_{\omega_3}\label{cut3tau134}\\
\tau_1 \tau_2 \tau_3\tau_4: &   -2i \frac{V_L(a)}{\Omega_2} W_{\omega_3} \label{cut3tau1234}.
\ea
Next we turn to the long cut which is contained in the two triplets: $LLR(1)$, $LLR(2)$, $LRR(1)$, and $LRR(2)$.
Their contributions to the scattering amplitude are:
\ba
\tau_1\tau_4:& F_{LLR(1)}^{\omega_2-\omega_3-cut} + F_{LLR(2)}^{\omega_2-\omega_3-cut}\rightarrow 
2i e^{-i\pi \omega_3} \frac{W_{\omega_2\omega_3:L}}{\Omega_{32}}\\
\tau_1\tau_2\tau_4:& F_{LLR(1)}^{\omega_2-\omega_3-cut} + F_{LLR(2)}^{\omega_2-\omega_3-cut}\rightarrow 
2i e^{-i\pi \omega_3} e^{i\pi \omega_2} \frac{W_{\omega_2\omega_3:L}}{\Omega_{32}}\\
\tau_1\tau_3\tau_4:& F_{LLR(1)}^{\omega_2-\omega_3-cut} + F_{LLR(2)}^{\omega_2-\omega_3-cut}\rightarrow 
2i  \frac{W_{\omega_2\omega_3:L}}{\Omega_{32}}\\
\tau_1\tau_2\tau_3\tau_4:& F_{LLR(1)}^{\omega_2-\omega_3-cut} + F_{LLR(2)}^{\omega_2-\omega_3-cut}\rightarrow 
2i e^{-i\pi \omega_2} \frac{W_{\omega_2\omega_3:L}}{\Omega_{32}}
\ea   
and 
\ba
\tau_1\tau_4:& F_{LRR(1)}^{\omega_2-\omega_3-cut} + F_{LRR(2)}^{\omega_2-\omega_3-cut}\rightarrow 
2i e^{-i\pi \omega_2} \frac{W_{\omega_2\omega_3:R}}{\Omega_{23}}\\
\tau_1\tau_2\tau_4:& F_{LRR(1)}^{\omega_2-\omega_3-cut} + F_{LRR(2)}^{\omega_2-\omega_3-cut}\rightarrow 
2i  \frac{W_{\omega_2\omega_3:R}}{\Omega_{23}}\\
\tau_1\tau_3\tau_4:& F_{LRR(1)}^{\omega_2-\omega_3-cut} + F_{LRR(2)}^{\omega_2-\omega_3-cut}\rightarrow 
2i e^{-i\pi \omega_2} e^{i\pi \omega_3} \frac{W_{\omega_2\omega_3:R}}{\Omega_{23}}\\
\tau_1\tau_2\tau_3\tau_4:& F_{LRR(1)}^{\omega_2-\omega_3-cut} + F_{LRR(2)}^{\omega_2-\omega_3-cut}\rightarrow 
2i e^{-i\pi \omega_3} \frac{W_{\omega_2\omega_3:R}}{\Omega_{23}}.
\ea   
The sum of all four contributions, $F_{LLR(1)}^{\omega_2-\omega_3-cut} + F_{LLR(2)}^{\omega_2-\omega_3-cut} + F_{LRR(1)}^{\omega_2-\omega_3-cut} + F_{LRR(2)}^{\omega_2-\omega_3-cut}$ can be combined into: 
\ba
\tau_1\tau_4:&\rightarrow& 2i e^{-i\pi(\omega_2+\omega_3)}  \left[ e^{i\pi \omega_2} \frac{W_{\omega_2\omega_3;L}}{\Omega_{32}}+
e^{i\pi \omega_3} \frac{W_{\omega_2\omega_3;R}}{\Omega_{23}}\right]\label{longcuttau14}\\
\tau_1\tau_2\tau_4:&\rightarrow& 2ie^{-i\pi \omega_3}  \left[ e^{i\pi \omega_2} \frac{W_{\omega_2\omega_3;L}}{\Omega_{32}}+
e^{i\pi \omega_3} \frac{W_{\omega_2\omega_3;R}}{\Omega_{23}}\right]\label{longcuttau124}\\
\tau_1\tau_3\tau_4:&\rightarrow& 2ie^{-i\pi\omega_2} \left[ e^{i\pi \omega_2} \frac{W_{\omega_2\omega_3;L}}{\Omega_{32}}+
e^{i\pi \omega_3} \frac{W_{\omega_2\omega_3;R}}{\Omega_{23}}\right]\label{longcuttau134}\\
\tau_1\tau_2\tau_3\tau_4:&\rightarrow&  2i  \left[ e^{-i\pi \omega_2} \frac{W_{\omega_2\omega_3;L}}{\Omega_{32}}+
e^{-i\pi \omega_3} \frac{W_{\omega_2\omega_3;R}}{\Omega_{23}}\right].
\label{longcuttau1234}
\ea 
One easily verifies that in fact, that the regions $\tau_2\tau_4$ and $\tau_2\tau_3\tau_4$ which contain the short cut in $\omega_3$ receive no contribution from the long cut. Similarly, the regions $\tau_1\tau_3$ and $\tau_1\tau_2\tau_3$ contain only the Regge poles and the short $\omega_2$-cut.

Before we combine the contributions of the long cut with short cuts and Regge poles, let us pause for a moment 
and take a closer look at the long cut contributions. The long cut is obtained from the discontinuity in $s_{123}$ (or the $s$-discontinuity). The structure of the long cut expression (square brackets in (\ref{longcuttau14}) - (\ref{longcuttau1234})) is illustrated in Fig.\ref{longcut} (for the weak coupling see also Fig.\ref{s123-discontinuity-2to5} ):
 \begin{figure}[H]
 \centering
 \epsfig{file=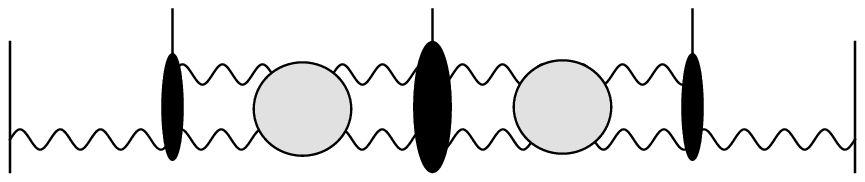, scale=0.8}
 \caption{structure of the long Regge cut}
 \label{longcut} 
 \end{figure}
 \noindent
It consists of impact factors on the left and on the right sides of the two reggeon cut which,  because of  Regge factorization, are the same as in the $2\to4$ scattering amplitude  (in the leading approximation they are illustrated in Fig.\ref{RPRR-vertex}) and the $RRPRR$ production vertex in the center.  The latter one is a new element which, in the leading order, has been calculated in \cite{Bartels:2011ge}. The phase structure contained in (\ref{longcuttau14}) -  (\ref{longcuttau1234}) indicates that, beyond the leading approximation, this production vertex must become  complex-valued. It is instructive to recapitulate the $RPR$ production vertex in the BDS formula for $2\to3$ \cite{Bartels:2013jna}. For the different kinematic regions labeled by the $\tau$-factors the relevant phase factors are:
\ba
1:  \rightarrow& e^{-i\pi (\omega_1 + \omega_2)} \left( e^{i\pi \omega_1} \frac{V_L}{\Omega_{21}} 
+ e^{i\pi \omega_2} \frac{V_R}{\Omega_{12}}\right) &=  e^{-i\pi (\omega_1 + \omega_2)} e^{i\pi \omega_a}\nonumber\\
\tau_1:  \rightarrow &e^{-i\pi \omega_2} \left( e^{i\pi \omega_1} \frac{V_L}{\Omega_{21}} 
+ e^{i\pi \omega_2} \frac{V_R}{\Omega_{12}}\right) &=  e^{-i\pi \omega_2} e^{i\pi \omega_a}\nonumber\\
\tau_2:  \rightarrow &e^{-i\pi \omega_1} \left( e^{i\pi \omega_1} \frac{V_L}{\Omega_{21}} 
+ e^{i\pi \omega_2} \frac{V_R}{\Omega_{12}}\right) &=  e^{-i\pi \omega_1} e^{i\pi \omega_a}\nonumber\\
\tau_1\tau_2:  \rightarrow & \left( e^{-i\pi \omega_1} \frac{V_L}{\Omega_{21}} 
+ e^{-i\pi \omega_2} \frac{V_R}{\Omega_{12}}\right) &=   e^{-i\pi \omega_a}.
\ea        
This phase structure allows for two equivalent descriptions: either we write a sum of two terms with real-valued 
coefficients $V_L(a)$ and $V_R(a)$ (in agreement with the Steinmann relations) or, alternatively,  we use a factorized representation with the complex-valued production vertex $e^{i\pi \omega_a}$. Comparing the bracket expressions with the square brackets in  (\ref{longcuttau14}) - (\ref{longcuttau1234}) we find the same phase structure. Therefore, for the RRPRR-vertex in the center of Fig.\ref{longcut}, we either retain the sum of the two terms with real-valued coefficients contained in $W_{\omega_2-\omega_3;L}$ and  $W_{\omega_2-\omega_3;R}$, or we introduce a complex-valued production vertex. In leading order, this vertex is real. In contrast to the Regge pole case in $2\to3$, we do not yet know the complex-valued RRPRR vertex function beyond the leading order. It is tempting to expect, again, some form of exponentiation.   

Finally we combine the contributions of the long cut, the contributions of the short cut in  with the Regge pole contributions which are taken from the appendix of  \cite{Bartels:2013jna}. The latter ones are:
\ba
\tau_1\tau_4: &e^{-i\pi\left(\omega_2+\omega_3\right)}\left[e^{i\pi\left(\omega_a+\omega_b+\omega_c\right)} - 2ie^{i\pi\left(\omega_2+\omega_3\right)} \frac{\Omega_a\Omega_b\Omega_c}{\Omega_2\Omega_3}\right]\\
\tau_1\tau_2\tau_4:&-e^{-i\pi\omega_3}\left[e^{i\pi\left(-\omega_a+\omega_b+\omega_c\right)} - 2ie^{i\pi\omega_3}
\frac{\Omega_{2a}\Omega_b\Omega_c}{\Omega_2\Omega_3} \right]\\
\tau_1\tau_3\tau_4:&-e^{-i\pi\omega_2}\left[e^{i\pi\left(\omega_a+\omega_b-\omega_c\right)} - 2ie^{i\pi\omega_2}
\frac{\Omega_a\Omega_b\Omega_{3c}}{\Omega_2\Omega_3} \right]\\
\tau_1\tau_2\tau_3\tau_4:&\left[e^{i\pi\left(-\omega_a+\omega_b-\omega_c\right)} - 
2i\frac{\Omega_{2a}\Omega_b\Omega_{3c}}{\Omega_2\Omega_3} \right].
\ea
The contributions of the long cuts (\ref{longcuttau14}) - (\ref{longcuttau1234}), together with those of the 
short cut in $\omega_3$ (\ref{cut3tau14}) -  (\ref{cut3tau1234}) (and analogous expressions for the short cut in $\omega_2$) are 
\ba
\label{cutpaper-14}
\tau_1\tau_4: \rightarrow& 2i e^{-i\pi(\omega_2+\omega_3)} \left( \left[ e^{i\pi \omega_2} 
\frac{W_{\omega_2-\omega_3;L}}{\Omega_{32}} +e^{i\pi \omega_3}  \frac{W_{\omega_2-\omega_3;R}}{\Omega_{23}}\right]
+ e^{i\pi \omega_2} \frac{\Omega_{a} W_{\omega_3}}{\Omega_2}
 + e^{i\pi \omega_3} \frac{W_{\omega_2} \Omega_c }{\Omega_3}
\right)\nonumber\\
=&2i e^{-i\pi(\omega_2+\omega_3)} \left( e^{i\pi \omega_2} \left[\frac{W_{\omega_2-\omega_3;L}}{\Omega_{32}} + 
\frac{\Omega_{a} W_{\omega_3} }  {\Omega_2} \right] 
+e^{i\pi \omega_3} \left[ \frac{W_{\omega_2-\omega_3;R}}{\Omega_{23}}
+ \frac{W_{\omega_2} \Omega_c }{\Omega_3}\right]
\right)
\nonumber\\
& \\
\label{cutpaper-124}
\tau_1\tau_2\tau_4: \rightarrow&  2ie^{-i\pi\omega_3}  \left( \left[ e^{i\pi \omega_2} 
\frac{W_{\omega_2-\omega_3;L}}{\Omega_{32}} +e^{i\pi \omega_3}  \frac{W_{\omega_2-\omega_3;R}}{\Omega_{23}}\right]
-  \frac{\Omega_{2a} W_{\omega_3}}{\Omega_2}
 + e^{i\pi \omega_3} \frac{W_{\omega_2} \Omega_c }{\Omega_3}
\right) \nonumber\\
 =& 2ie^{-i\pi\omega_3} \left( e^{i\pi \omega_2} \left[\frac{W_{\omega_2-\omega_3;L}}{\Omega_{32}} + 
\frac{\Omega_{a} W_{\omega_3} }  {\Omega_2} \right] 
+e^{i\pi \omega_3} \left[ \frac{W_{\omega_2-\omega_3;R}}{\Omega_{23}}
+ \frac{W_{\omega_2} \Omega_c }{\Omega_3}\right]- e^{i\omega_a} W_{\omega_3}
\right)\nonumber\\
& \\
\label{cutpaper-134}
\tau_1\tau_3\tau_4: \rightarrow&  2ie^{-i\pi\omega_2}  \left( \left[ e^{i\pi \omega_2} 
\frac{W_{\omega_2-\omega_3;L}}{\Omega_{32}} +e^{i\pi \omega_3}  \frac{W_{\omega_2-\omega_3;R}}{\Omega_{23}}\right]
+ e^{i\pi \omega_2} \frac{\Omega_a W_{\omega_3}}{\Omega_2}
 -  \frac{W_{\omega_2} \Omega_{3c} }{\Omega_{3}}
\right) \nonumber\\
=& 2ie^{-i\pi\omega_2} \left( e^{i\pi \omega_2} \left[\frac{W_{\omega_2-\omega_3;L}}{\Omega_{32}} + 
\frac{\Omega_{a} W_{\omega_3} }  {\Omega_2} \right] 
+e^{i\pi \omega_3} \left[ \frac{W_{\omega_2-\omega_3;R}}{\Omega_{23}}
+ \frac{W_{\omega_2} \Omega_c }{\Omega_3}\right]- W_{\omega_2} e^{i\omega_c}
\right)\nonumber\\
& \\
\label{cutpaper-1234}
\tau_1\tau_2\tau_3\tau_4: \rightarrow&2i   \left( \left[ e^{-i\pi \omega_2} 
\frac{W_{\omega_2-\omega_3;L}}{\Omega_{32}} +e^{-i\pi \omega_3}  \frac{W_{\omega_2-\omega_3;R}}{\Omega_{23}}\right]
-  \frac{\Omega_{2a} W_{\omega_3}}{\Omega_2}
 -  \frac{W_{\omega_2} \Omega_{3c} }{\Omega_3}
\right)\nonumber\\
=& 2i \left( e^{-i\pi \omega_2} \left[\frac{W_{\omega_2-\omega_3;L}}{\Omega_{32}} + 
\frac{\Omega_{a} W_{\omega_3} }  {\Omega_2} \right] 
+e^{-i\pi \omega_3} \left[ \frac{W_{\omega_2-\omega_3;R}}{\Omega_{23}}
+ \frac{W_{\omega_2} \Omega_c }{\Omega_3}\right]\right.
\nonumber\\
&\left. - W_{\omega_2} e^{-i\omega_c}-e^{-i\omega_a} W_{\omega_3}\right). 
\ea
For the expressions on the rhs of these equations we have only weak coupling limit results. Disregarding the phases and using (\ref{W-long-L-R1}) we find for the region $\tau_1\tau_4$:
\be
 e^{-i\pi(\omega_2+\omega_3)} \left( i \delta_{14} + 2 if_{\omega_2\omega_3} + 
i \pi \frac{\omega_{a} \omega_b \omega_{c}}{\omega_2 \omega_3}-i \pi (\omega_a+\omega_c) \right).
\ee
After combination with the Regge pole term, we find the expected cancellation between the subtraction term and the Regge pole piece, and we arrive at:  
\be
\tau_1\tau_4: \rightarrow e^{-i\pi(\omega_2+\omega_3)} \left(
1+ i\pi\omega_b +i\delta_{14} +2i f_{\omega_2\omega_3}
\right).
\ee
In the same way we compute the other regions and obtain
\ba
\tau_1\tau_2\tau_4: &\rightarrow -e^{-i\pi\omega_3} \Big( 1+ i\pi \omega_c - i\delta_{124} -2 i \left( f_{\omega_2\omega_3}  - f_{\omega_3}\right) \Big) \\
\tau_1\tau_3\tau_4: &\rightarrow  -e^{-i\pi\omega_2} \Big( 1+ i\pi \omega_a - i\delta_{134} -2 i \left( f_{\omega_2\omega_3}  - f_{\omega_2}\right) \Big) \\
\tau_1\tau_2\tau_3\tau_4: &\rightarrow 1+ i \delta_{1234} - i\pi \omega_b+ 2i \left(f_{\omega_2\omega_3}  - f_{\omega_2} - f_{\omega_3}     \right),
\ea
where 
\ba
\delta_{124}&=& \pi \left( V_{14} - V_{24} +\omega_a -\omega_b \right) \\
\delta_{134}&=& \pi \left( V_{14} - V_{13} +\omega_c -\omega_b \right) \\
\delta_{1234} &=& \pi \left( V_{14} - V_{13} - V_{24} -\omega_a -\omega_c \right) .
\ea
Note that, in analogy with our remark at the end of the $2\to4$ section, the term $V_{14}$ is the one loop approximation of the long Regge cut and is contained in the BDS formula; the same holds for the terms  and $V_{13}$ and $V_{24}$  which represent the one-loop approximations of the short cuts in the $\omega_2$ and $\omega_3$ channels.  

Making use of the results from \cite{Bartels:2013jna} we can slightly generalize our results. As discussed in 
Appendix B, our weak coupling results for the partial waves are in agreement with the subtractions predicted in \cite{Bartels:2013jna}, and in this paper it was shown that they remove all the unwanted pieces of the Regge pole terms. Therefore, this part of our results - the combination of subtraction terms with the Regge pole terms - can be 
generalized to all orders, and our restriction to the leading logarithmic approximation only applies to the calculation of the Regge cut contributions. From the second lines in (\ref{cutpaper-14}) - (\ref{cutpaper-1234}) we infer that the partial waves  $f_{\omega_3}$ ($f_{\omega_2}$) contained in $W_{\omega_3}$ ($W_{\omega_2}$) are multiplied by phase factors $e^{i \pi \omega_a}$ or $e^{-i \pi \omega_a}$ ($e^{i \pi \omega_c}$ or $e^{-i \pi \omega_c}$ ). We therefore write:
\ba
\label{Tfinal-14}
\tau_1\tau_4: & \rightarrow e^{-i\pi(\omega_2+\omega_3)}   \left[
e^{i\pi \omega_b}\cos \pi \omega_{ac}  +i\delta_{14}+2i f_{\omega_2\omega_3}
\right],\\
\label{Tfinal-124}
\tau_1\tau_2\tau_4: &\rightarrow -e^{-i\pi\omega_3} \left[ e^{i\pi \omega_c} \cos \pi \omega_{ab} -i\delta_{124}-2 i \left( f_{\omega_2\omega_3}  - e^{i \pi \omega_a} f_{\omega_3}\right) \right] \\
\label{Tfinal-134}
\tau_1\tau_3\tau_4: &\rightarrow  -e^{-i\pi\omega_2} \left[ e^{i\pi \omega_a}\cos \pi \omega_{bc} -i\delta_{134} -2 i \left( f_{\omega_2\omega_3}  - f_{\omega_2}e^{i \pi \omega_c}\right) \right] \\
\label{Tfinal-1234}
\tau_1\tau_2\tau_3\tau_4: &\rightarrow  
\left[ e^{-i\pi \omega_b}e^{i\pi \omega_{ba}} e^{i\pi \omega_{bc}}+2i \left(f_{\omega_2\omega_3} +i\delta_{1234} - f_{\omega_2}e^{-i \pi \omega_a} - e^{-i \pi \omega_a}f_{\omega_3}     \right)\right].
\ea
In order to pass to the conformally invariant remainder functions $R_{7;\tau_i\tau_j...\tau_k}$, we first recapitulate the relation between our scattering amplitude, the BDS amplitude, and the remainder function:   
\be
T_{2\to 5} = T_{2\to45}^{Born} \times T^{BDS} \times R. 
\ee
Here the BDS amplitude contains kinematic phases (e.g. $e^{-i\pi(\omega_2+\omega_3)}$ 
for the region $\tau_1\tau_4$),  the exponentials of production vertices: $e^{i\pi \omega_b}$, $e^{i\pi \omega_c}$, $e^{i\pi \omega_a}$, $e^{-i\pi \omega_b}$ for the regions $\tau_1\tau_4$, $\tau_1\tau_2,\tau_4$ $\tau_1\tau_3\tau_4$, $\tau_1\tau_2\tau_3\tau_4$, resp., and the phases $e^{i \delta_{ij...k}}$. Finally, the Born amplitude $T^{Born}$ is proportional  to $s$ which, when introducing a further twist in a t-channel, produces a minus sign (for example, when going from 
$\tau_1\tau_4$ to $\tau_1\tau_2\tau_4$).  These factors, therefore, have to be taken into account in our expressions for the scattering amplitudes in (\ref{Tfinal-14}) - (\ref{Tfinal-1234}), before we arrive at the remainder functions $R_7$. 

Before we write down our results for the remainder functions we want to make a further comment on the Regge cut amplitude $f_{\omega_2\omega_2}$. Since this amplitude is known only to leading order accuracy, we will not be able to write all the phase factors for this term. Beyond the leading order however, we know from our discussion after (\ref{longcuttau1234}) and from  (\ref{cutpaper-14}) - (\ref{cutpaper-1234}) that the RRPRR vertex becomes complex. As a result, also the amplitude $f_{\omega_2\omega_3}$ will become complex and the exponential of the production vertices can no longer be disregarded. For the region $\tau_1\tau_4$ this means:      
\be
\label{RRPRRvertex}
f_{\omega_2\omega_3} \rightarrow  e^{-i\pi \omega_b}\frac{e^{i\pi\omega_2} f_{\omega_2\omega_3;L}- e^{i\pi\omega_2} f_{\omega_2\omega_3:R}}{\Omega_{32}}.
\ee
Our prediction for higher order, therefore, is that the rhs of (\ref{RRPRRvertex}) must be conformally invariant. Finally, it is  customary to present results for the 
product of the remainder functions and the the phases $e^{i \delta_{ij...k}}$ which are part of the BDS formula.

With these modifications our final results for the remainder function become\footnote{
In our previous paper \cite{Bartels:2013jna} the remainder function was defined to include the sign changes due to the $s$ factors of the Born term. As a result, kinematic regions belonging to an odd number of $\tau$-factors have a global minus sign.}      
\begin{eqnarray}
\label{final-14}
R_{7;\tau_1\tau_4}e^{i\pi\delta_{14}}&=&\cos\pi\omega_{ac}+i\pi\delta_{14}+2i  f_{\omega_2\omega_3}\\
\label{final-124}
R_{7;\tau_1\tau_2\tau_4}e^{-i\pi\delta_{124}}&=&\cos\pi\omega_{ab}-i\pi\delta_{124}-2i (
f_{\omega_2\omega_3} - e^{i\pi \omega_{ac}}f_{\omega_3})\\
\label{final-134}
R_{7;\tau_1\tau_3\tau_4}e^{-i\pi\delta_{134}}
&=&\cos\pi\omega_{bc}-i\pi\delta_{134} -2if_{\omega_2\omega_3}-e^{i\pi \omega_{ca}} f_{\omega_2})\\
\label{final-1234}
R_{7;\tau_1\tau_2\tau_3\tau_4}e^{i\pi\delta_{1234}} &=&
e^{i\pi \omega_{ba}} e^{i\pi \omega_{bc}}+ i\pi\delta_{1234}+2i(f_{\omega_2\omega_3}
-e^{i\pi \omega_{bc}}f_{\omega_2}-e^{i\pi \omega_{ba}}f_{\omega_3})
\end{eqnarray}
Eqs.(\ref{final-24}),(\ref{final-234}), (\ref{final-14}), (\ref{final-124}), (\ref{final-134}), and (\ref{final-1234}), represent our final results. All  unphysical  singularities have been cancelled, and the final expressions consist of conformally invariant Regge pole and Regge cut contributions.

\section{Summary and conclusions}

In this paper we have completed our analysis of the $n=7$ BDS scattering amplitude in the multi-Regge limit. To summarize the result of this work, we once more list the final results for the remainder function $R_{7;\tau_i\tau_j...\tau_k}$ in the different Mandelstam kinematic regions labeled by $\tau_i\tau_j...\tau_k$. We follow the definitions given in our previous paper \cite{Bartels:2013jna}: from the expressions listed in the previous section, we remove the kinematic phase factors and exponents of the production vertices, $e^{\pm i\pi \omega_a}, e^{\pm i\pi \omega_b}, e^{\pm i\pi \omega_c}$ which are already part of the BDS formula. The final expressions are sums of the  conformally invariant contributions of Regge poles and Regge cuts:
\begin{eqnarray}
R_{7;\tau_2\tau_4}e^{i\delta_{24}}&=&\cos\pi\omega_{bc}+i\delta_{24}+2if_{\omega_3}\\
R_{7;\tau_2\tau_3\tau_4}e^{-i\delta_{24}}&=&\cos\pi\omega_{bc}-i\delta_{24}-2if_{\omega_3}\\
R_{7\tau_1\tau_3}e^{i\delta_{13}}&=&\cos\pi\omega_{ab}+i\delta_{13}+2if_{\omega_2}\\
R_{7;\tau_1\tau_2\tau_3}e^{-i\delta_{13}}&=&\cos\pi\omega_{ab}-i\delta_{13}-2if_{\omega_2}\\
R_{7;\tau_1\tau_4}e^{i\pi\delta_{14}}&=&\cos\pi\omega_{ac}+i\pi\delta_{14}+2i  f_{\omega_2\omega_3}\\
R_{7;\tau_1\tau_2\tau_4}e^{-i\pi\delta_{124}}&=&\cos\pi\omega_{ab}-i\pi\delta_{124}-2i (f_{\omega_2\omega_3} - e^{i\pi \omega_{ac}}f_{\omega_3})\\
R_{7;\tau_1\tau_3\tau_4}e^{-i\pi\delta_{134}}
&=&\cos\pi\omega_{bc}-i\pi\delta_{134} -2if_{\omega_2\omega_3}-e^{i\pi \omega_{ca}} f_{\omega_2})\\
R_{7;\tau_1\tau_2\tau_3\tau_4}e^{i\pi\delta_{1234}} &=&
e^{i\pi \omega_{ba}} e^{i\pi \omega_{bc}}+ i\pi\delta_{1234}+2i(f_{\omega_2\omega_3}
-e^{i\pi \omega_{bc}}f_{\omega_2}-e^{i\pi \omega_{ba}}f_{\omega_3})
\end{eqnarray}
The Regge cut amplitides $f_{\omega_2}, f_{\omega_3}$ and $f_{\omega_2\omega_3}$ are explicitly given in \eqref{W-omega2-cut} and in \eqref{f-long}, respectively. 
As stated before, these Regge cut contributionsare are valid only in the weak coupling approximation: this restriction comes from the calculation of unitarity integrals in which we have used leading order amplitudes $M_{n\to m}$ and from the use of bootstrap relations. As explained at the end 
of the previous section, in next-to-leading order the long cut amplitude $f_{\omega_2\omega_3}$ 
is expected to become complex. Since production vertices have been calculated in NLO \cite{Fadin:1993wh,Fadin:2014gra} and  bootstrap equations have been proven to be valid also in NLO \cite{Bartels:2003jq,Fadin:2002hz}, all ingredients for a  complete NLO analysis are available. 

It is important to note that recently both the $n=6$ \cite{Bartels:2010ej,Bartels:2013dja} and $n=7$ scattering amplitudes \cite{Bartels:2012gq,Bartels:2014ppa,BSS} in multi-Regge kinematics have been investigated in the strong coupling region. The results show a remarkable consistency between the structure at weak and strong coupling, thus providing strong support for the AdS/CFT duality hypothesis. \\ \\  

\section*{Acknowledgments}

Two of us (A. Kormilitzin and L. N. Lipatov) want to thank the II.Institute for Theoretical Physics, University of Hamburg, and DESY for their hospitality. L.N.Lipatov is grateful for the support of the Alexander von Humboldt Foundation, and  A. Kormilitzin gratefully acknowledges the support of the Minerva Fellowship.

\appendix
\section{Partial waves and multiparticle amplitudes}

In this appendix we review and present further details of the Regge pole analysis, derived from models \cite{Brower:1974yv} (scalar field theories and dual amplitudes) and from S-matrix theory \cite{White:1976qm, Stapp:1982mq}.  

We begin with a brief review of the simplest examples, namely the  $2\to3$ and $2\to4$ scattering amplitudes in the multi-Regge limit, which have been discussed before. The possible energy discontinuities of the $2\to3$ case are illustrated in Figs.10a. 
\begin{figure}[H]
\centering
\includegraphics[scale=0.8]{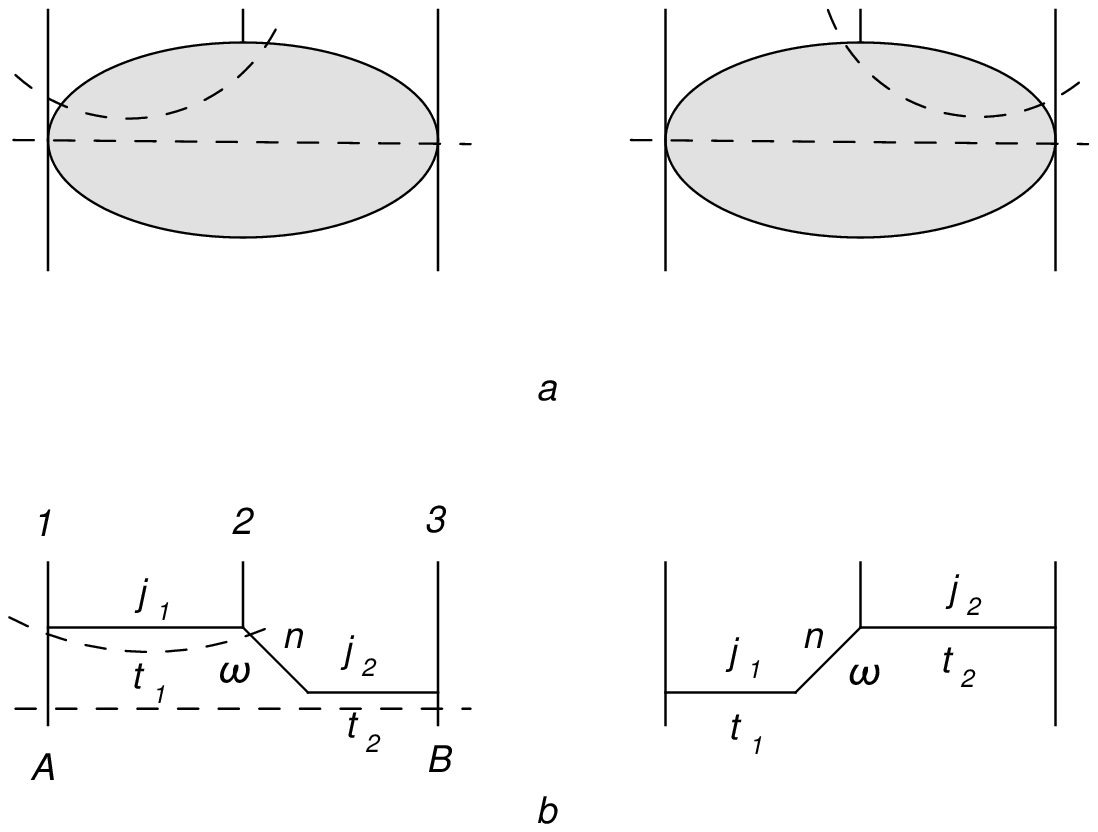}
\caption{(a) the two sets of energy discontinuities; (b) the corresponding hexagraphs}  
\label{2to3}
\end{figure}
\noindent
In a Regge pole description, one starts from multiple partial wave expansions in the crossed channels. For the $2\to3$ case, such an expansion contains the triple sum over the angular momenta in the $t_1$ and $t_2$ channels, $j_1$ and $j_2$, and the helicity variable $n$ conjugated to the Toller angle $\omega$ at the production vertex. As it was pointed out in \cite{White:1976qm}, the definition of the multiple partial wave and its subsequent analytic continuation to complex values of angular momenta and helicity requires a decomposition of the scattering amplitude into separate pieces (spectral components), which correspond to the two terms in Fig.\ref{2to3}. Each term allows us the construction of a Froissart-Gribov partial wave, and the analytic continuation can be done in two of the three angular momentum variables $j_1$, $j_2$ and $n$. The coupling of these variables is illustrated in hexagraphs, shown in Fig.\ref{2to3}b. Disregarding all complications which are unnecessary for the present discussion, we have      
\be
T_{2\to3} =\sum_n \sum_{j_1\ge n} \sum_ {j_2 \ge n} d_{0n}^{j_1}(\cos \theta_1) u^n d_{n0}^{j_2}(\cos \theta_2) F(j_1,j_2,n;t_1,t_2),
\ee
where $u=e^{i\omega}$, and $\theta_1$ and $\theta_2$ denote the scattering angles in the $t_1$ and $t_2$ channels respectively. As mentioned before, the Sommerfeld-Watson transformation and analytic continuation in $j_1$, $j_2$ and $n$ requires the decomposition into two terms. The first one (left parts of Fig.\ref{2to3}a and b) reads as:
\be
T^{(1)}_{2\to3} = \frac{1}{(2\pi i)^2} \int \int \frac{dj_1}{\sin \pi (j_1-n)} \frac{dn}{\sin\pi n} \sum_{N=0}^{\infty}  
d_{0n}^{j_1}(\cos \theta_1) u^n d_{n0}^{n+N}(\cos \theta_2)F^{(1)}(j_1,j_2,n;t_1,t_2),
\ee 
i.e. we have put $j_2=n+N$, where $N\in\mathbb{Z}$. Assuming the existence of Regge poles at $j_1=\alpha_1$ and $n=\alpha_2 -N$ with factorizing residues, we have for $s_1\sim \cos \theta_1 \to \infty$ and $s_2\sim \cos \theta_2 \to \infty$, the 
Regge form 
\be
 T^{(1)}_{2\to3} = s_1^{\alpha_1} s_2^{\alpha_2} u^{\alpha_2}  
 \beta(t_1) \Gamma(-\alpha_1) \frac{\sin \pi \alpha_1 \tilde{V}_R (t_1,t_2,u)}{\sin \pi (\alpha_1 -\alpha_2)}  \Gamma(-\alpha_2)\beta(t_2).
 \ee
Here we have used the fact that for large $z$,  
\be 
\label{d-function}
d_{on}^{j}(z) \sim \frac{z^j}{\Gamma(1+j)}
\ee
and 
\be
\label{Gamma-function}
\frac{1}{\Gamma(1+j)} = \frac{-\sin \pi j}{\pi} \Gamma(-j). 
\ee
The $\Gamma$-function on the rhs contains the particle pole in $t$-channel. The vertex function $V_R$ (for  a massive theory) is analytic in  $\eta=u^{-1}$ near $\eta=0$. Moreover, since in the multi-Regge limit $\eta=u^{-1} \sim \frac{s_1 s_2}{s}$, we can also write
 \be
 T^{(1)}_{2\to3} = s_1^{\alpha_1-\alpha_2} s^{\alpha_2} \beta(t_1) \Gamma(-\alpha_1) 
   \frac{\sin \pi \alpha_1 \tilde{V}_R (t_1,t_2,u)}{\sin \pi (\alpha_1 -\alpha_2)} \Gamma(-\alpha_2) \beta(t_2).
 \ee
Here the energy factors are in accordance with the singularity structure illustrated in the upper line of 
Fig.\ref{2to3}a (left part). In the same way, the right part of Fig.\ref{2to3} corresponds to the second part of the scattering amplitude:
\be
T^{(2)}_{2\to3} = \frac{1}{(2\pi i)^2} \int \int \frac{dj_2}{\sin \pi (j_2-n)} \frac{dn}{\sin\pi n} \sum_{N=0}^{\infty}  
d_{0n}^{n+N}(\cos \theta_1) u^n d_{n0}^{j_2}(\cos \theta_2)F^{(2)}(j_1,j_2,n;t_1,t_2).
\ee 
With Regge poles at $j_2=\alpha_2$ and $n=\alpha_1-N$ we arrive at
\be
 T^{(2)}_{2\to3} = s_2^{\alpha_2-\alpha_1} s^{\alpha_1} \beta(t_1) \Gamma(-\alpha_1)
   \frac{\sin \pi \alpha_2 \tilde{V}_L (t_1,t_2,u)}{\sin \pi (\alpha_2 -\alpha_1)} \Gamma(-\alpha_2)  \beta(t_2).
 \ee
In the following it will be convenient to define
\be
\sin \pi \alpha_1 \tilde{V}_R (t_1,t_2,u) = V_R(t_1,t_2,u)
\label{defVR}
\ee
and
\be
\sin \pi \alpha_2 \tilde{V}_L (t_1,t_2,u) = V_L(t_1,t_2,u).
\label{defVL}
\ee
Note that this definition of the production vertices (apart from constant factors) is in accordance with the notation used in the main part of our paper.

We generalize this to higher order amplitudes. Let us consider the $2\to4$ case. Turning to the $2\to 4$ process, we only emphasize the new feature. Obviously, we now have five different ways of drawing maximal sets of non-overlapping energy variables, and each such diagram has its own hexagraph: 
\begin{figure}[H]
\centering
\includegraphics[scale=0.8]{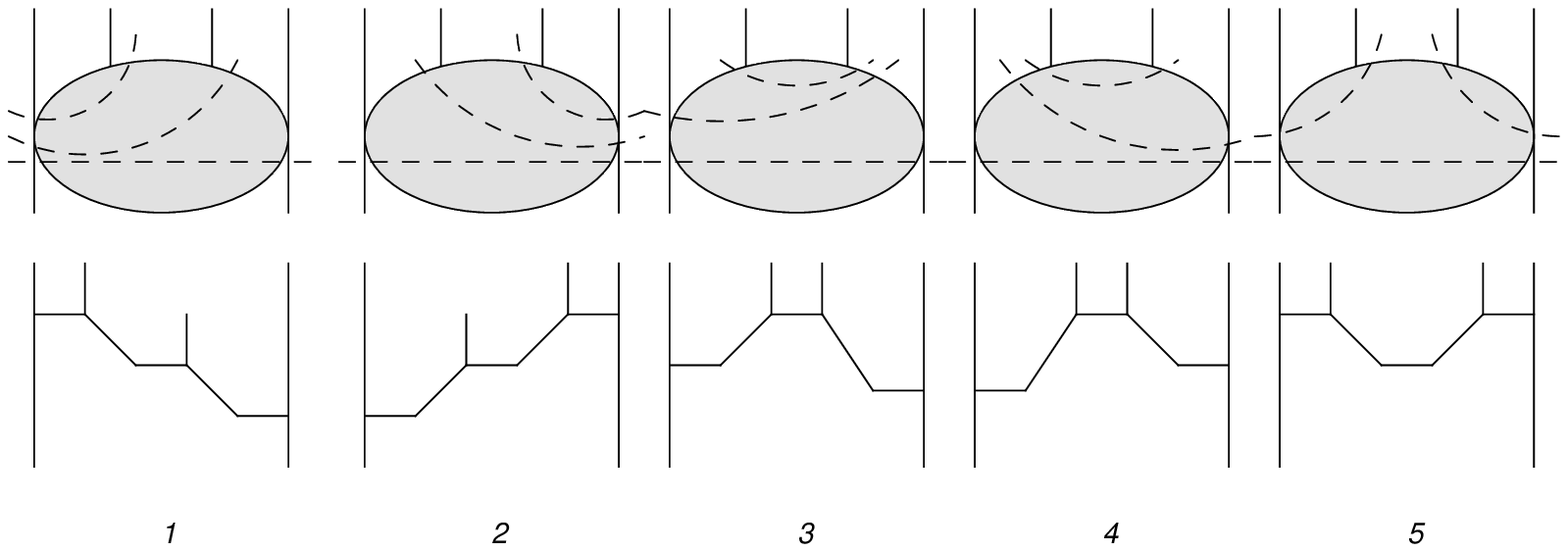}
\caption{the five sets of energy discontinuities (upper row) and the corresponding hexagraphs (lower row)}  
\label{2to4}
\end{figure}
\noindent
Focusing on the terms "3" and "4", we have:
\ba
T^{(3)}_{2\to 4} = &&\frac{1}{(2\pi i)^3} \int \int \int \frac{dj_2}{\sin \pi (j_2-n_1)}  \frac{dn_1}{\sin \pi (n_1-n_2)}  \frac{dn_2}{\sin \pi n_2}\times
\\ &&\times\sum_{N_1} \sum_{N_2}  u_1^{n_1} u_2^{n_2} 
d_{0n_1}^{n_1+N_1}(\cos \theta_1)d_{n_1}^{j_2}(\cos \theta_2)d_{n_2}^{n_2+N_2}(\cos \theta_3) F^{(3)}(j_1,j_2,j_3,t_1,t_2,t_3,n_1,n_2).\nonumber
\ea
Assuming the existence of Regge poles at
\be
j_2=\alpha_2, \;\;\,n_1=\alpha_1-N_1,\;\;\;n_2=\alpha_3-N_2
\ee
we obtain: 
\be
 T^{(3)}_{2\to4} = s_1^{\alpha_1} s_2^{\alpha_2} s_3^{\alpha_3} u_1^{\alpha_1} u_2^{\alpha_3}\nonumber 
 \ee
 \be   
 \times \beta(t_1) \Gamma(-\alpha_1) \frac{\sin \pi \alpha_1 \sin\pi\alpha_2 \tilde{V}_L (t_1,t_2,u_1) \Gamma(-\alpha_2)\tilde{V}_R (t_2,t_3,u_2)}
  {\sin \pi (\alpha_2 -\alpha_1)\sin\pi(\alpha_1-\alpha_3)} 
   \Gamma(-\alpha_3)
 \beta(t_3).
  \label{2to4piece3pole}
 \ee
Here the energy factors in the first line can also be written as
 \be
 s_1^{\alpha_1} s_2^{\alpha_2} s_3^{\alpha_3} u_1^{\alpha_1} u_2^{\alpha_3} = s_2^{\alpha_2-\alpha_1} s_{123}^{\alpha_1-\alpha_3} s^{\alpha_3},
 \ee 
 which is in agreement with the energy singularity structure of this term. Similarly, the fourth term corresponds to  
\ba
T^{(4)}_{2\to4} = &&\frac{1}{(2\pi i)^3} \int \int \int \frac{dj_2}{\sin \pi (j_2-n_2)} \frac{dn_1}{\sin \pi (n_2-n_1)} \frac{dn_2}{\sin \pi n_1}
\times \\ \times &&\sum_{N_1} \sum_{N_2} 
 \times u_1^{n_1} u_2^{n_2} 
d_{0n_1}^{n_1+N_1}(\cos \theta_1)d_{n_1}^{j_2}(\cos \theta_2)d_{n_2}^{n_2+N_2}(\cos \theta_3) F^{(4)}(j_1,j_2,j_3,t_1,t_2,t_3,n_1,n_2)\nonumber  
\ea
and
\ba\label{2to4piece4pole}
T^{(4)}_{2\to4} = && s_1^{\alpha_1} s_2^{\alpha_2} s_3^{\alpha_3} u_1^{\alpha_1} u_2^{\alpha_3}\times \\ &&\times \beta(t_1) \Gamma(-\alpha_1) \frac{\sin \pi \alpha_2 \sin\pi\alpha_3 \tilde{V}_L (t_1,t_2,u_1) \Gamma(-\alpha_2)\tilde{V}_R (t_2,t_3,u_2)} {\sin \pi (\alpha_2 -\alpha_3)\sin\pi(\alpha_3-\alpha_1)} \Gamma(-\alpha_3) \beta(t_3)\nonumber 
\ea
with
\be
 s_1^{\alpha_1} s_2^{\alpha_2} s_3^{\alpha_3} u_1^{\alpha_1} u_2^{\alpha_3} = s_2^{\alpha_2-\alpha_3} s_{234}^{\alpha_3-\alpha_1} s^{\alpha_1}.
 \ee 
In order to exhibit Regge factorization we  use the definitions (\ref{defVR}) and  (\ref{defVL}) and rewrite   (\ref{2to4piece3pole}) and (\ref{2to4piece4pole}) as follows
\ba\label{2to4piece3polefac}   
T^{(3)}_{2\to4}  &&=s_1^{\alpha_1} s_2^{\alpha_2} s_3^{\alpha_3} u_1^{\alpha_1} u_2^{\alpha_3}\times \\ &&\times\beta(t_1) \Gamma(-\alpha_1) \frac{\sin \pi \alpha_1 \sin\pi(\alpha_2-\alpha_3)} {\sin \pi \alpha_2 \sin\pi(\alpha_1-\alpha_3)} \frac{V_L (t_1,t_2,u_1)}{\sin \pi (\alpha_2 -\alpha_1)}  \Gamma(-\alpha_2)\frac{V_R (t_2,t_3,u_2)}{\sin\pi(\alpha_2-\alpha_3)}\Gamma(-\alpha_3)\beta(t_3)\nonumber
\ea
and 
\ba \label{2to4piece4polefac}  
T^{(4)}_{2\to4}  &&= s_1^{\alpha_1} s_2^{\alpha_2} s_3^{\alpha_3} u_1^{\alpha_1} u_2^{\alpha_3}\times \\ &&\times\beta(t_1) \Gamma(-\alpha_1) \frac{\sin \pi \alpha_3 \sin\pi(\alpha_2-\alpha_1)} {\sin \pi \alpha_2 \sin\pi(\alpha_3-\alpha_1)} \frac{V_L (t_1,t_2,u_1)}{\sin \pi (\alpha_2 -\alpha_1)}  \Gamma(-\alpha_2)\frac{V_R (t_2,t_3,u_2)}{\sin\pi(\alpha_2-\alpha_3)}\Gamma(-\alpha_3)\beta(t_3).\nonumber
\ea
The trigonometric prefactors in (\ref{2to4piece3polefac}) and (\ref{2to4piece4polefac}) agree with those of (\ref{2to4-partialwavesLR1}) and (\ref{2to4-partialwavesLR2}). The denominators $ \sin \pi \alpha_i$ result from the definitions (\ref{defVR}) and  (\ref{defVL}), i.e from the requirement that each production vertex can be written in the form  
\be
\label{leftvertex}
\frac{V_L (t_1,t_2,u_1)}{\sin \pi (\alpha_2 -\alpha_1)}= \frac{V_L (t_1,t_2,u_1)}{\Omega_{21}}
\ee
or 
\be
\label{rightvertex}
\frac{V_R (t_1,t_2,u_1)}{\sin \pi (\alpha_1 -\alpha_2)}= \frac{V_R (t_1,t_2,u_1)}{\Omega_{12}},
\ee
where $\Omega_{ij} = \sin \pi (\omega_i - \omega_j)$. Let us generalize the construction of these trigonometric prefactors to general $2\to n-2$ amplitudes.
We find it convenient to first draw the hexagraphs. For the example of the $2\to5$ case, the 14 terms with energy discontinuities have been presented in Figs.\ref{f3_to_w3} - \ref{f3_to_w3triplets}. Here we list the corresponding hexagraphs:    
\begin{figure}[H]
\centering
\includegraphics[scale=0.8]{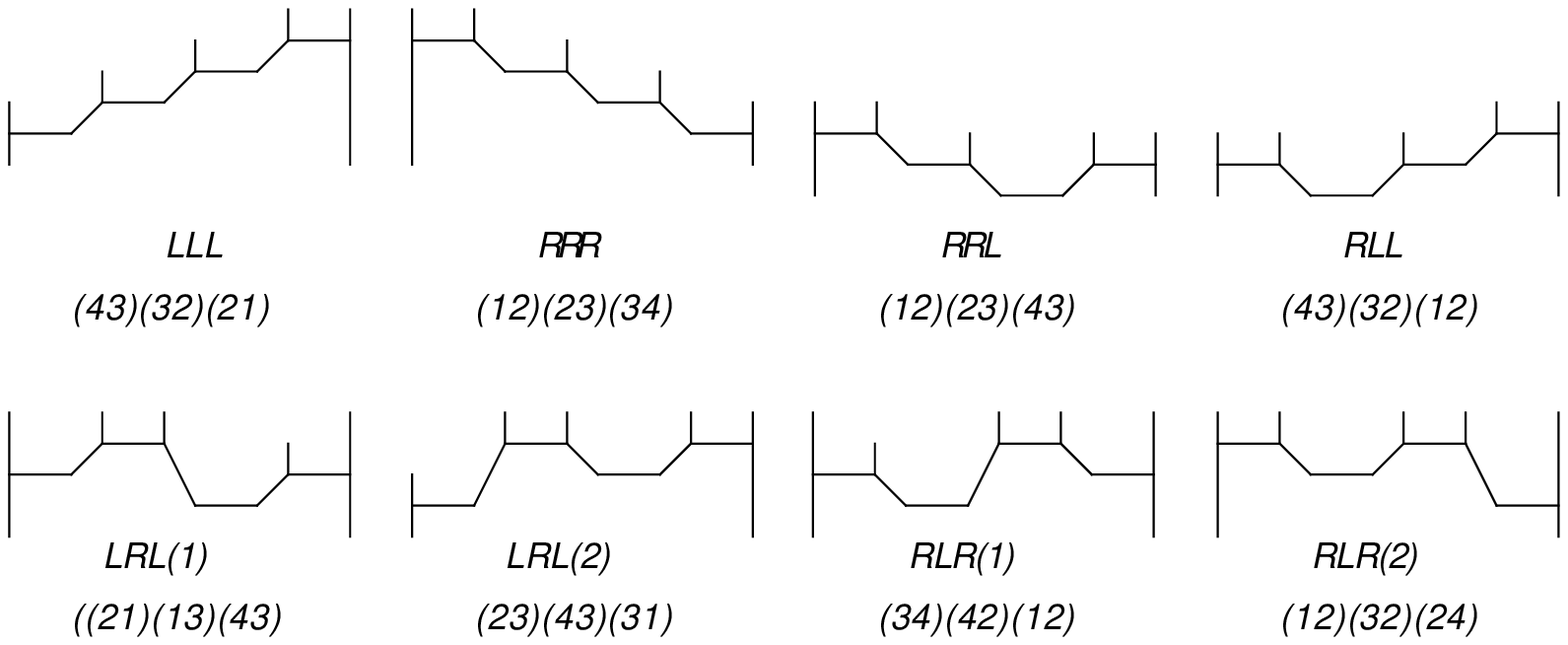}
\centering
\includegraphics[scale=0.8]{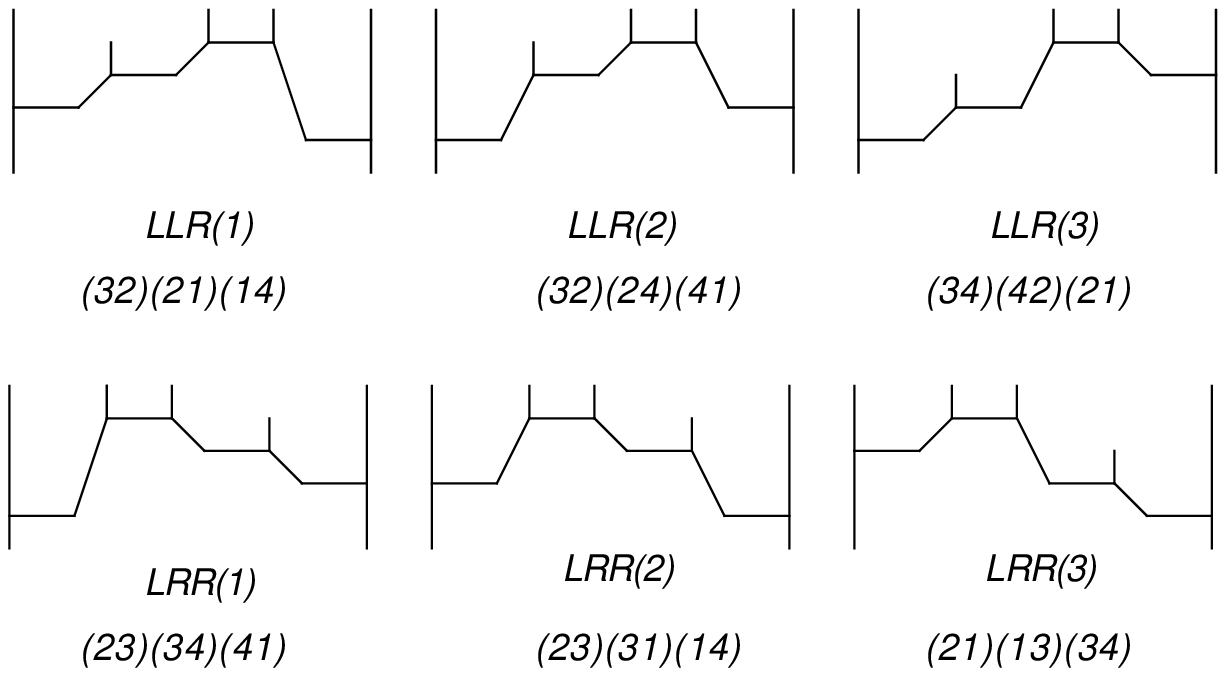}
\caption{the helicity structure of the 14 terms in Figs.\ref{f3_to_w3} - \ref{f3_to_w3triplets}}
\label{helicity}
\end{figure}
\noindent
As suggested in the discussion of A.R.White \cite{White:1976qm, Stapp:1982mq}, we note a one-to-one correspondence between the decomposition (\ref{decomp}) illustrated in Figs.\ref{f3_to_w3} - \ref{f3_to_w3triplets} and different sequences of analytic continuation in the complex helicity variables.  A connection between these two seemingly different arguments can be seen as follows. As an example, we consider the first hexagraph graph in Fig.\ref{helicity} which we redraw in Fig.\ref{helicitygraph}:
\begin{figure}[H]
\centering
\includegraphics[scale=0.8]{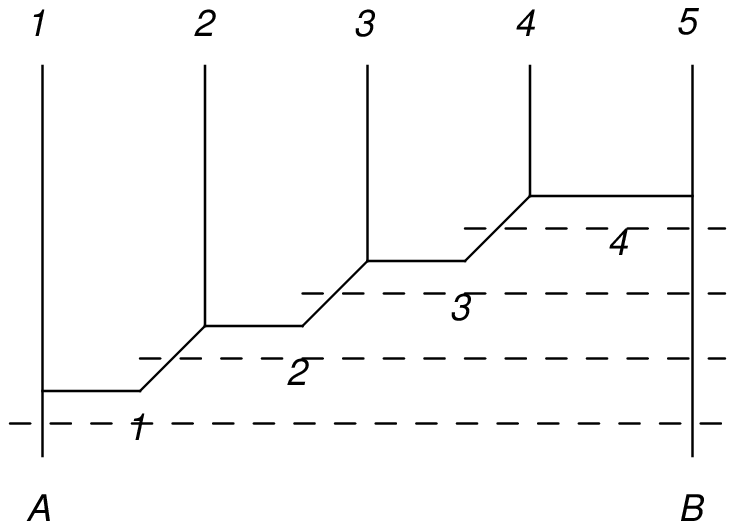}
\caption{ interpretation of the helicity graphs in Fig.4}
\label{helicitygraph}. 
\end{figure}
\noindent
We can interpret Fig.\ref{helicitygraph} as a sequence of reggeon scattering subprocesses:\\
(i) the lowest horizontal line "1" can be attributed to the reggeon exchange between the incoming particle "1" and the outing cluster "2+3+4+5"
At the same time, this exchanged reggeon can be viewed as an "incoming reggeon" for the subprocess: reggeon 1 + particle B   $\to$ cluster "2+3+4+5". Next, within this subprocess, the horizontal line "2" denotes the exchange between the incoming reggeon "1" and the cluster "3+4+5"; at the same 
time, it describes the "incoming reggeon" for the subprocess:  reggeon 2 + particle B   $\to$ particles "3+4+5".\\  
(ii) for each production vertex, it is either the left or the right reggeon which plays the role of the "incoming" reggeon: correspondingly, the vertex carries the subscript "L" or "R". Fig.\ref{helicitygraph} has only vertices of a single type "L". One easily sees that, for the assignments LRL and RLR there exist two possibilities, whereas for LLR and LRR we have even three terms.\\     
(iii) Each horizontal dashed lines denotes an "intermediate" state which belongs to a certain energy variable. In our example, the sequence of energies corresponds to $s$, $s_{2345}$, $s_{345}$, and $s_{45}=s_4$, in agreement with the 
energy discontinuity structure of the first graph of Fig.\ref{f3_to_w3}.

These hexagraphs allow for an easy understanding of the labeling "LLL" etc. As discussed before,  for the Regge poles we have two types of production vertices, denoted by $V_L$ and $V_R$. As it can be seen easily from the "hexagraphs" in fig.\ref{helicity}, each production vertex has a sloped "incoming" line and horizontal "exchange" line: a vertex $V_L$ has the incoming line on the left, the "exchange" line on the rhs. In our example of fig.\ref{helicitygraph}, all vertices are of the type "L". One easily translates this also into the other diagrams of Fig.\ref{2to4}: for a production vertex of the type "L" , energy discontinuity line enters to the left of the produced particle. In this way, each of the 14 terms has a uniquely defined sequence of subscripts. On the other hand, a given sequence "LR" may belong to several terms.

Next one writes down the corresponding multiple Sommerfeld-Watson integrals; the examples of the $2 \to 3$ and the $2 \to 4$ processes suggest a correspondence between a given hexagraph and the trigonometric denominators of the Sommerfeld-Watson integral.  Let us, once more, consider the term "LR(1)" of the $2\to4$ process. For the Regge pole contribution to this partial wave we had the following collection of trigonometric factors:
\be
\label{polefactors}
\frac{1}{\Omega_{21} \Omega_{13}\Omega_3}\,\,\, \Omega_1\Omega_2\Omega_3\,\,\, \frac{1}{\Omega_2^2} 
\,\,\,V_L(a) V_R(b)
= \frac{\Omega_1}{\Omega_2} \frac{\Omega_{23}}{\Omega_{13}} \,\frac{V_L(a)}{\Omega_{21}} \frac{V_R(b)}{\Omega_{23}}.
\ee 
Here the first group of trigonometric factors on the lhs results from the three  $1/ \sin$ factors in the Sommerfeld-Watson integrals, the second group from the three $d$ functions.  The third group arises from the production vertices, if we agree to write each production vertex in the form (\ref{leftvertex}) or (\ref{rightvertex}): for a vertex of type "L" we insert a factor $1/\sin \pi \alpha_{right}$, for a vertex of type "R" a factor $1/\sin \pi \alpha_{left}$.  In this way we obtain the trigonometric factors of Regge pole factors used in Sec. III.  
   
For the Regge cut in the $t_2$ channel we modify (\ref{polefactors}) as follows. Since there is no particle pole in the  $t_2$ channel, we leave the $d$-function of the $t_2$-channel as in (\ref{d-function}) and do make use of (\ref{Gamma-function}): this eliminates the factor $\Omega_2$ in the second group. Next, instead of the two production vertices of the particles "a" and "b" (which led to the factor $1/\Omega_2^2$), we put a new factor 
\be
\frac{\Omega_{2i}}{\Omega_i},\,\, i=1.
\ee
Here the label "i" refers to one of the two $t$-channel neighboring the $t_2$-channel containing the Regge cut: 
it is the $t$-channel to which,  in the Sommerfeld-Watson integral, the angular momentum $j_2$ couples. In our case: the $t_1$ channel. Combining all these factors we arrive at 
 \ba
F_{LR(1)}&=& \frac{1}{\Omega_{21} \Omega_{13}\Omega_3}\,\,\, \Omega_1 \Omega_3\,\,\, \frac{\Omega_{21}}{\Omega_1} 
\,\,\,W_{\Omega_2}\nonumber\\ 
&=&  \frac{W_{\Omega_2}}{\Omega_{13}}.
\ea      
This leads to the trigonometric factor of the Regge cut used in Sec. III. 

Let us generalize our rules for the Regge poles to the $2 \to 5$ case. Fig.\ref{helicity} contains those trigonometric factors which follow from the Sommerfeld-Watson integral. As an example, in the first term the notation $(43)(32)(21)$ stands for the factors
\be
\frac{1}{\Omega_{43}\Omega_{32}\Omega_{21}},
\ee
where $\Omega_{lm} = \sin \pi (\omega_l - \omega_m) = \sin \pi(j_l- j_m)$. To make contact with (\ref{polefactors}), we still need to add the last factor $1/\Omega_1$. The remaining groups in  (\ref{polefactors}) are easily generalized to the  $2 \to 5$ case. In this way one derives the trigonometric factors for the Regge poles listed in Sec. II. 

Turning to Regge cuts, the above rules for  $2\to4$ case can be used directly. As an example, we consider in Fig.\ref{helicity} the term "LLR(1)" and derive the factors for the short Regge cut in the $\omega_3$ channel. We find:
\ba
\frac{1}{\Omega_{32} \Omega_{21} \Omega_{14}\Omega_4} \Omega_1 \Omega_2 \Omega_4 \frac{1}{\Omega_2} \frac{\Omega_{32}}{\Omega_2}V_L(a) W_{\omega_3} &=& \frac{\Omega_1}{\Omega_2 \Omega_{21}\Omega_{14}} V_L(a) W_{\omega_3} \nonumber\\
&=& \frac{\Omega_1}{\Omega_2} \frac{\Omega_{24}}{\Omega_{14}} \frac{V_L(a)} {\Omega_{21}} \frac{W_{\omega_3}}
{\Omega_{24}},
\ea 
in agreement with (\ref{LLR-cut}) in Sec. II.

This completes our formulation of rules for the determination of the trigonometric factors. It is straightforward to apply these rules to the $2\to5$ amplitude and to verify all the trigonometric factors listed in Sec. II. In a forthcoming paper we will make use of these rules for the investigation of the $2\to6$ scattering amplitude. As we have said before, the rules for the Regge cuts are partly heuristic and a more systematic derivation is needed. At present, their justification comes from the results which are obtained with these rules.    

\section{Comparison with the results of the previous paper}

In our previous paper we have started from the Regge pole expressions, and we have determined the phases and the analytic expressions of the subtraction terms inside the Regge cut contribution. They were derived from the condition  that the remaining Regge pole terms are finite and conformal invariant. Following these requirements we were led to introduce, for the long Regge cuts, linear combinations of partial waves which slighty differ from the one used in the present paper.  

In order to see the connection with our present paper, we summarize a few results. For the short cut in $\omega_3$, for which we used the same partial wave decomposition  as in the present paper, we found the subtraction 
\be
\delta f_{\omega_3} = -\sin \pi(\omega_b+\omega_c) + 2 \cos \pi \omega_3 
\frac{\Omega_b \Omega_c}{\Omega_3}. 
\label{delta-f3}
\ee
This should be compared with the subtraction in $2 W_{\omega_3}$ found in (\ref{W_omega_3}): the latter  coincides with the weak coupling limit of (\ref{delta-f3}). 

For the long cut we found it convenient to use a decomposition which differs from the one used in the present paper. Let us list the phase structures of the four participating kinematic regions:
\ba 
\label{polepaper-14}
\tau_1\tau_4:  &   i e^{- \pi (\omega_2+\omega_3)} \left( e^{i\pi \omega_a} \delta f^a_{\omega_2\omega_3} + e^{i\pi \omega_c} \delta f^c_{\omega_2\omega_3} \right)
\\ 
\label{polepaper-124}
\tau_1\tau_2\tau_4: &i e^{-i\pi \omega_3} \left( e^{i\pi \omega_a} \delta f^a_{\omega_2\omega_3} + e^{i\pi \omega_c} \delta f^c_{\omega_2\omega_3} - e^{i\pi \omega_a} \delta f_{\omega_3} \right)\\
\label{polepaper-134}
\tau_1\tau_3\tau_4; & i e^{-i\pi \omega_2}\left( e^{i\pi \omega_a} \delta f^a_{\omega_2\omega_3} + e^{i\pi \omega_c} \delta f^c_{\omega_2\omega_3} - e^{i\pi \omega_c} \delta f_{\omega_2} \right)\\
\label{polepaper-1234}
\tau_1\tau_2\tau_3\tau_4: &i \left( e^{-i\pi \omega_a} \delta f^a_{\omega_2\omega_3} + e^{-i\pi \omega_c} \delta f^c_{\omega_2\omega_3}  - e^{-i\pi \omega_a} \delta f_{\omega_3} - e^{-i\pi \omega_c} \delta f_{\omega_2}\right)  .
\ea
For the subtractions we found the expressions:
\be 
\label{delta-fa}
\delta f^a_{\omega_2\omega_3} = -\frac{\Omega_c}{\Omega_{ac}} \frac{\Omega_{2a}\Omega_b\Omega_{3c}}{\Omega_2\Omega_3}
+\delta f_{\omega_3}
\ee
and
\be
\delta f^c_{\omega_2\omega_3} = -\frac{\Omega_c}{\Omega_{ac}} \frac{\Omega_{2a}\Omega_b\Omega_{3c}}{\Omega_2\Omega_3}
+\delta f_{\omega_2}.
\label{delta-fc}
\ee
With these subtractions we have shown that, after combination with the Regge pole terms, all unwanted pole contributions cancel, and for the remainder function we were left with the conformally invariant Regge pole terms:
\ba
\tau_1\tau_4:&\cos\pi \omega_{ac}\\
\tau_1\tau_2\tau_4:&- \cos\pi \omega_{ab}\\
\tau_1\tau_3\tau_4:&- \cos\pi \omega_{bc}\\
\tau_1\tau_2\tau_3\tau_4:&e^{i\pi \omega_{ba}}e^{i\pi \omega_{bc}}
\ea 
All these expressions are valid to all orders in the coupling constant. Let us now compare (\ref{polepaper-14}) -  (\ref{polepaper-1234}) with (\ref{cutpaper-14}) -  (\ref{cutpaper-1234}). We introduce
\ba
\frac{\tilde{W}_ {\omega_2\omega_3;L}}{\Omega_{32}} =\frac{W_ {\omega_2\omega_3;L}}{\Omega_{32}} +\frac{\Omega_a W_{\omega_3}}{\Omega_2}\\
\frac{\tilde{W}_ {\omega_2\omega_3;L}}{\Omega_{23}} = \frac{W_ {\omega_2\omega_3;L}}{\Omega_{32}} + \frac{W_{\omega_2}\Omega_c}{\Omega_3}
\ea
and denote the subtraction terms inside $\tilde{W}_ {\omega_2\omega_3;L}$, $\tilde{W}_ {\omega_2\omega_3;R}$ by $\delta \tilde{W}_{\omega_2\omega_3;L}$, $\delta \tilde{W}_{\omega_2\omega_3;R}$ .
Obviously we need the identity:
\be
2\left( e^{i\pi \omega_2} \frac{\delta \tilde{W}_{\omega_2\omega_3;L}}{\Omega_{32}} 
+ e^{i\pi \omega_3} \frac{\delta \tilde{W}_{\omega_2\omega_3;R}}{\Omega_{23}}
\right) = e^{i\pi \omega_a} \delta f^a_{\omega_2\omega_3} + e^{i\pi \omega_c} \delta f^c_{\omega_2\omega_3}. 
\ee
One easily verifies that this equation is fulfilled if we impose the following relations between the subtraction terms 
$\delta \tilde{W}_{\omega_2\omega_3;L}$, $\delta \tilde{W}_{\omega_2\omega_3;R}$
and $\delta f^a_{\omega_2\omega_3}$, $\delta f^c_{\omega_2\omega_3}$:
\ba
2 \delta \tilde{W}_{\omega_2\omega_3;L} &=& \Omega_{3a} \delta f^a_{\omega_2\omega_3}
+ \Omega_{3c} \delta f^c_{\omega_2\omega_3}\\
2 \delta \tilde{W}_{\omega_2\omega_3;R} &=& \Omega_{2a} \delta f^a_{\omega_2\omega_3}
+ \Omega_{2c} \delta f^c_{\omega_2\omega_3}.
\ea
Inserting ({\ref{delta-fa}) and (\ref{delta-fc}) we find:
\be
2  \delta \tilde{W}_{\omega_2\omega_3;L} = \Omega_{ba} \Omega_{a} - \Omega_{3a} \sin\pi(\omega_b+\omega_c) + 2 \cos \pi \omega_2 \frac{\Omega_{3a}\Omega_b\Omega_c}{\Omega_2 \Omega_3},   
\ee
which in the weak coupling limit becomes:
\be
2  \delta \tilde{W}_{\omega_2\omega_3;L}  \approx \pi^2 \left( (\omega_b-\omega_3)(\omega_a+\omega_c) + 2\omega_a \omega_c 
- 2 \frac{\omega_a\omega_b\omega_c }{\omega_3} \right).
\ee  
For the subtraction $2  \delta \tilde{W}_{\omega_2\omega_3;R}$ we find analogous
results, and in the weak coupling limit the combination becomes  
\be
2 \frac{\delta \tilde{W}_{\omega_2\omega_3;L} - \delta \tilde{W}_{\omega_2\omega_3;R}}{\pi \omega_{32}} =    \pi \left( -(\omega_a+\omega_c) +2  \frac{\omega_a\omega_b\omega_c }{\omega_2\omega_3} \right).
\ee
It agrees with the subtraction term obtained in  (\ref{W-long-L-R1}).

We thus have shown that the results of our previous paper are fully consistent with those of the present paper. Moreover, as discussed in Sec. VI, they can be used to generalize some of our weak coupling results beyond leading order.

%\section{Phases for the $2 \to5$ scattering amplitude} 
\begin{sidewaystable}[ht]
\begin{center}
Phases for the $2 \to5$ scattering amplitude
\end{center}
\begin{tabular}{|c|c|c|c|c|c|}
\hline
 & $a_1$ & $a_2$ & $c_1$ & $c_2$ & $c_3$\\
 \hline \\
\raisebox{-0.5\height}{\epsfig{file=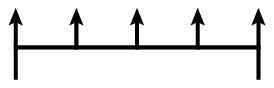}} &  $e^{-i\pi(\omega_1-\omega_2+\omega_3)}$ & $e^{-i\pi(\omega_1-\omega_2+\omega_3)}$ & $e^{-i\pi\omega_3}$ & $e^{-i\pi\omega_3}$ &  $e^{-i\pi\omega_3}$ \\
\hline\hline
\raisebox{-0.5\height}{\epsfig{file=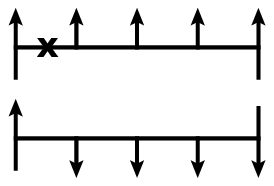}} & $e^{-i\pi(-\omega_2+\omega_3)}$ & $e^{-i\pi(-\omega_2+\omega_3)}$ & $e^{-i\pi(-\omega_1+\omega_3)}$ & $e^{-i\pi(-\omega_1+\omega_3)}$ &  $e^{-i\pi(-\omega_1+\omega_3)}$ \\
\hline\hline
\raisebox{-0.5\height}{\epsfig{file=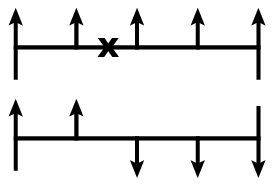}} & $e^{-i\pi(\omega_1-2\omega_2+\omega_3)}$ & $e^{-i\pi(\omega_1-2\omega_2+\omega_3)}$ & $e^{-i\pi(-\omega_2+\omega_3)}$ & $e^{-i\pi(-\omega_2+\omega_3)}$ &  $e^{-i\pi(-\omega_2+\omega_3)}$ \\
\hline\hline
\raisebox{-0.5\height}{\epsfig{file=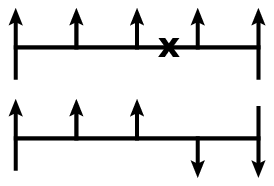}} & $e^{-i\pi(\omega_1-\omega_2)}$ & $e^{-i\pi(\omega_1-\omega_2)}$ & 1 & 1 &  1 \\
\hline\hline
\raisebox{-0.5\height}{\epsfig{file=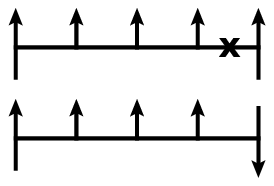}} & $e^{-i\pi(\omega_1-\omega_2+\omega_3-\omega_4)}$ & $e^{-i\pi(\omega_1-\omega_2+\omega_3-\omega_4)}$ & $e^{-i\pi(\omega_3-\omega_4)}$ & $e^{-i\pi(\omega_3-\omega_4)}$ &  $e^{-i\pi(\omega_3-\omega_4)}$ \\
\hline\hline
\raisebox{-0.5\height}{\epsfig{file=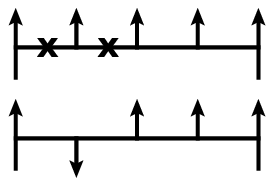}} & $e^{-i\pi(\omega_1-\omega_2+\omega_3-\omega_4)}$ & $e^{-i\pi(\omega_1-\omega_2+\omega_3-\omega_4)}$ & $e^{-i\pi(\omega_3-\omega_4)}$ & $e^{-i\pi(\omega_3-\omega_4)}$ &  $e^{-i\pi(\omega_3-\omega_4)}$ \\
\hline\hline
\raisebox{-0.5\height}{\epsfig{file=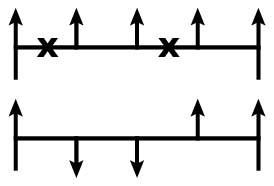}} & $e^{-i\pi\omega_2}$ & $e^{-i\pi\omega_2}$ & $e^{-i\pi\omega_1}$ & $e^{-i\pi\omega_1}$ &  $e^{-i\pi\omega_1}$ \\
\hline\hline
\raisebox{-0.5\height}{\epsfig{file=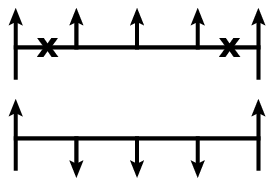}} & $e^{-i\pi\omega_3}e^{-i\pi(\omega_2-\omega_4)}$ & $e^{-i\pi\omega_3}e^{i\pi(\omega_2-\omega_4)}$ & $e^{-i\pi\omega_3}e^{-i\pi(-\omega_1+\omega_4)}$ & $e^{-i\pi\omega_3}e^{-i\pi(\omega_1-\omega_4)}$ &  $e^{-i\pi\omega_3}e^{-i\pi(\omega_1-\omega_4)}$ \\
\hline\hline
\end{tabular}
\end{sidewaystable}

\begin{sidewaystable}[ht]
\begin{tabular}{|c|c|c|c|c|c|}
\hline
 & $a_1$ & $a_2$ & $c_1$ & $c_2$ & $c_3$\\
\hline \\
\raisebox{-0.5\height}{\epsfig{file=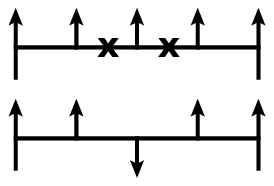}} & $e^{-i\pi\omega_1}$ & $e^{-i\pi\omega_1}$ & $e^{-i\pi\omega_2}$ & $e^{-i\pi\omega_2}$ &  $e^{-i\pi\omega_2}$ \\
\hline\hline
\raisebox{-0.5\height}{\epsfig{file=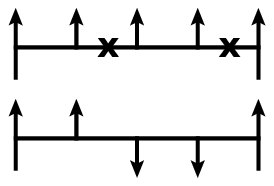}} & $e^{-i\pi\omega_3}e^{-i\pi(\omega_1-\omega_4)}$ & $e^{-i\pi\omega_3}e^{-i\pi(\omega_1-2\omega_2+\omega_4)}$ & $e^{-i\pi\omega_3}e^{-i\pi(-\omega_2+\omega_4)}$ & $e^{-i\pi\omega_3}e^{-i\pi(-\omega_2+\omega_4)}$& $e^{-i\pi\omega_3}e^{-i\pi(\omega_2-\omega_4)}$\\
\hline\hline
\raisebox{-0.5\height}{\epsfig{file=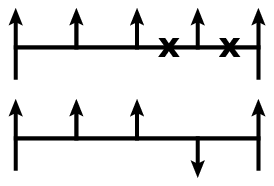}} & $e^{-i\pi(\omega_1-\omega_2+\omega_4)}$ & $e^{-i\pi(\omega_1-\omega_2+\omega_4)}$ & $e^{-i\pi\omega_4}$ & $e^{-i\pi\omega_4}$ &  $e^{-i\pi\omega_4}$ \\
\hline\hline
\raisebox{-0.5\height}{\epsfig{file=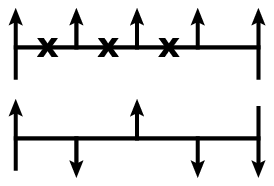}} & 1 & 1 & $e^{-i\pi(\omega_2-\omega_1)}$ & $e^{-i\pi(\omega_2-\omega_1)}$ &  $e^{-i\pi(\omega_2-\omega_1)}$ \\
\hline\hline
\raisebox{-0.5\height}{\epsfig{file=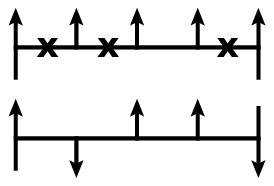}} & $e^{-i\pi(\omega_3-\omega_4)}$ & $e^{-i\pi(\omega_3-\omega_4)}$ &  $e^{-i\pi\omega_3}e^{-i\pi(\omega_1-\omega_2-\omega_4)}$ &$e^{-i\pi\omega_3}e^{-i\pi(-\omega_1-\omega_2+\omega_4)}$ &$e^{-i\pi\omega_3}e^{-i\pi(-\omega_1+\omega_2-\omega_4)}$\\
\hline\hline
\raisebox{-0.5\height}{\epsfig{file=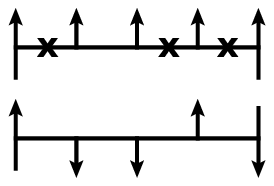}} & $e^{i\pi(\omega_2-\omega_4)}$ & $e^{-i\pi(\omega_2-\omega_4)}$ &  $e^{-i\pi(\omega_1-\omega_4)}$ &$e^{i\pi(\omega_1-\omega_4)}$ &$e^{i\pi(\omega_1-\omega_4)}$\\
\hline\hline
\raisebox{-0.5\height}{\epsfig{file=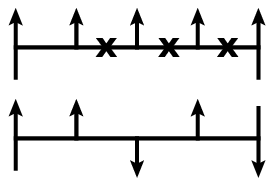}} & $e^{-i\pi(\omega_1-2\omega_2+\omega_4)}$ & $e^{-i\pi(\omega_1-\omega_4)}$ &  $e^{-i\pi(\omega_2-\omega_4)}$ &$e^{-i\pi(\omega_2-\omega_4)}$ &$e^{i\pi(\omega_2-\omega_4)}$\\
\hline\hline
\raisebox{-0.5\height}{\epsfig{file=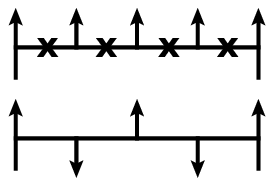}} & $e^{-i\pi\omega_4}$ & $e^{-i\pi\omega_4}$ &  $e^{-i\pi(-\omega_1+\omega_2+\omega_4)}$ &$e^{-i\pi(\omega_1+\omega_2-\omega_4)}$ &$e^{-i\pi(\omega_1-\omega_2+\omega_4)}$\\
\hline\hline
\end{tabular}
\end{sidewaystable}


\begin{thebibliography}{99}
\bibitem{Bartels:2013jna}
  J.~Bartels, A.~Kormilitzin and L.~Lipatov,
  %``Analytic structure of the $n=7$ scattering amplitude in $\mathcal{N}=4$ SYM theory at multi-Regge kinematics: Conformal Regge pole contribution,''
  Phys.\ Rev.\ D {\bf 89} (2014) 065002
  [arXiv:1311.2061 [hep-th]].

\bibitem{Bern:2005iz}
  Z.~Bern, L.~J.~Dixon and V.~A.~Smirnov,
  %``Iteration of planar amplitudes in maximally supersymmetric Yang-Mills theory at three loops and beyond,''
  Phys.\ Rev.\ D {\bf 72} (2005) 085001
  [hep-th/0505205].
  

  
 \bibitem{Bartels:2008ce}
  J.~Bartels, L.~N.~Lipatov and A.~Sabio Vera,
  %``BFKL Pomeron, Reggeized gluons and Bern-Dixon-Smirnov amplitudes,''
  Phys.\ Rev.\ D {\bf 80} (2009) 045002
  [arXiv:0802.2065 [hep-th]].
  %%CITATION = ARXIV:0802.2065;%%
\bibitem{Bartels:2008sc}
  J.~Bartels, L.~N.~Lipatov and A.~Sabio Vera,
  %``N=4 supersymmetric Yang Mills scattering amplitudes at high energies: The Regge cut contribution,''
  Eur.\ Phys.\ J.\ C {\bf 65} (2010) 587
  [arXiv:0807.0894 [hep-th]].
%\cite{Alday:2007he}
\bibitem{Alday:2007he}
  L.~F.~Alday and J.~Maldacena,
  %``Comments on gluon scattering amplitudes via AdS/CFT,''
  JHEP {\bf 0711} (2007) 068
  [arXiv:0710.1060 [hep-th]].
  %%CITATION = ARXIV:0710.1060;%%
  
  %%CITATION = ARXIV:0807.0894;%%  

  %%CITATION = ARXIV:1008.1015;%%
 %\cite{Goncharov:2010jf}
\bibitem{Goncharov:2010jf}
  A.~B.~Goncharov, M.~Spradlin, C.~Vergu and A.~Volovich,
  %``Classical Polylogarithms for Amplitudes and Wilson Loops,''
  Phys.\ Rev.\ Lett.\  {\bf 105} (2010) 151605
  [arXiv:1006.5703 [hep-th]].
  %\cite{DelDuca:2009au}
\bibitem{DelDuca:2009au}
  V.~Del Duca, C.~Duhr and V.~A.~Smirnov,
  %``An Analytic Result for the Two-Loop Hexagon Wilson Loop in N = 4 SYM,''
  JHEP {\bf 1003} (2010) 099
  [arXiv:0911.5332 [hep-ph]]. 
%\cite{DelDuca:2010zg}

\bibitem{DelDuca:2010zg}
  V.~Del Duca, C.~Duhr and V.~A.~Smirnov,
  %``The Two-Loop Hexagon Wilson Loop in N = 4 SYM,''
  JHEP {\bf 1005} (2010) 084
  [arXiv:1003.1702 [hep-th]].
  
%\cite{Dixon:2011pw}
\bibitem{Dixon:2011pw}
  L.~J.~Dixon, J.~M.~Drummond and J.~M.~Henn,
  %``Bootstrapping the three-loop hexagon,''
  JHEP {\bf 1111} (2011) 023
  [arXiv:1108.4461 [hep-th]].
%\cite{Dixon:2011nj}
\bibitem{Dixon:2011nj}
  L.~J.~Dixon, J.~M.~Drummond and J.~M.~Henn,
  %``Analytic result for the two-loop six-point NMHV amplitude in N=4 super Yang-Mills theory,''
  JHEP {\bf 1201} (2012) 024
  [arXiv:1111.1704 [hep-th]].
%\cite{Dixon:2012yy}
\bibitem{Dixon:2012yy}
  L.~J.~Dixon, C.~Duhr and J.~Pennington,
  %``Single-valued harmonic polylogarithms and the multi-Regge limit,''
  JHEP {\bf 1210} (2012) 074
  [arXiv:1207.0186 [hep-th]].
%\cite{Pennington:2012zj}
\bibitem{Pennington:2012zj}
  J.~Pennington,
  %``The six-point remainder function to all loop orders in the multi-Regge limit,''
  JHEP {\bf 1301} (2013) 059
  [arXiv:1209.5357 [hep-th]].  
%\cite{Dixon:2013eka}
\bibitem{Dixon:2013eka}
  L.~J.~Dixon, J.~M.~Drummond, M.~von Hippel and J.~Pennington,
  %``Hexagon functions and the three-loop remainder function,''
  arXiv:1308.2276 [hep-th]. 
\bibitem{Lipstein:2013xra}
  A.~E.~Lipstein and L.~Mason,
  %``From dlogs to dilogs; the super Yang-Mills MHV amplitude revisited,''
  arXiv:1307.1443 [hep-th].
%\cite{Golden:2013xva}
\bibitem{Golden:2013xva}
  J.~Golden, A.~B.~Goncharov, M.~Spradlin, C.~Vergu and A.~Volovich,
  %``Motivic Amplitudes and Cluster Coordinates,''
  arXiv:1305.1617 [hep-th].    
%\cite{DelDuca:2013lma}   
\bibitem{DelDuca:2013lma}
  V.~Del Duca, L.~J.~Dixon, C.~Duhr and J.~Pennington,
  %``The BFKL equation, Mueller-Navelet jets and single-valued harmonic polylogarithms,''
  arXiv:1309.6647 [hep-ph].
%\cite{Dixon:2014xca}
\bibitem{Dixon:2014xca}
  L.~J.~Dixon, J.~M.~Drummond, C.~Duhr, M.~von Hippel and J.~Pennington,
  %``Bootstrapping six-gluon scattering in planar ${\cal N}=4$ super-Yang-Mills theory,''
  arXiv:1407.4724 [hep-th].
  %%CITATION = ARXIV:1407.4724;%%
%\cite{Dixon:2014iba}
\bibitem{Dixon:2014iba}
  L.~J.~Dixon and M.~von Hippel,
  %``Bootstrapping an NMHV amplitude through three loops,''
  arXiv:1408.1505 [hep-th].
  %%CITATION = ARXIV:1408.1505;%%
%\cite{Dixon:2014voa}
\bibitem{Dixon:2014voa}
  L.~J.~Dixon, J.~M.~Drummond, C.~Duhr and J.~Pennington,
  %``The four-loop remainder function and multi-Regge behavior at NNLLA in planar N = 4 super-Yang-Mills theory,''
  JHEP {\bf 1406} (2014) 116
  [arXiv:1402.3300 [hep-th]].
  %%CITATION = ARXIV:1402.3300;%%
 
%\cite{Golden:2014xqf}
\bibitem{Golden:2014xqf} 
  J.~Golden and M.~Spradlin,
  %``An analytic result for the two-loop seven-point MHV amplitude in $ \mathcal{N} $ = 4 SYM,''
  JHEP {\bf 1408}, 154 (2014)
  [arXiv:1406.2055 [hep-th]].
  %%CITATION = ARXIV:1406.2055;%%
%\cite{Golden:2014xqa}
\bibitem{Golden:2014xqa} 
  J.~Golden, M.~F.~Paulos, M.~Spradlin and A.~Volovich,
  %``Cluster Polylogarithms for Scattering Amplitudes,''
  arXiv:1401.6446 [hep-th].
  %%CITATION = ARXIV:1401.6446;%%

%\cite{Basso:2014koa}
\bibitem{Basso:2014koa}
  B.~Basso, A.~Sever and P.~Vieira,
  %``Space-time S-matrix and Flux-tube S-matrix III. The two-particle contributions,''
  JHEP {\bf 1408} (2014) 085
  [arXiv:1402.3307 [hep-th]].
%\cite{Basso:2014nra}
\bibitem{Basso:2014nra}
  B.~Basso, A.~Sever and P.~Vieira,
  %``Space-time S-matrix and Flux-tube S-matrix IV. Gluons and Fusion,''
  arXiv:1407.1736 [hep-th].
  %%CITATION = ARXIV:1407.1736;%%

  

\bibitem{Lipatov:2010qf}
  L.~N.~Lipatov,
  %``Analytic properties of high energy production amplitudes in N=4 SUSY,''
  Theor.\ Math.\ Phys.\  {\bf 170} (2012) 166
  [arXiv:1008.1015 [hep-th]].


%\cite{Lipatov:2010qf}



%\cite{Brower:1974yv}
\bibitem{Brower:1974yv}
  R.~C.~Brower, C.~E.~DeTar and J.~H.~Weis,
  %``Regge Theory for Multiparticle Amplitudes,''
  Phys.\ Rept.\  {\bf 14} (1974) 257.
  
\bibitem{Weis:1972ir}
  J.~H.~Weis,
  %``Factorization of multi-regge amplitudes,''
  Phys.\ Rev.\ D {\bf 4} (1971) 1777.
  %%CITATION = PHRVA,D4,1777;%%

\bibitem{White:1976qm}
  A.~R.~White,
  %``The Analytic Foundations of Regge Theory,''
  CERN-TH-2136.
 
 \bibitem{White:1990ch}
  A.~R.~White,
  %``Analytic multi-Regge theory and the pomeron in QCD: Part 1,''
  Int.\ J.\ Mod.\ Phys.\ A {\bf 6} (1991) 1859.  

\bibitem{Stapp:1982mq}
  H.~P.~Stapp and A.~R.~White,
  %``An Asymptotic Dispersion Relation for the Six Particle Amplitude,''
  Phys.\ Rev.\ D {\bf 26} (1982) 2145.
\bibitem{Weis:1972tn}
  J.~H.~Weis,
  %``Factorization of multi-regge amplitudes. ii,''
  Phys.\ Rev.\ D {\bf 5} (1972) 1043.
  %%CITATION = PHRVA,D5,1043;%% 
 
 %\cite{Mandelstam:1963cw}
\bibitem{Mandelstam:1963cw}
  S.~Mandelstam,
  %``Cuts in the Angular Momentum Plane. 2,''
  Nuovo Cim.\  {\bf 30} (1963) 1148. 

%\cite{Bartels:2010ej}
\bibitem{Bartels:2010ej}
  J.~Bartels, J.~Kotanski and V.~Schomerus,
  %``Excited Hexagon Wilson Loops for Strongly Coupled N=4 SYM,''
  JHEP {\bf 1101} (2011) 096
  [arXiv:1009.3938 [hep-th]].
\bibitem{Fadin:2011we}
  V.~S.~Fadin and L.~N.~Lipatov,
  %``BFKL equation for the adjoint representation of the gauge group in the next-to-leading approximation at N=4 SUSY,''
  Phys.\ Lett.\ B {\bf 706} (2012) 470
  [arXiv:1111.0782 [hep-th]].  

\bibitem{Bartels:2011ge}
  J.~Bartels, A.~Kormilitzin, L.~N.~Lipatov and A.~Prygarin,
  %``BFKL approach and $2 \to 5$ maximally helicity violating amplitude in ${\cal N}=4$ super-Yang-Mills theory,''
  Phys.\ Rev.\ D {\bf 86} (2012) 065026
  [arXiv:1112.6366 [hep-th]].
  %%CITATION = ARXIV:1112.6366;%%
%\cite{Fadin:1993wh}
\bibitem{Fadin:1993wh}
  V.~S.~Fadin and L.~N.~Lipatov,
  %``Radiative corrections to QCD scattering amplitudes in a multi - Regge kinematics,''
  Nucl.\ Phys.\ B {\bf 406} (1993) 259.

%\cite{Fadin:2014gra}
\bibitem{Fadin:2014gra}
  V.~S.~Fadin and R.~Fiore,
  %``Impact factors for Reggeon-gluon transition in N = 4 SYM with large number of colours,''
  arXiv:1402.5260 [hep-th].  
 
 %\cite{Bartels:2003jq}
\bibitem{Bartels:2003jq}
  J.~Bartels, V.~S.~Fadin and R.~Fiore,
  %``The Bootstrap conditions for the gluon reggeization,''
  Nucl.\ Phys.\ B {\bf 672} (2003) 329
  [hep-ph/0307076]. 
  
 %\cite{Fadin:2002hz}
\bibitem{Fadin:2002hz}
  V.~S.~Fadin and A.~Papa,
  %``A Proof of fulfillment of the strong bootstrap condition,''
  Nucl.\ Phys.\ B {\bf 640} (2002) 309
  [hep-ph/0206079].


\bibitem{Bartels:2013dja}
  J.~Bartels, J.~Kotanski, V.~Schomerus and M.~Sprenger,
  %``The Excited Hexagon Reloaded,''
  arXiv:1311.1512 [hep-th].
  %%CITATION = ARXIV:1311.1512;%%
  
%\cite{Bartels:2012gq}
\bibitem{Bartels:2012gq}
  J.~Bartels, V.~Schomerus and M.~Sprenger,
  %``Multi-Regge Limit of the n-Gluon Bubble Ansatz,''
  JHEP {\bf 1211} (2012) 145
  [arXiv:1207.4204 [hep-th]].
  

\bibitem{Bartels:2014ppa}
  J.~Bartels, V.~Schomerus and M.~Sprenger,
  %``Heptagon Amplitude in the Multi-Regge Regime,''
  JHEP {\bf 1410} (2014) 67
  [arXiv:1405.3658 [hep-th]].

\bibitem{BSS}
J.~Bartels, V.~Schomerus and M.~Sprenger, to appear

\bibitem{Drummond:1969ft}
  I.~T.~Drummond, P.~V.~Landshoff and W.~J.~Zakrzewski,
  %``The two-reggeon/particle coupling,''
  Nucl.\ Phys.\ B {\bf 11} (1969) 383.
\end{thebibliography}
 \end{document}